\documentclass[fleqn,usenatbib]{mnras}
\usepackage[T1]{fontenc}
\usepackage{comment}
\DeclareRobustCommand{\VAN}[3]{#2}
\let\VANthebibliography\thebibliography
\def\thebibliography{\DeclareRobustCommand{\VAN}[3]{##3}\VANthebibliography}

\usepackage{graphicx}	
\usepackage{amsmath}	
\usepackage{amssymb}	
\usepackage{subfigure}
\usepackage{float}

\title[Diffusive rotational instabilities]{ Local stability of differential rotation in magnetised radiation zones and the solar tachocline}

\author[R. W. Dymott et al.]{
R.W. Dymott$^{1}$\thanks{E-mail: mmrwd@leeds.ac.uk},
A. J. Barker$^{1}$\thanks{E-mail: A.J.Barker@leeds.ac.uk},
C.A. Jones$^{1}$
and S.M. Tobias$^{1}$
\\
$^{1}$Department of Applied Mathematics, School of Mathematics, University of Leeds, Leeds, LS2 9JT, UK\\
}

\date{Accepted Nov 6 666. Received Nov 6 666; in original form Nov 6 666}
\pubyear{2024}

\begin{document}
\label{firstpage}
\pagerange{\pageref{firstpage}--\pageref{lastpage}}
\maketitle

\begin{abstract}
We study local magnetohydrodynamical (MHD) instabilities of differential rotation in magnetised, stably-stratified regions of stars and planets using a Cartesian Boussinesq model. We consider arbitrary latitudes and general shears (with gravity direction misaligned from this by an angle $\phi$), to model radial ($\phi=0$), latitudinal ($\phi=\pm 90^\circ$), and mixed differential rotations, and study both non-diffusive (including magnetorotational, MRI, and Solberg-H\o iland instabilities) and diffusive instabilities (including Goldreich-Schubert-Fricke, GSF, and MRI with diffusion). These instabilities could drive turbulent transport and mixing in radiative regions, including the solar tachocline and the cores of red giant stars, but their  dynamics are incompletely understood. We revisit linear axisymmetric instabilities with and without diffusion and analyse their properties in the presence of magnetic fields, including deriving stability criteria and computing growth rates, wavevectors and energetics, both analytically and numerically. We present a more comprehensive analysis of axisymmetric local instabilities than prior work, exploring arbitrary differential rotations and diffusive processes. The presence of a magnetic field leads to stability criteria depending upon angular velocity rather than angular momentum gradients. We find MRI operates for much weaker differential rotations than the hydrodynamic GSF instability, and that it typically prefers much larger lengthscales, while the GSF instability is impeded by realistic strength magnetic fields. We anticipate MRI to be more important for turbulent transport in the solar tachocline than the GSF instability when $\phi>0$ in the northern (and vice versa in the southern) hemisphere, though the latter could operate just below the convection zone when MRI is absent for $\phi<0$.
\end{abstract} 



\section{Introduction}

The evolution of angular momentum (AM) and its profile in a star is one of the most fundamental -- yet poorly understood -- aspects of stellar evolution \citep[e.g.][]{MM2000,2009Book}. Many unsolved problems in solar physics, including the development, and maintenance of, the differentially-rotating solar convection zone, tachocline, and solar dynamo, as well as those of other stars, all likely require a better understanding of the mechanisms that shape the AM transport and hence profile of star throughout its life-cycle. Current stellar evolution codes are too simplistic and do not correctly model the evolution of AM, as is evident from red and sub-giant stars for example, whose core-envelope differential rotations inferred from asteroseismology are not well explained by existing stellar evolution models \citep[e.g.][]{aerts2019}.

Some of the most important proposed mechanisms for AM transport and turbulent mixing in stars are magnetohydrodynamical (MHD) instabilities, which have not yet been fully explored \citep[e.g.][]{Zahn1974,MM2000} and their nonlinear evolution is particularly poorly understood. Here we analyse local MHD instabilities in stellar radiative zones, focussing on the stability properties of stably-stratified, differentially-rotating and magnetised shear flows -- relevant, for example, to modelling the solar tachocline or the cores of red giant stars. Radiative zones are often thought of as quiescent regions, but they have already been shown to exhibit a range of local and global MHD instabilities that transport AM and lead to turbulent mixing. Examples of relevant instabilities in such regions include the hydrodynamic Goldreich-Schuburt-Fricke (GSF) instability, a double-diffusive centrifugal instability of differential rotation enabled by thermal diffusion \citep{GSF,Fricke1968,knobloch.et.al.1982,Rashid2008,barker2019,barker2020,Park2020,Park2021,Dymott2023,Bindesh2024}, and the more violent instabilities excited when the Solberg-H\o iland (SH) criteria for non-diffusive instabilities are violated \citep[][]{Solberg1936,Hoiland1941}. Not only do these linear instabilities lead to turbulence with enhanced transport properties in their nonlinear evolution, they also exhibit the emergence of long-term anisotropic quasi-stable structures such as `zonal jets' (or layering in the AM) \citep{barker2020,Dymott2023}. The complex anisotropic nature of the long-term evolution of many of these instabilities suggests that modelling them in one-dimensional stellar evolution models simply as a one-dimensional diffusive process is probably insufficient. Indeed, in certain cases turbulent transport is known to be anti-diffusive.

The presence of even a weak magnetic field is known to drastically modify the stability of differentially-rotating flows \citep[e.g.][]{C1961,Acheson1978,BH1991}. Stability criteria with magnetic fields tend to involve angular velocity gradients -- which typically require much weaker differential rotations to predict instability -- rather than the angular momentum gradients without fields; this is because the field can act as a tether between fluid particles and allow them to exchange angular momentum. The magneto-rotational instability (MRI) is one such manifestation when a weak magnetic field is introduced into a differentially-rotating flow \citep[e.g.][]{C1961,Acheson1978,BH1991,Balbus1994,Balbus1995,Spruit1999,Ogilvie2007,Balbus2009,Oishi2020,V2024}. This can operate and drive turbulence even in many hydrodynamically stable flows. Its operation in stably-stratified stellar interiors (i.e.~radiation zones) in the presence of diffusive processes has been studied in some prior works \citep{Menou2004,Menou2006,ParfreyMenou2007,Guilet2015,Caleo2016,Caleo2016a}, but much remains to be explored of its linear properties, and especially its nonlinear evolution in stars. \cite{Guilet2015} performed linear analysis and numerical simulations of the MRI in a local stably-stratified model of a proto-neutron star (with extra neutrino cooling). Our approach is broadly similar to theirs but we will study arbitrary local differential rotations. Global simulations in spherical geometry of the MRI (or Tayler instability, which is a current-driven instability that is also present in these) in stellar radiative zones have also been performed \citep{Jouve2015,Gaurat2015,Meduri2019,Jouve2020}, though these kinds of studies may not adequately capture all of the possible local instabilities. We choose to adopt a local model here, partly for simplicity and because such models are appropriate for studying small-scale MHD instabilities, and also because they can explore more realistic parameter regimes with numerical simulations (particularly with regards to smaller diffusivities) than global models would allow in nonlinear regimes.

Here we introduce magnetic fields to build directly upon \citet[][hereafter paper 1]{barker2019}, \citet[][hereafter paper 2]{barker2020} and \citet[][hereafter paper 3]{Dymott2023}{}{} that studied hydrodynamical instabilities in a local Cartesian representation of a small patch of a stably-stratified, differentially-rotating stellar or planetary radiation zone. A global ``shellular'' (radial) differential rotation varying only with spherical radius was considered at the equator in paper 1 \citep[and an axisymmetric turbulence closure model was developed and verified for this case by][]{Bindesh2024}, and at a general latitude in paper 2. In paper 3, we generalised the model to consider an arbitrary differential rotation profile, which varies with both radius and latitude. Here we incorporate a poloidal magnetic field into this more general model. Following a similar approach, we perform an axisymmetric linear stability analysis here, which we will follow with complementary three-dimensional nonlinear numerical simulations (that can consider the effects of more general field orientations) in future work. Our primary goals are to understand the properties of the GSF instability in the magnetic system, as well as the operation the MRI, and to determine their potential roles in angular momentum transport, chemical mixing, and dynamo generation. Our linear study is related to the one undertaken by \citet{LatterPap2018} for the Vertical Shear Instability in astrophysical discs \citep[e.g.][]{UB1998,Nelson2013,BL2015}, which is the name used for the GSF instability in that context.

The goal of this paper is to gain insights into how the presence of a locally uniform magnetic field affects the linear
properties of local instabilities of differential rotation in stellar and planetary radiative zones. We do this by investigating the axisymmetric linear stability of the system (that we define in \S~\ref{model}), both analytically and numerically in \S~\ref{LinearTheory}--\ref{energy}. We determine how the properties of the unstable modes depend on magnetic field strength $B_0$ and magnetic Prandtl number $\mathrm{Pm}=\nu/\eta$ (the ratio of kinematic viscosity $\nu$ to ohmic diffusivity $\eta$). We will analyse the energetics of the various instabilities in our model and derive several new results before applying them to the solar tachocline and red giant stars in \S~\ref{Applications}.

\section{Local Cartesian model}\label{model}

\subsection{The model and governing equations}
\label{sec:maths}

\begin{figure}
\centering
\subfigure{\includegraphics[width= 0.48\textwidth]{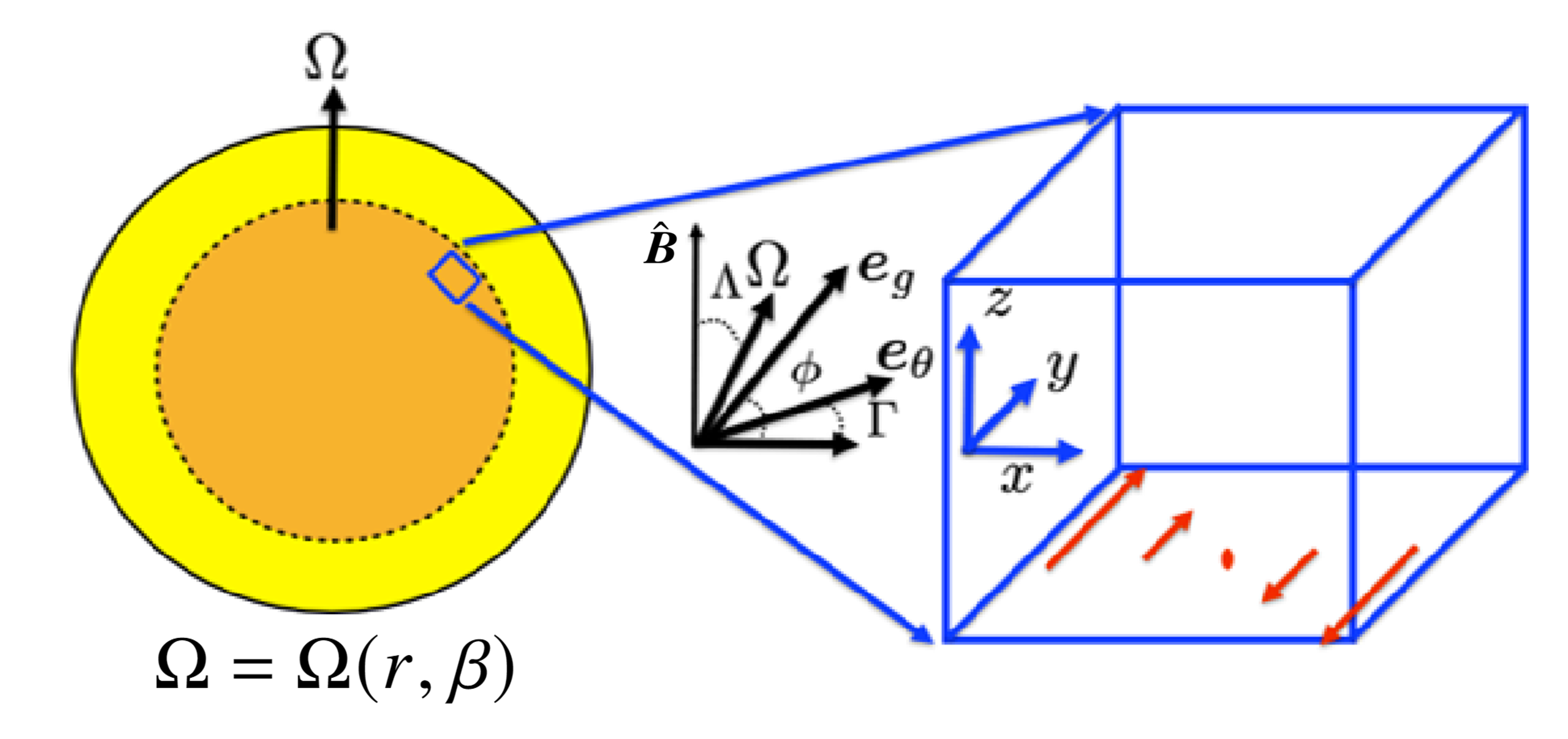}}
\subfigure{\includegraphics[width= 0.45\textwidth]{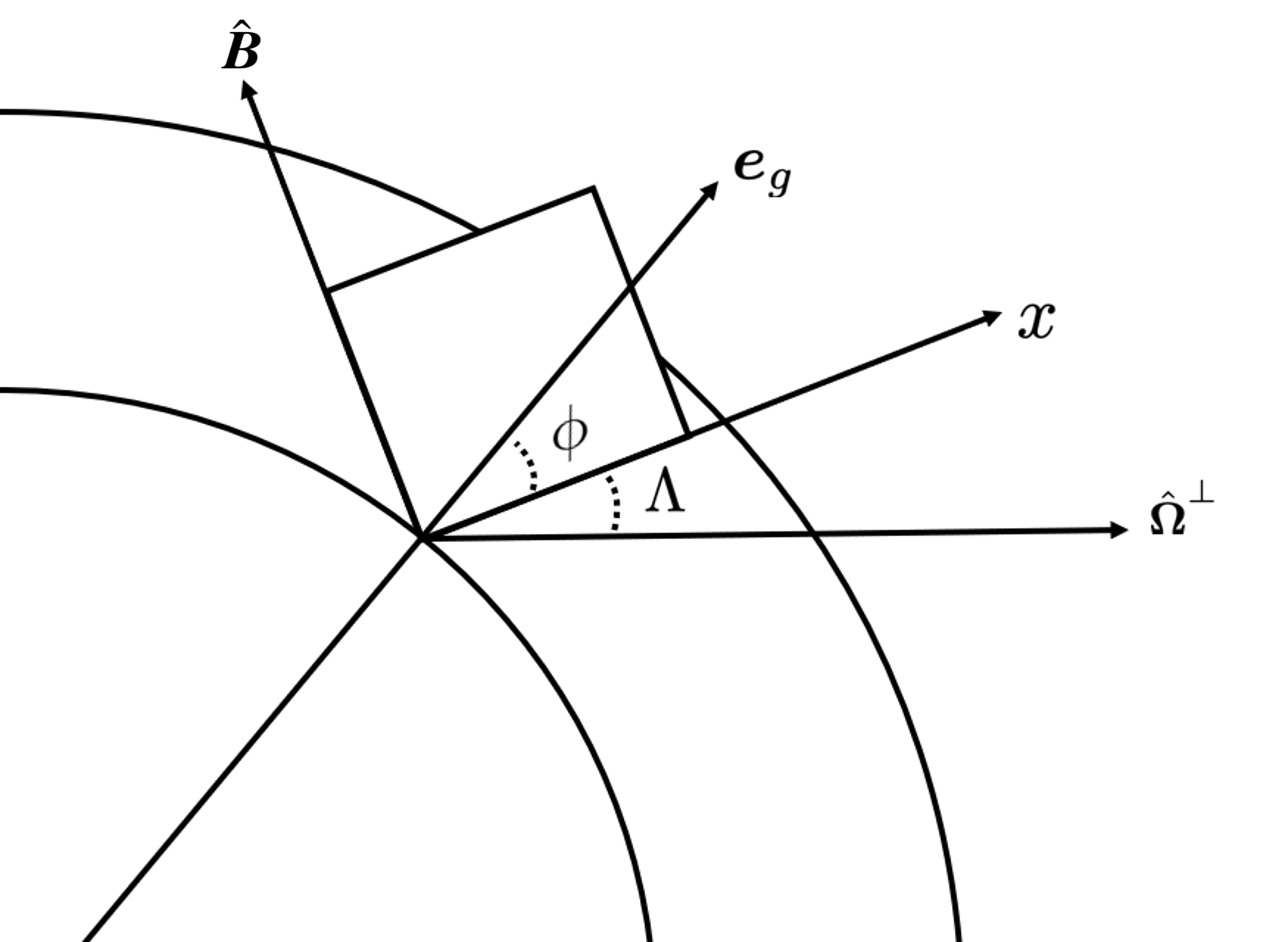}}
\caption{Illustration of the various vectors and corresponding angles in the $(x,z)$-plane. The cylindrical radial direction (along the equator) is $\hat{\boldsymbol{\Omega}}^\perp$, and the rotation axis is $\hat{\boldsymbol{\Omega}}$. The local radial direction is (approximately) along the effective gravity direction $\boldsymbol{e}_g$, which is misaligned with respect to the $x$-direction when $\phi$ is nonzero. The magnetic field in linear theory is always along $z$ (therefore perpendicular to the shear in $x$), which is the only direction in the meridional plane in which an equilibrium exists.}
    \label{fig:Boxmodel}
\end{figure}

We follow papers 1--3 and employ a local Cartesian model to study small-scale instabilities of differential rotation in a stably-stratified region of a star or planet. We use coordinates $(x,y,z)$, where $y$ is the local azimuthal coordinate, and $x$ and $z$ are two coordinates in the meridional plane. We adopt the Bousinessq approximation \citep{Spiegel1960}, which is expected to be valid for the local instabilities we study \citep[see e.g.][for justification regarding studying GSF modes]{barker2019}. The differential rotation is represented by a linear shear flow $\boldsymbol{U}_0 = -\mathcal{S}x\boldsymbol{e}_y$, which in general varies with both spherical radius $r$ and latitude $\beta$, and we have defined $x$ to be aligned with the axis of variation of $\boldsymbol{U}_0$. $\mathcal{S}$ is the constant value locally of $-\varpi|\nabla\Omega(r,\beta)|$ (where $\Omega(r,\beta)$ is the angular velocity), and $\varpi$ is the distance from the axis of rotation (cylindrical radius). The local effective gravity vector $\boldsymbol{e}_g=(\cos \phi,0,\sin\phi)$ defines the angle $\phi$, and the rotation axis lies along the vector $\hat{\boldsymbol{\Omega}} = (\sin{\Lambda},0,\cos{\Lambda})$, thereby defining the angle $\Lambda$.  We note that both of these are defined locally with respect to $x$. The latitude angle is then given by $\beta = \Lambda+\phi$, which measures the angle between the equator $\hat{\boldsymbol{\Omega}}^{\perp}= (\cos{\Lambda},0,-\sin{\Lambda})$, and the spherical radial direction $\boldsymbol{e}_g$. Our model is illustrated in Fig.~\ref{fig:Boxmodel}, which shows the various angles involved.

We build upon the hydrodynamical studies of papers 1-3 by introducing a uniform static background poloidal magnetic field $\boldsymbol{B}_0=B_0\hat{\boldsymbol{B}}$ that is in equilibrium, satisfying the local analogue of Ferraro's law of isorotation \citep{Ferraro1937}. For this flow to be in equilibrium in the meridional/poloidal $(x,z)$ plane, it must lie along $z$ with\footnote{The MHD equations are invariant under the transformation $\boldsymbol{B}\to -\boldsymbol{B}$, so there is no loss of generality in considering $B_0\geq 0$.} $\hat{\boldsymbol{B}}=(0,0,1)$, being always perpendicular to variation of the shear flow $\boldsymbol{U}_0$ locally. This permits a well-defined equilibrium state even if it may complicate interpretation of our model because the field is not purely radial or horizontal, depending on the value of $\phi$. The field is radial if $\phi=\pm 90^\circ$ and it is latitudinal if $\phi=0^\circ, \pm 180^\circ$. We do not consider toroidal/azimuthal fields in our local analysis, which are typically thought to be dominant in the solar tachocline, because they play no role for linear incompressible axisymmetric perturbations. A toroidal field would affect non-axisymmetric perturbations \citep[e.g.][]{Ogilvie1996} but analysing those (and their non-modal growth) is less straightforward, and it is likely that axisymmetric instabilities are the fastest growing ones in any case \citep[e.g.][]{LatterPap2018}. Since we have adopted a local Cartesian model, as appropriate to explore small-scale instabilities in stellar radiative zones, we do not capture the effects of azimuthal magnetic fields on axisymmetric modes via hoop stresses, leading to azimuthal magnetorotational instability \citep{HR2005,KS2010,Guseva2017,M2019,Meduri2024}. This could be important on larger length-scales than those that we consider but it would require us to adopt a global model. An initially purely toroidal field can play a role nonlinearly in local models even if the linear instability is axisymmetric however, so future nonlinear simulations should explore these fields, also because of their possible role in driving non-axisymmetric instabilities.

The incompressible MHD equations governing perturbations to the shear flow $\boldsymbol{U}_0$ and background stable stratification in the Boussinesq approximation, in the frame rotating at the rate $\boldsymbol{\Omega}$, are
\begin{align}
\label{eq1}
   & D\boldsymbol{u}+2\boldsymbol{\Omega}\times\boldsymbol{u}+\boldsymbol{u}\cdot\nabla{\boldsymbol{U}_0} = -\nabla{p}+\theta\boldsymbol{e}_g+\boldsymbol{B} \cdot {\nabla{\boldsymbol{B}}}+\nu\nabla^2{\boldsymbol{u}}, \\  
\label{eq2}
    & D{\theta}+\mathcal{N}^2 \boldsymbol{u} \cdot \boldsymbol{e}_\theta = \kappa\nabla^2{\theta}, \\
\label{eq3}
    & {D}\boldsymbol{B} = \boldsymbol{B}\cdot\nabla \boldsymbol{u} + \boldsymbol{B}\cdot\nabla \boldsymbol{U}_{0} + \eta {\nabla}^2 \boldsymbol{B}, \\
\label{eq4}
    & \nabla\cdot{\boldsymbol{B}}=0, \\
\label{eq5}
    & \nabla\cdot{\boldsymbol{u}}=0, \\
\label{eq6}
    & D \equiv \partial_{t}+\boldsymbol{u}\cdot\nabla+ \boldsymbol{U}_0\cdot\nabla.
\end{align}
Here $\boldsymbol{u}$ is a velocity perturbation and $\boldsymbol{B}$ is the total magnetic field. We define a temperature perturbation $\theta$ having units of acceleration and related to the standard temperature perturbation $\tilde{T}$ via $\theta = \alpha{g}\tilde{T}$, where $\alpha$ is the thermal expansion coefficient and $g$ is the local gravitational acceleration. We use Alfv\'{e}n speed units for the magnetic field, such that the dimensional magnetic field is $\boldsymbol{B}/\sqrt{\mu_0 \rho}$, where $\rho$ is the constant reference density that we henceforth set to unity and $\mu_0$ is the vacuum permeability. Magnetic pressure is contained within the total pressure $p$. We consider constant kinematic viscosity $\nu$, thermal diffusivity $\kappa$ and ohmic diffusivity $\eta$. For reference, the basic state satisfies:
\begin{align}
\label{BS1}
    2\boldsymbol{\Omega}\times\boldsymbol{U}_0 &= -\nabla p_0 + \alpha g T \boldsymbol{e}_g, \\
    0&=\kappa \nabla^2 T, 
\label{BS2}
\end{align}
since $\partial_t \boldsymbol{U}_0=\boldsymbol{U}_0\cdot \nabla \boldsymbol{U}_0=\nabla^2\boldsymbol{U}_0=\boldsymbol{B}_0\cdot \nabla \boldsymbol{B}_0=\boldsymbol{0}$ and the equivalents of Eqs.~\ref{eq3}--\ref{eq5} are trivially satisfied.

A background temperature (entropy) profile $T(x,z)$ has also been adopted, with uniform gradient (hence satisfying Eq.~\ref{BS2}) $\alpha{g}\nabla{T} = \mathcal{N}^{2}\boldsymbol{e}_{\theta}$, where $\boldsymbol{e}_\theta = (\cos{\Gamma},0,\sin{\Gamma})$, where the buoyancy frequency $\mathcal{N}^2 > 0$ in radiative zones. The effective gravity vector $\boldsymbol{e}_g$ lies approximately in the spherical radial direction, and is inclined to $x$ by an angle $\phi$. For clarity, we consider sufficiently slowly rotating stars that $\boldsymbol{e}_g$ lies approximately along the spherical radial direction, and hence ``radial'' will be assumed to be along $\boldsymbol{e}_g$, though the model itself does not require this restriction (and it would not be the correct interpretation in very rapidly rotating stars). 

\begin{figure}
    \centering
    \includegraphics[width = 8cm , height = 7cm]{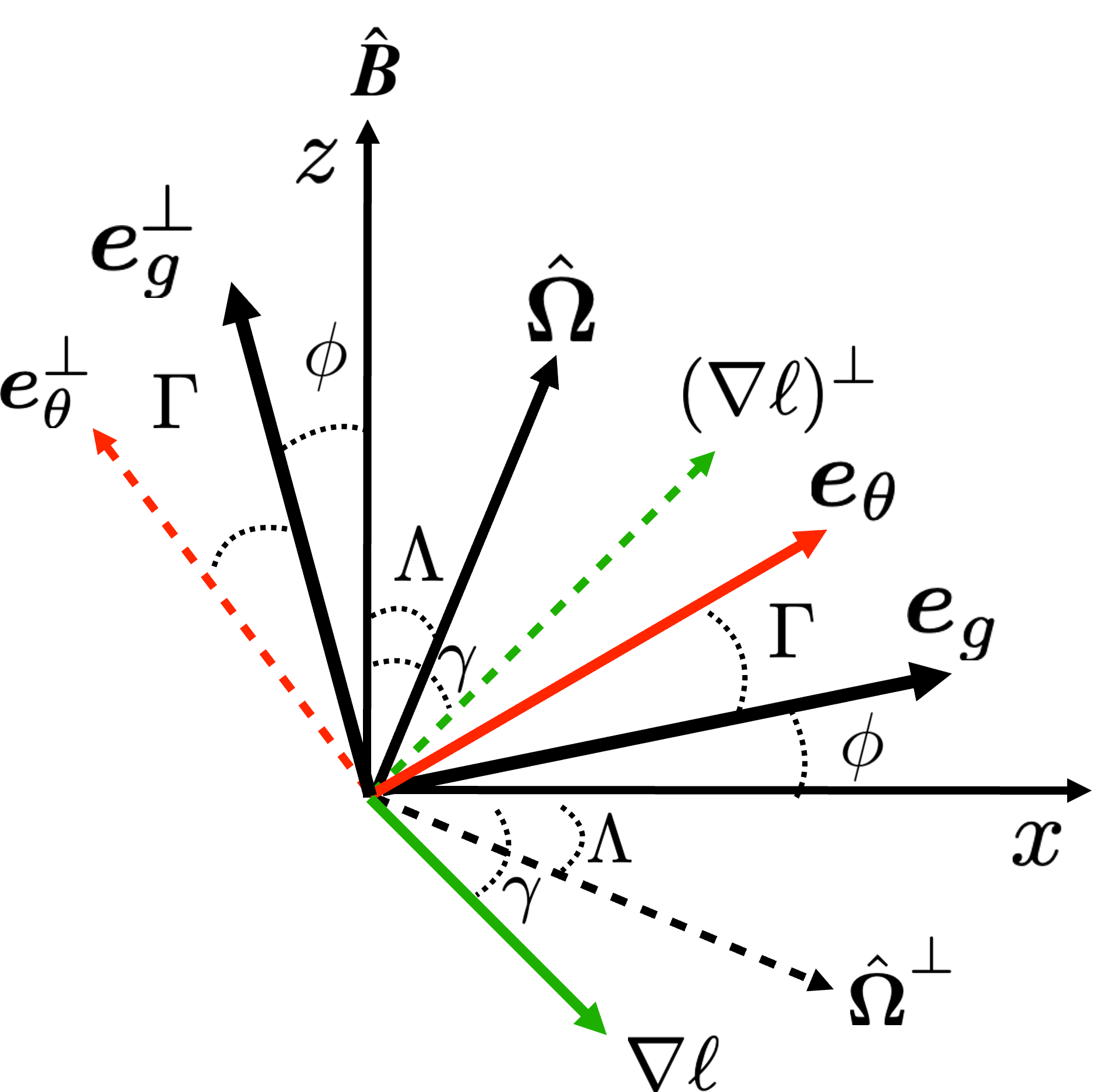}
    \caption{Illustration of the key vectors and corresponding angles in the $(x,z)$-plane. The cylindrical radial direction (along the equator) is along $\hat{\boldsymbol{\Omega}}^\perp$, and the rotation axis is along $\hat{\boldsymbol{\Omega}}$. The local radial direction is (approximately) along the effective gravity direction $\boldsymbol{e}_g$, which is misaligned with respect to the $x$-direction when $\phi$ is nonzero. The magnetic field is along $z$.}
    \label{Angles}
\end{figure}

We expect the star to adjust rapidly to satisfy thermal wind balance, and enforcing this requirement eliminates the angle $\Gamma$ as a free parameter. This means that $\boldsymbol{U}_0$ and its thermal state satisfy the thermal wind equation (TWE)
\begin{equation}
\label{TWE}
    2\Omega \mathcal{S}\sin{\Lambda}=\mathcal{N}^2\sin(\Gamma-\phi),
\end{equation}
which is derived from the azimuthal component of the vorticity equation for the basic state, i.e., the curl of Eq.~\ref{BS1}. This is unaffected by our magnetic field $\boldsymbol{B}_0$. The representation of various global differential rotation profiles and magnetic field orientations in our local model are summarised in Table.~\ref{tableangles}. We also illustrate the various angles in our problem in Fig.~\ref{Angles}. See paper 3 for further details of the non-magnetic model.

\begin{table}
\addtolength{\tabcolsep}{-5pt}
\begin{flushleft}
\begin{tabular}{|c|c|c|c|} 
 \hline
 $\Lambda$ & $\phi$ & Differential rotation & Magnetic field \\ 
 \hline
 $0$ & - & $\Omega(\varpi)$ (cylindrical) & arbitrary \\ 
 $\pm 90^\circ$ & - & $\Omega(z)$ (axial variation) & arbitrary \\ 
 - & 0 & $\Omega(r)$ (spherical/shellular) & horizontal/latitudinal \\ 
 - & $\pm90^\circ$ & $\Omega(\beta)$ (horizontal/latitudinal) & radial \\ 
  - & - & $\Omega(r,\beta)$ (arbitrary) & arbitrary \\ 
 \hline
\end{tabular}
\caption{Table of differential rotation profiles and magnetic field orientations (in the meridional plane) as $\Lambda$ and $\phi$ are varied. Here $\beta$ is co-latitude, $z$ is distance along rotation axis, $r$ is spherical radius and $\varpi$ is cylindrical radius.}
\end{flushleft}
 \label{tableangles}
\end{table}

We use units defined by the rotational timescale, $\Omega^{-1}$, and the lengthscale 
\begin{equation}
    d = \left(\frac{\nu \kappa}{\mathcal{N}^2}\right)^{\frac{1}{4}}.
\end{equation}
The fastest growing hydrodynamic (GSF) modes typically have wavelengths $O(d)$. With this choice of length, the buoyancy timescale $\mathcal{N}^{-1}$ is equal to the geometric mean of the viscous ($d^2/\nu$) and thermal ($d^2/\kappa$) diffusion timescales \citep[see e.g.][for other double-diffusive problems]{Radko2013}. Note that with the addition of a magnetic field it is not at all clear that unstable modes will necessarily have lengthscales $O(d)$, and in fact we will show that MRI modes may have much larger scales. However, for comparison with papers 1-3 and for comparing GSF and MRI modes, we continue to adopt this choice of units here.


\section{Linear theory}\label{LinearTheory}

We consider linear perturbations to our flow $\boldsymbol{U}_0$, thermal state, and magnetic field $\boldsymbol{B}_0$. From Eqs.~\ref{eq1}--\ref{eq6}, such velocity ($\boldsymbol{u}$), magnetic ($\boldsymbol{B}$), pressure ($p$) and temperature ($\theta$) perturbations are described by (where we have avoided introducing hats on perturbations)
\begin{align}
\label{eq1a}
   & D\boldsymbol{u}+2\boldsymbol{\Omega}\times\boldsymbol{u}+\boldsymbol{u}\cdot\nabla{\boldsymbol{U}_0} = -\nabla{p}+\theta\boldsymbol{e}_g+\boldsymbol{B}_0 \cdot {\nabla{\boldsymbol{B}}}+\nu\nabla^2{\boldsymbol{u}}, \\  
\label{eq2a}
    & D{\theta}+\mathcal{N}^2 \boldsymbol{u} \cdot \boldsymbol{e}_\theta = \kappa\nabla^2{\theta}, \\
\label{eq3a}
    & {D}\boldsymbol{B} = \boldsymbol{B}_0\cdot\nabla \boldsymbol{u} + \boldsymbol{B}\cdot\nabla \boldsymbol{U}_{0} + \eta {\nabla}^2 \boldsymbol{B}, \\
\label{eq4a}
    & \nabla\cdot{\boldsymbol{B}}=0, \\
\label{eq5a}
    & \nabla\cdot{\boldsymbol{u}}=0, \\
\label{eq6a}
    & D \equiv \partial_{t}+ \boldsymbol{U}_0\cdot\nabla.
\end{align}
Note that we have defined our field and flow to satisfy $\boldsymbol{B}_0\cdot\nabla\boldsymbol{U}_0=0$ so that the basic state is an equilibrium configuration. Note that this restriction was \textit{not} made in many prior works, including \cite{Balbus1994,Menou2004,Menou2006}, but it is necessary to have a well-defined steady basic state. It is unclear whether results obtained for any other poloidal field configuration (with a time-dependent basic state) are valid. We might expect results in such cases to only be approximately valid if growth times are sufficiently small compared with the timescale for the evolution of the basic state but not when the instability grows weakly.

\subsection{Dispersion relation for axisymmetric modes}

We consider axisymmetric modes with meridional wavevectors $\boldsymbol{k}=(k_x,0,k_z)=k(\cos\theta_k,0,-\sin\theta_k)$ with magnitudes $k=\sqrt{k_x^2+k_z^2}$ and angles $\theta_k$, since axisymmetric modes are likely to be the fastest growing \citep[e.g.][]{LatterPap2018}, and we define $\hat{\boldsymbol{k}}=\boldsymbol{k}/k$. These permit complex exponential solutions proportional to $\mathrm{exp}(\mathrm{i}k_x{x} + \mathrm{i}k_z{z} + st)$. We define the complex growth rate $s=\sigma + \mathrm{i} \omega$, where the growth (decay) rate $\sigma\in\mathbb{R}$ and the oscillation frequency $\omega\in\mathbb{R}$. We manipulate Eqs.~\ref{eq1a}--\ref{eq6a} for such perturbations. and define the modified growth rates $s_{\nu}=s+\nu{k^2}$, $s_{\kappa}=s+\kappa {k^2}$ and $s_{\eta}=s+\eta {k^2}$, to obtain the quintic dispersion relation 
\begin{equation}
\label{DR}
    s^{2}_{\eta}s^{2}_{\nu}s_{\kappa}+ 2s_{\eta}s_{\nu}s_{\kappa}{\omega^{2}_A} + s_{\kappa}{\omega^{4}_A} + a s^{2}_{\eta}s_{\kappa}+ s_{\kappa}\xi +
     b (s^{2}_{\eta}s_{\nu}+s_{\eta}\omega^{2}_{A})  =0,
\end{equation}
where
\begin{align}
a &= \frac{2}{\varpi}(\hat{\boldsymbol{k}}\cdot\boldsymbol{\Omega})(\hat{\boldsymbol{k}}\cdot (\nabla\ell)^\perp), \label{eq_defa_1}\\
&=\frac{2\Omega}{k^2}(s_{\Lambda}k_x+c_{\Lambda}k_z)(2\Omega k_x s_{\Lambda}+(2\Omega c_{\Lambda}-\mathcal{S})k_z), \\
&= \frac{2 \Omega | \nabla \ell |}{\varpi} s_{\Lambda - \theta_k} s_{\gamma - \theta_k},
\label{eq_defa_2}
\end{align}
\begin{align}
b&= \mathcal{N}^2(\hat{\boldsymbol{k}}\cdot\boldsymbol{e}_{\theta}^\perp)(\hat{\boldsymbol{k}}\cdot{\boldsymbol{e}_{g}^\perp}), \\
 &= \frac{\mathcal{N}^2}{k^2}(k_z{c_{\Gamma}}-k_x{s_{\Gamma}})(k_{z}c_{\phi}-k_x{s_{\phi}}), \\
 &=  {\cal N}^2 s_{\theta_k  + \phi}s_{\theta_k+\Gamma},
 \label{eq_defb}
\end{align}
\begin{align}
\xi = -2 (\hat{\boldsymbol{k}}\cdot{\boldsymbol{\Omega}})\mathcal{S}\omega^{2}_{A} \hat{k}_z=2 S \Omega s_{\Lambda-\theta_k} s_{\theta_k}\omega_A^2,
\label{eq_def_xi}
\end{align}
and
\begin{align}
\omega^{2}_{A} = (\boldsymbol{B_0}\cdot{\boldsymbol{k}})^2=k^2 B_0^2 s^2_{\theta_k},
\end{align}
is the squared Alfv\'{e}n frequency.
In the above $c_\Lambda$ and $s_\Lambda$ refer to $\cos\Lambda$ and $\sin\Lambda$ for brevity, and similarly for trigonometric functions with other arguments (though $s_{\nu},s_{\kappa}$ and $s_{\eta}$ always represent modified growth rates instead). We have also defined the local angular momentum gradient
\begin{align}
 \nabla {\ell} =\varpi(2\Omega{c_{\Lambda}}-\mathcal{S},0,-2\Omega{s_{\Lambda}}),
          =|\nabla{\ell}|(c_{\gamma},0,-s_{\gamma}),
\end{align}
which has magnitude
\begin{equation}
   |\nabla{\ell}|^2=\varpi^2\left({\mathcal{S}^2}+4{\Omega}(\Omega-\mathcal{S}c_{\Lambda})\right).
\end{equation}
The normal to this is
\begin{equation}
 (\nabla{\ell})^\perp = \varpi(2\Omega s_{\Lambda},0,2\Omega c_\Lambda-\mathcal{S})
 =|\nabla\ell|(s_\gamma,0,c_\gamma).
 \label{eq_normal_ell}
\end{equation}
We also define the vector perpendicular to the effective gravity,
\begin{equation}
\boldsymbol{e}_{g}^\perp=(-s_{\phi},0,c_{\phi}),
\end{equation}
and the normal to stratification surfaces, 
\begin{equation}
\boldsymbol{e}_{\theta}^\perp=(-s_{\Gamma},0,c_{\Gamma}).
\end{equation}
The baroclinic shear (along the rotation axis) is 
\begin{align}
\label{baroshear}
\hat{\boldsymbol{\Omega}}\cdot(\nabla{\ell})=-\mathcal{S}\varpi s_{\Lambda}
=|\nabla{\ell}|s_{\gamma-\Lambda}.
\end{align}

The dispersion relation (\ref{DR}) can be expanded out as a quintic equation
\begin{equation}
s^5 + c_1 s^4 + c_2 s^3 +c_3 s^2 + c_4 s + c_5 =0, 
\label{eq_quintic}
\end{equation}
where the coefficients $c_1$ to $c_5$ are given by
\begin{align}
    c_1&=k^2(2 \eta +2 \nu +\kappa), \\
    c_2&=k^4(\eta^2 +2 \eta \kappa +4 \eta \nu + 2 \nu \kappa +\nu^2) + 2 \omega_A^2 + a + b, \\
    c_3&=k^6(\eta^2 \kappa + 2 \eta \nu^2 + 2 \eta^2 \nu +\kappa \nu^2 + 4 \eta \kappa \nu) \nonumber \\
    &+ 2 \omega_A^2 k^2(\eta+\nu +\kappa) +ak^2(2 \eta+\kappa) + bk^2(2\eta+\nu), \\
    c_4&=k^8(2\eta \nu^2 \kappa + 2 \eta^2 \nu \kappa+ \eta^2 \nu^2) + 2 \omega_a^2 k^4(\eta \nu + \eta \kappa + \nu \kappa) \nonumber \\
    &+ \omega_A^4 + a k^4(2 \eta\kappa+\eta^2) + \xi + bk^4(2 \eta \nu + \eta^2) + b \omega_A^2, \\
    c_5&= k^{10} \eta^2 \nu^2 \kappa + 2 \omega_A^2 k^6 \eta \nu \kappa + \omega_A^4 k^2 \kappa + a \eta^2 \kappa k^6 \nonumber \\
    &+ \xi k^2 \kappa + b k^6 \eta^2 \nu + b \omega_A^2 k^2 \eta.
\label{eq_c_coeffs}
\end{align}

\subsection{Non-diffusive (in)stability} \label{adiab}
Non-diffusive modes, i.e.~those with $\nu=\kappa=\eta=0$, are described by the reduced dispersion relation
\begin{equation}
    s^4 + (2{\omega^{2}_A}+a  + b)s^2 +  ({\omega^{4}_A + b{\omega^{2}_A}+\xi}) = 0,
\label{eq_adiab}
\end{equation}
ignoring neutral modes with $s=0$. Note that the only appearance of the magnetic field is through the combination $\boldsymbol{B}_0\cdot \boldsymbol{k}=B_0 k_z$ in $\omega_A$, therefore the adiabatic growth rate is independent of $B_0$ if arbitrary $k_z$ are permitted. This can be solved to give 
\begin{equation}
    s^2 = \frac{-(2{\omega^{2}_A+a+b)} \pm \sqrt{4a{\omega^{2}_A}-4\xi+(a+b)^2}}{2}.
\end{equation}
If we take $B_0 = 0$ then this reduces to the adiabatic dispersion relation in paper 3 (equation 27), $s^2 = -(a+b)$. We would also obtain $s^2 = -(a+b)$ in the diffusive case if we took $k\to 0$ in Eq.~\ref{eq_quintic}, since $\omega_A^2\to 0$ in this limit, thereby eliminating the influence of magnetic fields on such modes.

Since $s^2$ is always negative if $a+b$ is positive, with $s$ being purely imaginary, then the system is Solberg-H\o iland stable. The discriminant $\Delta=4a{\omega^{2}_A}-4\xi+(a+b)^2$ in (\ref{eq_adiab}) is always positive, so the roots for $s^2$ are always real. Hence, non-diffusive oscillatory instabilities cannot occur. To see this, note that from Eqs.~\ref{eq_defa_1} and \ref{eq_normal_ell} we have
\begin{equation}
    \hat{\boldsymbol{k}}\cdot (\nabla\ell)^\perp = \varpi (2 (\hat{\boldsymbol{k}}\cdot\boldsymbol{\Omega}) - \hat{k}_z\mathcal{S}),
\end{equation}
and using the definition of $\xi$ (Eq.~\ref{eq_def_xi}),
\begin{equation}
  \Delta=4a{\omega^{2}_A}-4\xi+(a+b)^2= (a+b)^2 + 16 \omega^{2}_A (\hat{\boldsymbol{k}}\cdot\boldsymbol{\Omega})^2,
\end{equation}
which being the sum of squares must be non-negative.
 
The criterion for onset of direct instability (real roots) occurs when (for neutral stability $s=0$)
  \begin{equation}
  \label{neutraladistab}
      \omega_A^4 +b \omega_A^2+\xi=0,
  \end{equation}
 and instabilities occur when this term is negative. So the only way to destabilise a hydrodynamically Solberg-H\o iland stable configuration without diffusion is for the left hand side of Eq.~\ref{neutraladistab} to be negative, which corresponds with a direct instability, the MRI. MRI works best with a weak field (or on large length-scales), meaning the stabilising term $\omega_A^4$ is small compared with the others, and when the fluid is neutrally rather than stably stratified $(b=0)$. Then, MRI just requires a mode with a $\boldsymbol{k}$ which makes $\xi$ negative. In the weak field or small wavenumber case, $\omega_A^4\to 0$ faster than the remaining terms  in Eq.~\ref{neutraladistab}, so for non-zero $B_0$, instability occurs if
 \begin{equation}
     b-2(\hat{\boldsymbol{k}}\cdot\boldsymbol{\Omega}) \mathcal{S} \hat{k}_z<0.
     \label{b_crit}
 \end{equation}
Hence in the special case of cylindrical differential rotation, $\Omega(\varpi)$, we have $\mathcal{N}^2-2\Omega \mathcal{S}<0$. If the stabilising effects of buoyancy are absent, the stability criterion in the weak field case is $-2\Omega \mathcal{S}<0$, which involves angular velocity rather than angular momentum gradients \citep[e.g.][]{BH1998}. So $\mathcal{S}>0$ is required for instability (to MRI), which is generally much easier to satisfy than Rayleigh's criterion for centrifugal instability, which requires $\mathcal{S}>2$ in the hydrodynamic case, implying outwardly decreasing angular momentum.

The three quantities $a$, $b$ and $\xi$  that appear in Eq. \ref{eq_adiab} correspond to three different 
instability mechanisms. If there is no stratification or magnetic field, angular momentum-driven instability occurs if $\boldsymbol{k}$ 
can be chosen so that $a<0$. In the case of cylindrical rotation this is just Rayleigh's criterion that instability occurs when the angular momentum decreases outwards. More generally, using Eq. \ref{eq_defa_2},  $a < 0$ when the wavevector $\boldsymbol{k}$
lies in the wedge between $\boldsymbol{\Omega}^\perp$ and $\nabla  \ell$. The quantity $b$ is associated with the stratification,
see Eq. \ref{eq_defb}, and baroclinic instability $b<0$  occurs when the wavevector $\boldsymbol{k}$ lies in the wedge between 
$\boldsymbol{e}_g$ and $\boldsymbol{e}_\theta$. When the stratification is very strong, the thermal wind equation makes this wedge angle small, so baroclinic instability is weak. Then in a convectively stable region $b$ will be large and positive unless the wavevector is nearly parallel to gravity, i.e. horizontal flow along the isobars. The quantity $\xi$, Eq. \ref{eq_def_xi}, is associated with MRI instability. From  Eq. \ref{neutraladistab} if $b>0$ we need $\xi <0$ for instability. This happens if the wavevector
$\boldsymbol{k}$ lies outside the wedge between $\hat{\boldsymbol{\Omega}}^\perp$ and the $x$-axis.

\subsubsection{Marginal stability to stratified non-diffusive MRI}

In order to find the non-diffusive unstable modes for weak fields we substitute $\boldsymbol{k}=k(\cos\theta_k,0,-\sin\theta_k)$ into Eq.~\ref{b_crit} and solve for the marginal stability lines, giving
\begin{equation}
    \mathcal{N}^2 \sin({\theta_k + \Gamma})\sin({\theta_k + \phi}) +2 \Omega \mathcal{S} \sin (\Lambda-\theta_k)\sin\theta_k=0.
    \label{marglines}
\end{equation}
In the strongly stratified limit, $N^2\gg |\Omega S|$ and $\Gamma\sim \phi$ (from Eq.~\ref{TWE}),
\begin{equation}
    \mathcal{N}^2 \sin^2({\theta_k + \phi}) - 2 \Omega\mathcal{S}\sin({\theta_k - \Lambda})\sin{\theta_k} = 0.
\end{equation}
If $\mathcal{N}^2\gg 2|\Omega S|$, this can only be satisfied when $\sin^2 (\theta_k+\phi)\sim 0$, hence $\theta_k+\phi\approx n \pi$ where $n\in \mathbb{N}$. Note that $\theta_k$ is defined below the $x$-axis so $-\theta_k$ is the angle above it. This means that $-\theta_k=\phi=\Gamma$ when $n=0$, indicating that $\boldsymbol{k}$ lies along $\boldsymbol{e}_\theta$ or $\boldsymbol{e}_g$, so that fluid motions are along stratification (or constant pressure) surfaces, i.e.~parallel to $\boldsymbol{e}_\theta^\perp\sim \boldsymbol{e}_g^\perp$. Hence instability is possible for a wedge of wavevector angles around $\boldsymbol{e}_\theta$ \citep[e.g.][]{Balbus1995}. If $\phi > 0$ and
the box is in the northern hemisphere, ${\boldsymbol{e}_g}$ lies outside the MRI-stable wedge between $\boldsymbol{\hat \Omega}^\perp$ and the $x$-axis, see Fig. \ref{Angles}, and we expect 
MRI instability. However, if $\phi<0$ in the northern hemisphere, ${\boldsymbol{e}_g}$ lies inside the MRI-stable wedge (recall $\Lambda + \phi > 0$ in the northern hemisphere so $\Lambda > |-\phi|$), and so no MRI will occur, though there may be unstable GSF modes.

\subsubsection{Fastest growing non-diffusive modes} \label{maxadmode}

We now find the wavevector magnitude $k$ and orientation $\theta_k$ corresponding to the maximum growth rate, and in turn identify the dominant mode. To find the fastest growing mode we first maximise over $k^2$, and then maximise over the angle $\theta_k$. Note $a$ and $b$ only depend on $\theta_k$ and not on the magnitude $k$, so $\partial a / \partial k^2$ and $\partial b / \partial k^2$ are both zero. We also have
\begin{align}
 \frac{\partial \omega_A^2}{ \partial k^2} =\frac{\omega_A^2}{k^2}, \quad 
\mathrm{and} \quad \frac{\partial \xi}{ \partial k^2}=\frac{\xi}{k^2}.
\end{align}
To obtain the fastest growing mode properties, we differentiate Eq.~\ref{eq_adiab} with respect to both $k^2$ and $\theta_k$ and require $\partial_{k^2}s=\partial_{\theta_k}s=0$. 
Setting the $k^2$ derivative of Eq.~\ref{eq_adiab} to zero gives:
\begin{align}
\label{diffk2}
    s^2=-\frac{\mathcal{N}^2}{2}s_{\phi+\theta_k}s_{\Gamma+\theta_k}-\Omega S s_{\theta_k} s_{\Lambda-\theta_k}- k^2 B^{2}_0 s^2_{\theta_k}.
\end{align}
This is clearly maximised for weak fields or for modes with $k\to 0$ where the last term vanishes, since that provides a stabilising effect, though we must have nonzero $B_0$ to obtain this result. Now we can maximise over $\theta_k$ to obtain
\begin{align}
    0=-\frac{\mathcal{N}^2}{2}s_{\Gamma+\phi+2\theta_k}-\Omega S s_{\Lambda-2\theta_k}-k^2B_0^2s_{2\theta_k}.
\end{align}
In the strongly stratified limit the dominant term is usually the first one involving $\mathcal{N}^2$, which is stabilising, unless we choose a specific range of $\theta_k$. This means that in order to maximise the growth rate we need to minimise this term. Indeed the magnitude of this term is smallest when $\theta_{k} \approx -\frac{\Gamma+\phi}{2}$ i.e.~when the wavevector is approximately halfway between $\boldsymbol{e}_g$ and $\boldsymbol{e}_\theta$. Note that in the strongly stratified limit the TWE implies that $\phi \approx \Gamma$ and hence this term approximately vanishes for $\theta_k\approx-\phi\approx-\Gamma$. For such wavevectors that minimise the stabilising effects of buoyancy, 
\begin{equation}
        s^2=\Omega S s_{\phi} s_{\Lambda+\phi} - B^{2}_0 \omega^2_{A} s^2_{\phi}.
\end{equation}
Note that the magnetic term that only occurs in the non-weak field limit is always negative, meaning that in the adiabatic regime a stronger magnetic field should decrease the maximum growth rate of the instability.
In the weak field case, where we can ignore the second term, we are left with 
\begin{equation}
        s^2=2\Omega S s_{\phi} s_{\Lambda+\phi}.
\end{equation}
Hence, we require both $\phi$ and $\Lambda+\phi$ to have the same sign, either both in the northern or southern hemisphere for onset of instability. 

\subsection{Diffusive instabilities}\label{diffusiveanalysis}

\subsubsection{Small $\mathrm{Pr}/\mathrm{Pm}$ limit: very efficient thermal diffusion}
In the limit of very efficient thermal diffusion relative to viscous and ohmic diffusion, we would expect Eq.~\ref{eq_adiab} to approximately apply for sufficiently large wavelength instabilities (with smallish $k$, for which viscous and ohmic diffusion are relatively unimportant) but with $b=0$. To show that this is indeed the case, if we consider Eq.~\ref{DR}, set $\nu=\eta=0$ and then consider the limit $\kappa\to \infty$. This means that in Eq.~\ref{DR} $\eta k^2 \ll s$, 
$\nu k^2 \ll s$ but $\kappa k^2 \gg s$, which is like considering the joint limits $\mathrm{Pr}/\mathrm{Pm}\to 0$ and $\mathrm{Pr}\to 0$ with all other quantities $O(1)$. We obtain the dispersion relation 
\begin{align}
    s^4+(2\omega_A^2+a)s^2+(\omega_A^4+\xi)=0.
\end{align}
This is the same as Eq.~\ref{eq_adiab} with $b=0$ and describes MRI modes satisfying the unstratified ($b=0$) non-diffusive dispersion relation with nonzero field\footnote{This is analogous to what was found for the hydrodynamic case in the limit $\mathrm{Pr}\to 0$ and $\mathrm{Ri}\mathrm{Pr}\to 0$, where the fastest growing mode growth rates were described by $s^2=-a$, the adiabatic unstratified dispersion relation \citep{barker2020,Dymott2023}. However, the dispersion relation here requires the presence of non-vanishing magnetic field.}. The fastest growing modes (maximising over $k^2$, i.e. setting $\partial_{k^2} s=0$) in the limit of weak fields or small $k$ (for which $\omega_A^4$ can be ignored relative to the other terms) satisfy
\begin{align}
    s^2 &= -\frac{\xi}{2\omega_A^2} = (\hat{\boldsymbol{k}}\cdot\boldsymbol{\Omega})S \frac{k_z}{k}=\mathcal{S} \Omega s_{\theta_k-\Lambda} s_{\theta_k}.
\end{align}
In this limit instability occurs for any $\mathcal{S}>0$ (though strictly the approximations for which this limit applies are then no longer valid).
The growth rate is maximised over $\theta_k$ when $\partial_{\theta_k} s^2=0$, giving
\begin{align}
    s_{\Lambda-2\theta_{k}}=0 \quad \Rightarrow \quad \theta_k=\frac{\Lambda}{2}-n\frac{\pi}{2},
\end{align}
for $n\in\mathbb{N}$, i.e.~for modes with orientations halfway between the rotation axis (along $\hat{\boldsymbol{\Omega}}$) and the angular velocity gradient (along $x$) when $n=1$. For cylindrical differential rotation ($\Lambda=0$), this implies $\theta_k=\pm\frac{\pi}{2}$, and hence wavevectors are along $z$, as expected \citep[e.g.][]{BH1991}. On the other hand, when $\Lambda=-30^\circ$, $\theta_k=-105^\circ$ (indicating $105^\circ$ \textit{above} the $x$ axis), and when $\Lambda=60^\circ$, $\theta_k=-60^\circ$ (indicating $60^\circ$ \textit{above} the $x$ axis). 
This is consistent with our later Figures~\ref{Lobes1}--\ref{Lobes3} for the largest Pm considered, and is most evident for the strongest magnetic fields plotted there.

For even stronger fields or larger wavenumbers, there is a stabilising effect of magnetic tension through the $\omega_A^4$ term in the dispersion relation. Fields are sufficiently strong when $\omega_A^2=B_0^2k^2s_{\theta_k}^2\sim 2\Omega \mathcal{S} s_{\theta_k-\Lambda}s_{\theta_k}$, and hence typically for $k^2\sim 2\Omega \mathcal{S}/B_0^2$.
In addition, larger $k$ modes would be increasingly affected by ohmic diffusion and viscosity.

\subsubsection{Diffusive modes in the small shear (small $\mathcal{S}$/$\Omega$) limit}\label{diffusiveanalysis2}

In paper 3, we computed the curves showing the lowest value of the shear $\mathcal{S}$ for which instability is possible as a function of
the angle $\phi$, the angle between the shear and gravity directions (see Figure 6 in paper 3). Apart from the exceptional case on the 
equator, there is a finite minimum $\mathcal{S}$ below which no instability occurs. This is no longer the case when a magnetic field is
added. There is then a whole range of $\phi$ for which the system is unstable for arbitrarily small $\mathcal{S}$. This is quite surprising, as both GSF and MRI instability are driven by the shear, so one might imagine that reducing the shear towards zero would eliminate them. What happens is that the growth rate does tend to zero as $\mathcal{S}$ is reduced, but it can always remain positive, so the critical value of $\mathcal{S}$ for instability can be zero. 

To establish this, we consider the case where $\mathcal{S}$/$\Omega \to 0$, and seek modes with small but positive growth rate $s \sim O(\mathcal{S})$. 
We now consider the ordering of the terms in the quintic dispersion relation Eq.~\ref{DR}. We retain the diffusive terms, choosing $O(\eta k^2) \sim O(\nu k^2) \sim O(\kappa k^2) \sim O(\mathcal{S})$. 
$a$ is $O(\Omega^2)$ and we choose the magnetic field strength so that $\omega_A^2 \sim O(\mathcal{S} \Omega)$, which makes $\xi \sim O(\mathcal{S}^2 \Omega^2)$.
If we take $\mathcal{N}^2 \sim O(\Omega^2)$ or larger, it appears we have an inconsistency, because then the dominant term in Eq. \ref{DR} is $b s_{\eta} \omega_A^2 \sim O(\mathcal{S}^2 \Omega^3)$ 
whereas the remaining terms are $O(\mathcal{S}^3 \Omega^2)$ or smaller. However, the thermal wind equation Eq.~\ref{TWE} implies that $\Gamma$ and $\phi$ are almost aligned if $\mathcal{S} / \Omega \ll 1$. If we choose our wavenumber 
${\boldsymbol{k}}= k(c_{\theta_k}, 0, -s_{\theta_k})$ so that $\theta_k = \pi - (\Gamma+\phi)/2$, aligned in the limit $\mathcal{S}/\Omega \to 0$ with both ${\boldsymbol{e}}_\theta$ and ${\boldsymbol{e}}_g$, 
then $b= \mathcal{N}^2 s_{\theta_k+ \Gamma} s_{\theta_k + \phi} = - \mathcal{N}^2 s_{\Gamma/2 - \phi/2}^2$, and now the thermal wind equation Eq.~\ref{TWE} implies that $b \sim O(\mathcal{S}^2)$, much smaller than $O(\Omega^2)$. Numerical solutions of 
Eq.~\ref{DR} show that the critical wavenumber $\boldsymbol{k}$  is indeed aligned with ${\boldsymbol{e}}_\theta$ and ${\boldsymbol{e}}_g$. Now the inconsistency in the $b$ term in Eq.~\ref{DR} is removed, because this term is now negligible compared with the $O(\mathcal{S}^3 \Omega^2)$ terms. We obtain  
\begin{align}
&a s_\eta^2 s_\kappa + (\omega_A^4 +\xi) s_\kappa=0.
\label{eq_cubica}
\end{align}
where 
\begin{eqnarray}
    \omega_A^2 &=& B_0^2 k^2 \sin^2 \phi, \qquad a= 4 \Omega^2 \sin^2 \beta, \nonumber \\ 
    \xi &=& - 2k^2 S \Omega B_0^2 \sin \beta \sin^3 \phi,
\label{smallSdefs}
\end{eqnarray}
and $\beta = \Lambda + \phi$ is the latitude. Dividing by $s_{\kappa}$, which must be positive for growing modes, 
\begin{align}
    a s_\eta^2 + \omega_A^4+\xi=0, \;\; \text{implying} \;\;
    s = -\eta k^2 + \sqrt{\frac{-\omega_A^4-\xi}{a}}.
    \label{eq_quada}
\end{align}
From Eq.~\ref{smallSdefs} $a>0$, so for instability we must have $\xi <0 $, and so for $\mathcal{S} > 0$ the angle $\phi$ and the latitude $\beta$ must have the same sign for instability, so in the northern hemisphere, $\beta>0$, we must have $\phi >0$
for our small $\mathcal{S}$ modes to be unstable. In the solution Eq.~\ref{eq_quada} the largest term as $k \to 0$ is the $\xi$ term, so that for negative $\xi$ there is always a small enough $k$ that makes $s > 0$, so that that however small $\mathcal{S}$ is, it is always possible to find a growing mode. This behaviour is illustrated below in Fig.~\ref{Scrit} for latitude $30^\circ$ N showing instability for very small $\mathcal{S}$ when $\phi > 0$.   
For very large $\mathcal{N}^2$, as we might expect in stars, and for small $\mathcal{S}$ (weak differential rotations), the inertial term $a$ is negligible.
Since $b$ would then in general dominate all other terms in Eq.~\ref{DR}, for instability we must choose modes for which fluid motion is along stratification surfaces, meaning $\boldsymbol{k}$ must be parallel or anti-parallel to gravity. For small $\mathcal{S}$, $\boldsymbol{e}_g$ is very close to $\boldsymbol{e}_\theta$. Why can’t you wipe out the effect of the stratification by thermal diffusion in this limit as you can with GSF (for example)? For thermal diffusion to work, $k$ must be reasonably large, and GSF modes are stabilised by the magnetic field at large (or moderate) $k$. At small $\mathcal{S}$, MRI is stabilised by the $\omega_A^4$ term too because $\xi$ is proportional to $\mathcal{S}$ unlike $\omega_A^4$. Only small $k$ (or small $B_0$) reduces $\omega_A^4$ relative to $\xi$, and having small $k$ doesn’t allow efficient enough thermal diffusion for instability. Hence, the fundamental MRI mechanism that operates for small $\mathcal{S}$ is typically non-diffusive.
 
As $\mathcal{N}^2$ is reduced, $\mathcal{S}$ is made larger, or the field is reduced, then GSF modes with a larger $k$ can operate, and there are more possibilities for instability. We will return to this point later when analysing instability in the solar tachocline.

\section{Numerical linear results} \label{linearDR}

\begin{figure*}
    \subfigure[]{\includegraphics[trim=0cm 0cm 0cm 0cm,clip=true,
    width=0.33\textwidth]{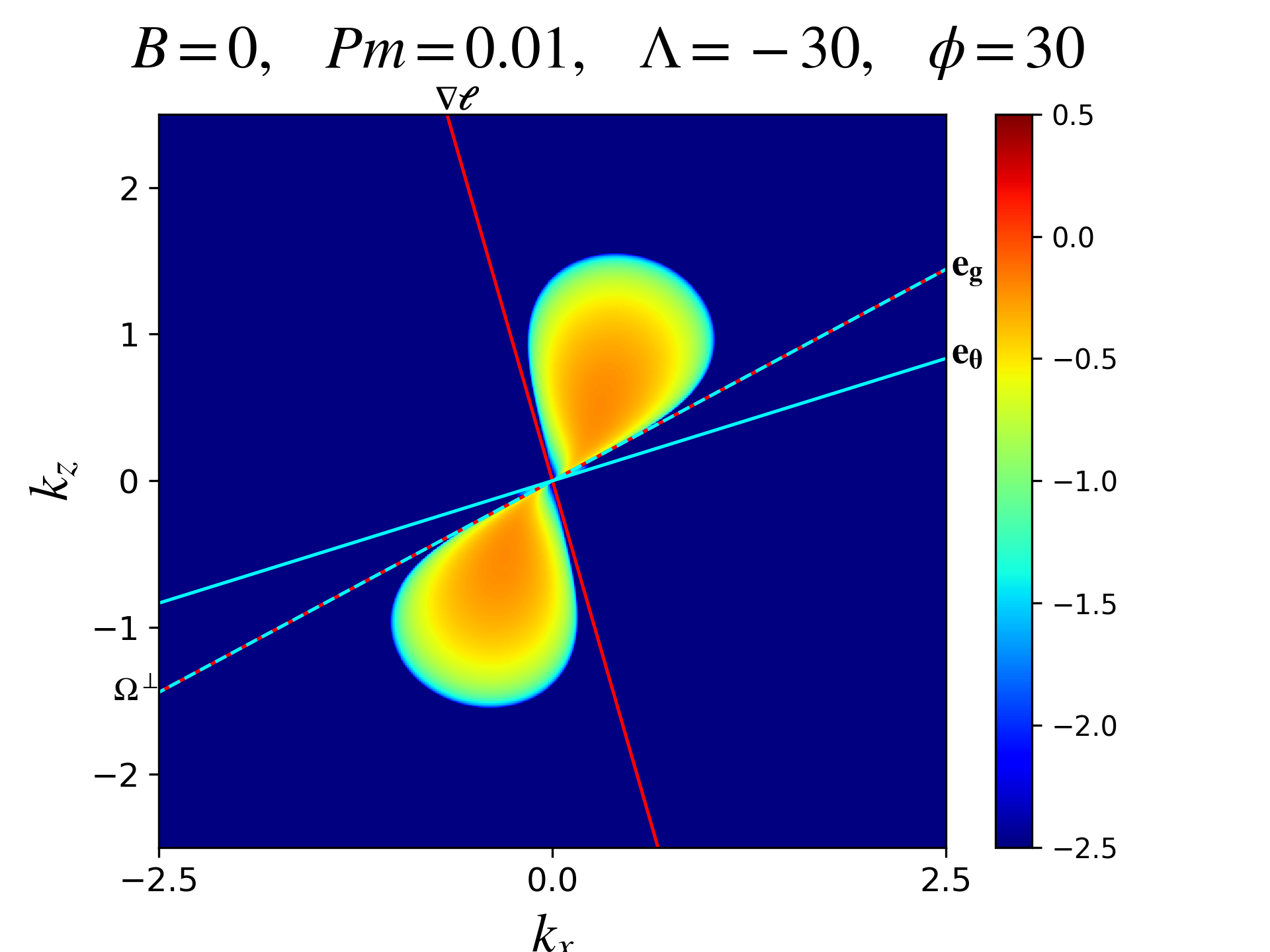}}
    \subfigure[]{\includegraphics[trim=0cm 0cm 0cm 0cm,clip=true,
    width=0.33\textwidth]{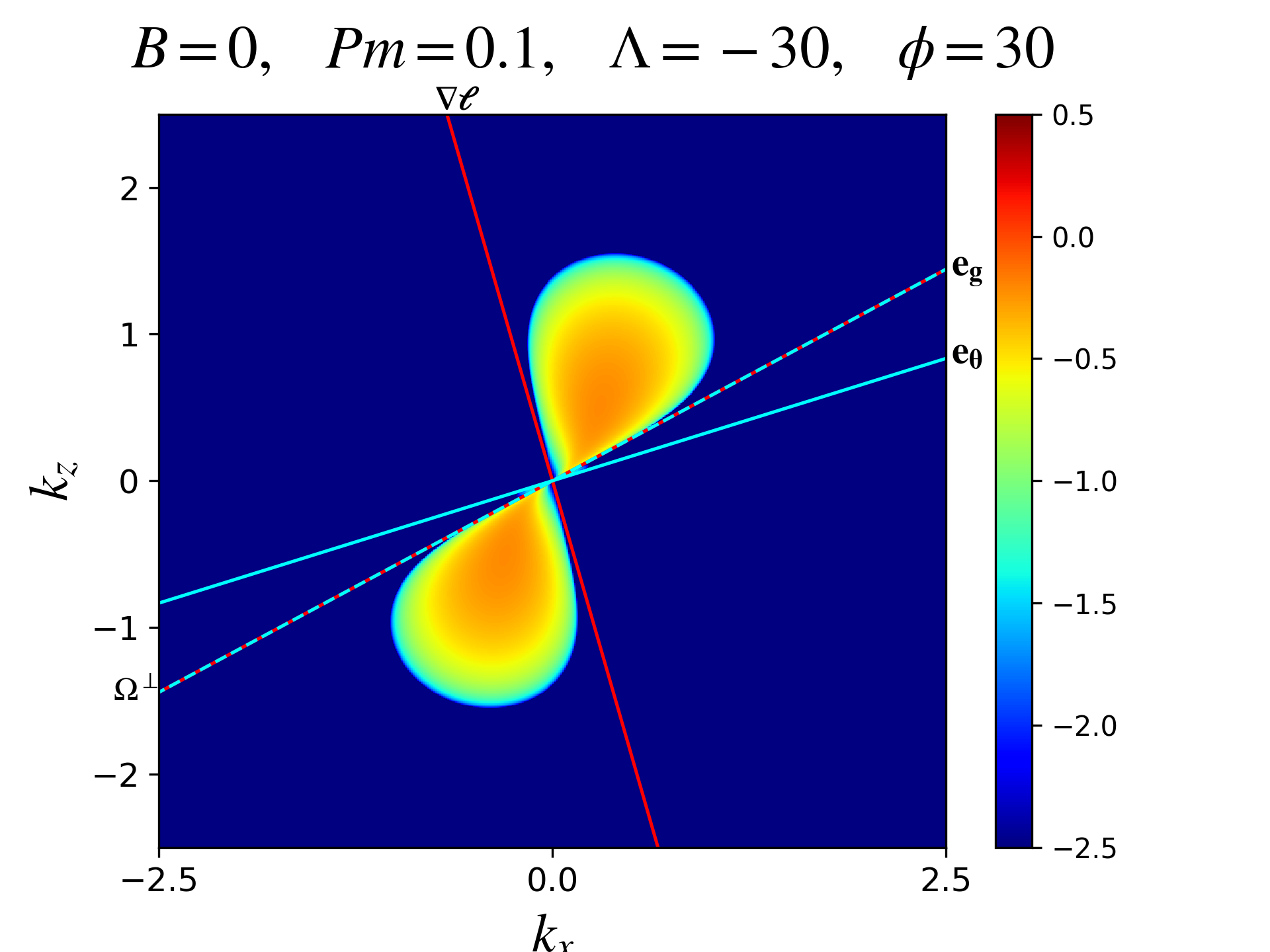}}
    \subfigure[]{\includegraphics[trim=0cm 0cm 0cm 0cm,clip=true,
    width=0.33\textwidth]{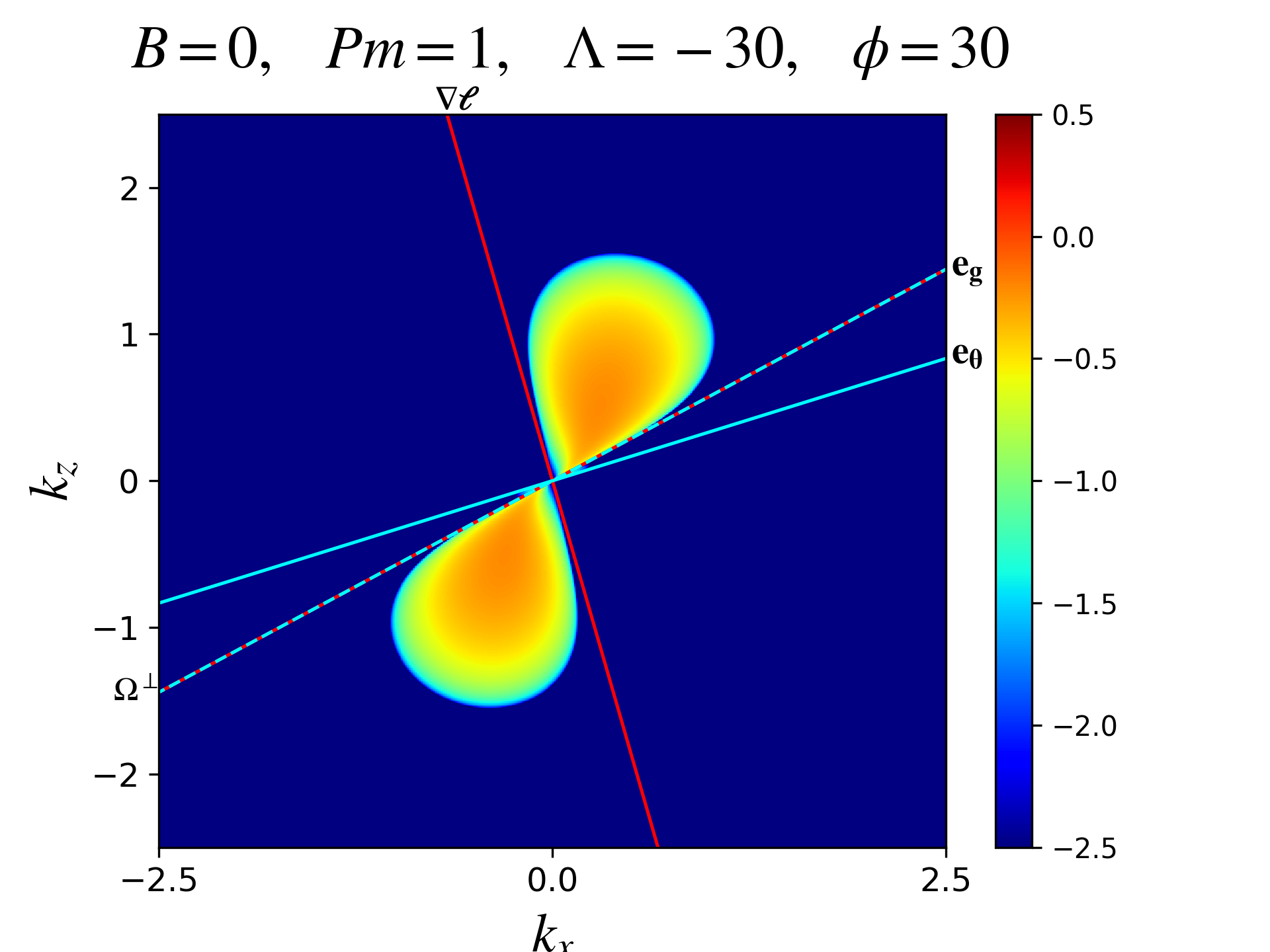}}
   
    \subfigure[]{\includegraphics[trim=0cm 0cm 0cm 0cm,clip=true,
    width=0.33\textwidth]{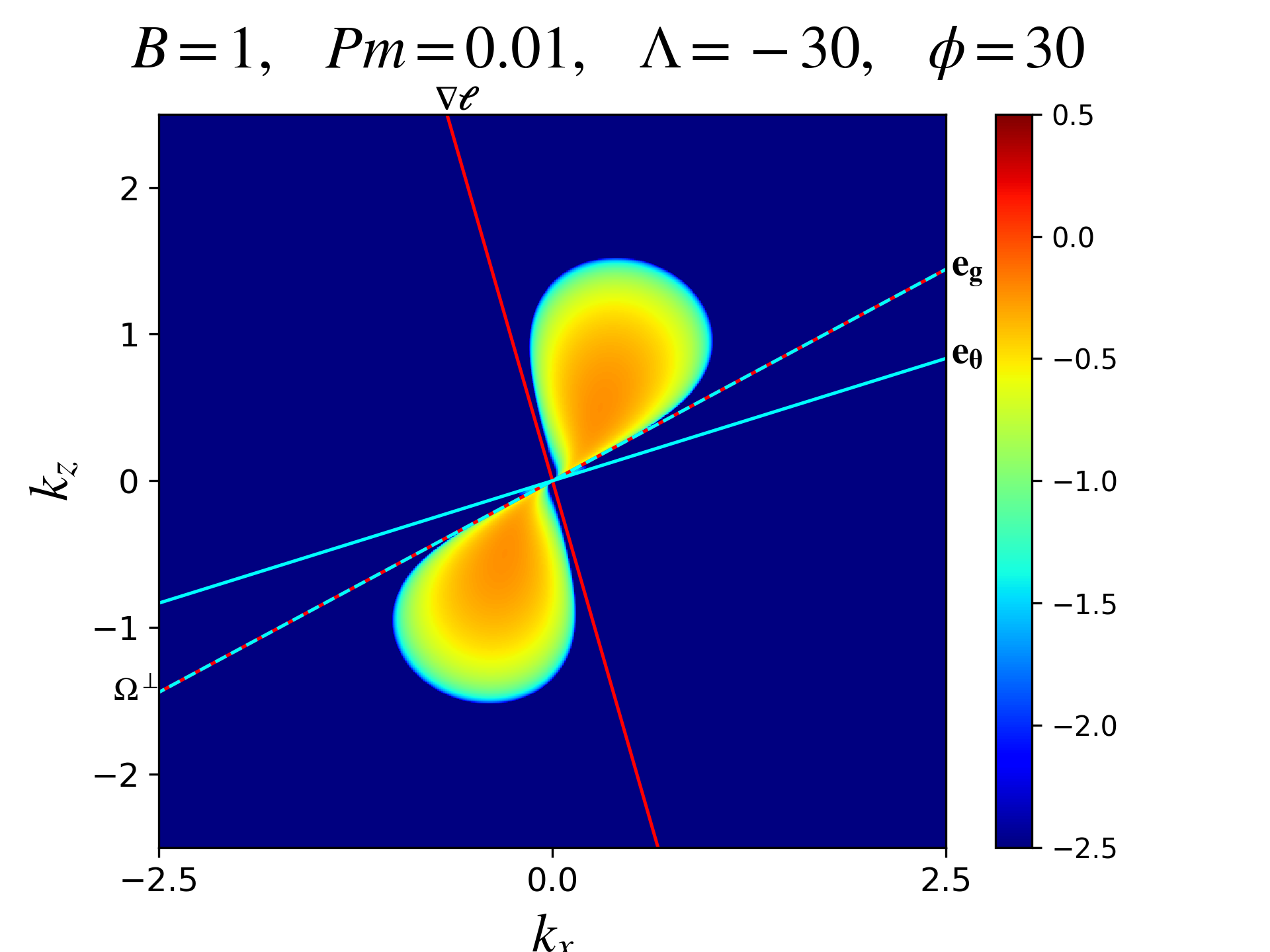}}
    \subfigure[]{\includegraphics[trim=0cm 0cm 0cm 0cm,clip=true,
    width=0.33\textwidth]{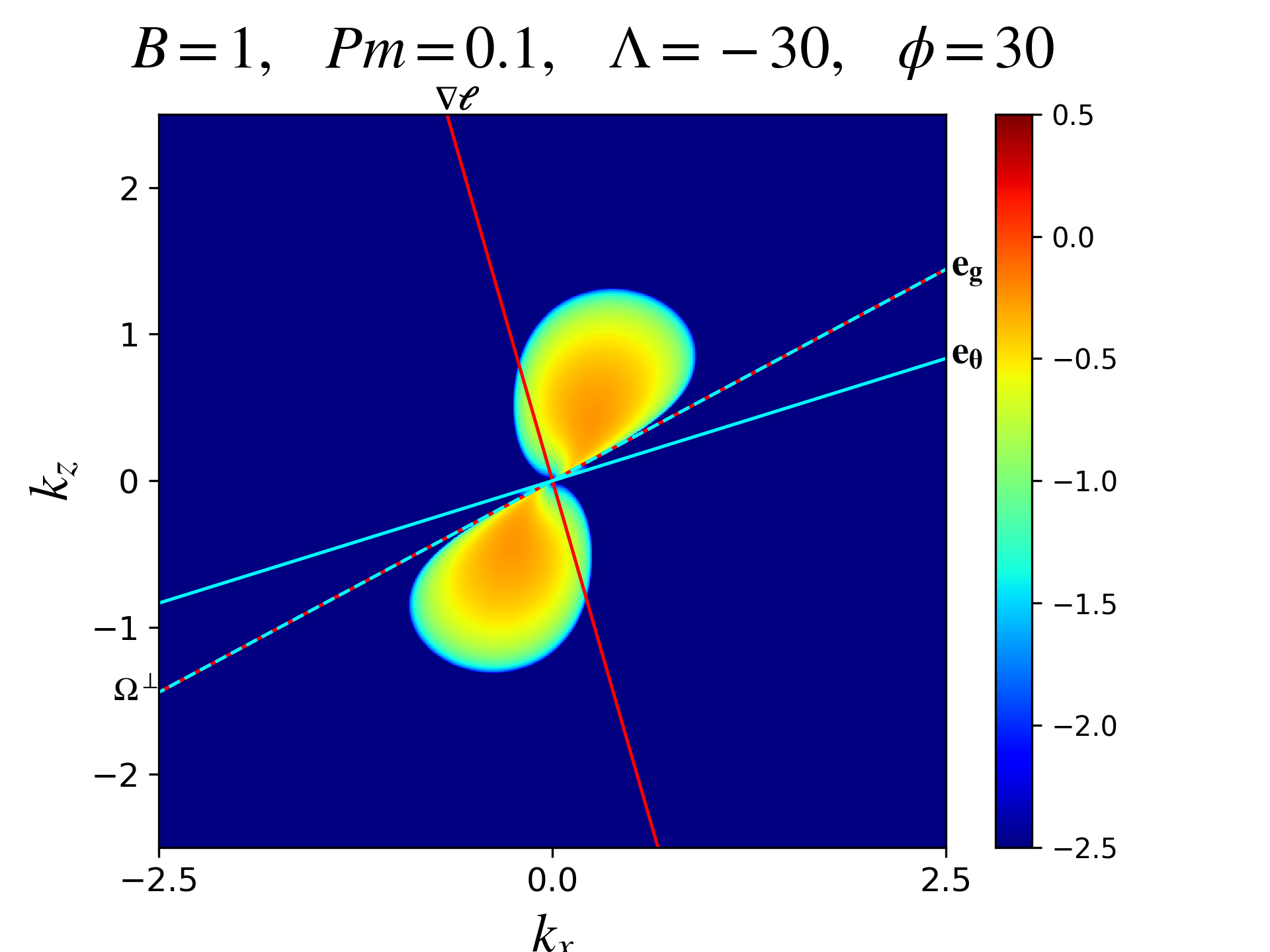}}
    \subfigure[]{\includegraphics[trim=0cm 0cm 0cm 0cm,clip=true,
    width=0.33\textwidth]{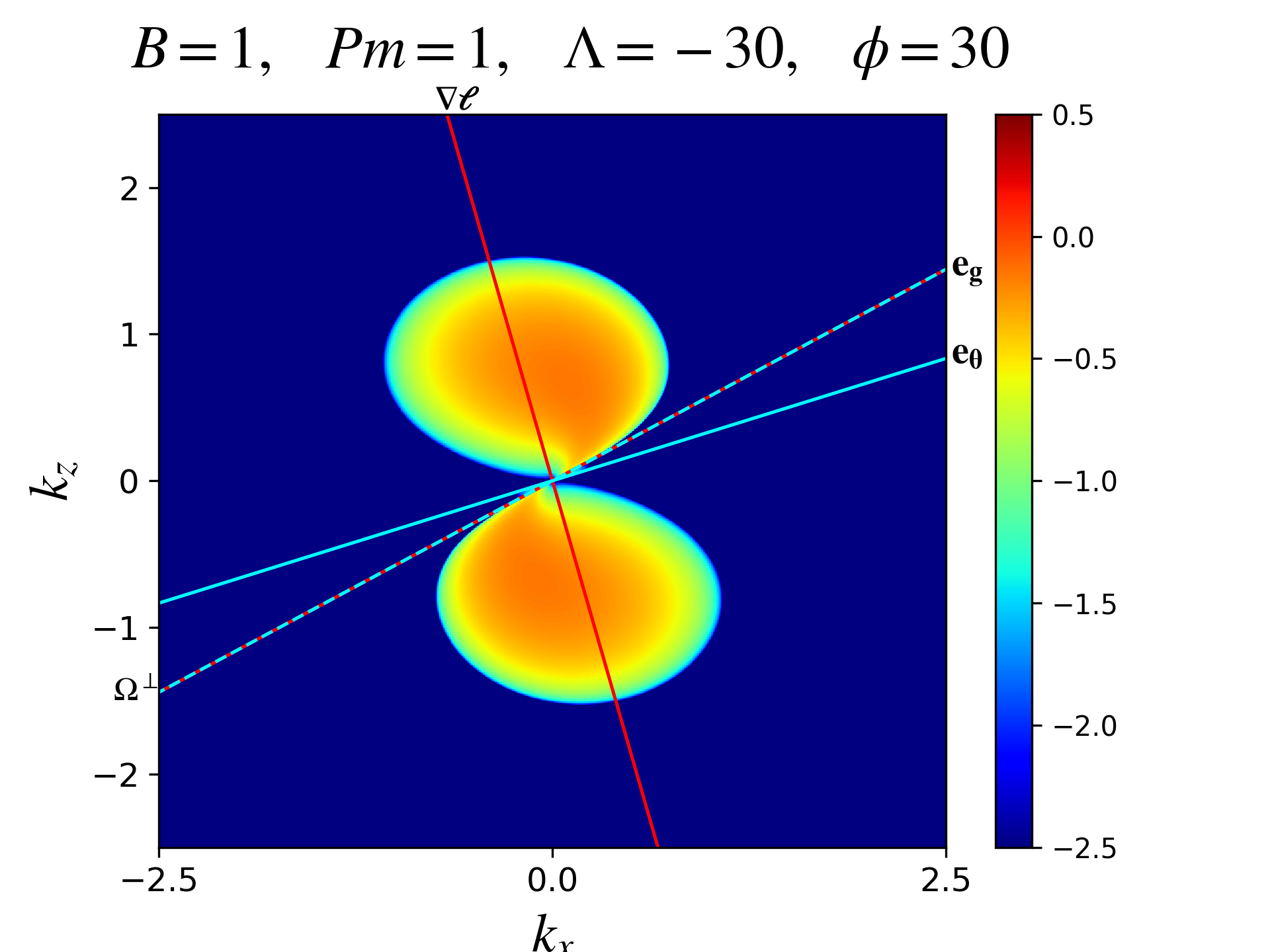}}
    
    \subfigure[]{\includegraphics[trim=0cm 0cm 0cm 0cm,clip=true,
    width=0.33\textwidth]{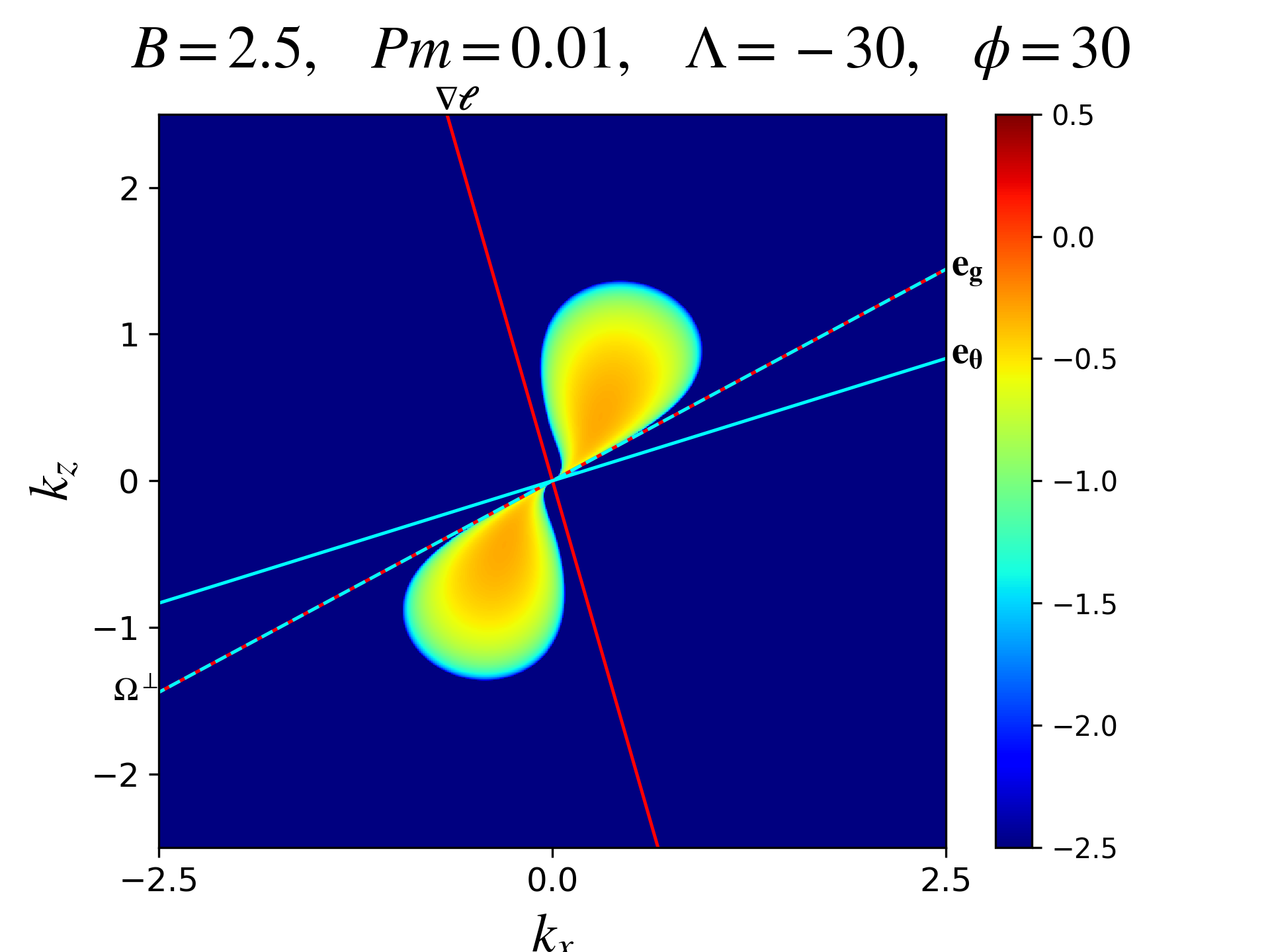}}
    \subfigure[]{\includegraphics[trim=0cm 0cm 0cm 0cm,clip=true,
    width=0.33\textwidth]{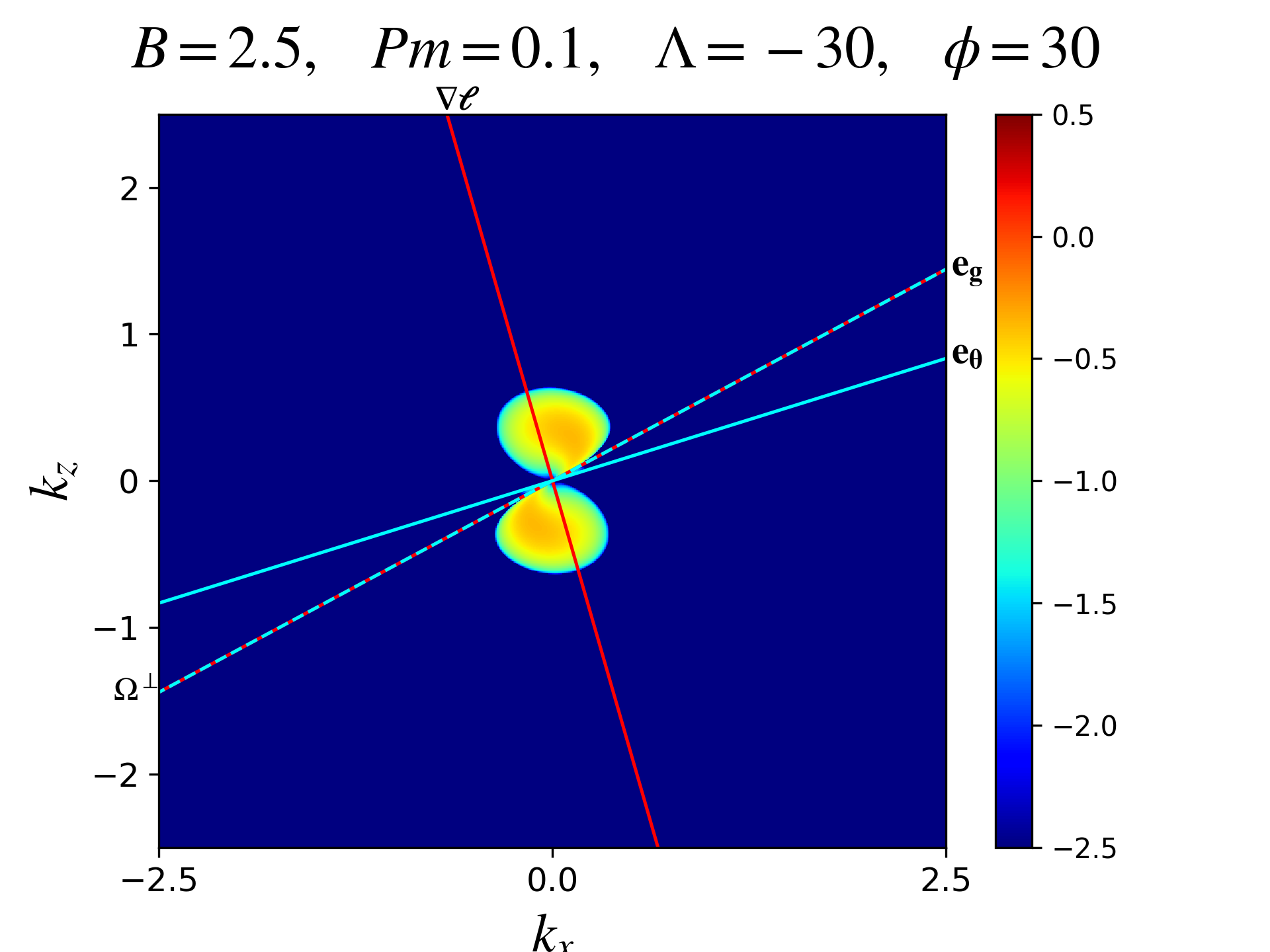}}
    \subfigure[]{\includegraphics[trim=0cm 0cm 0cm 0cm,clip=true,
    width=0.33\textwidth]{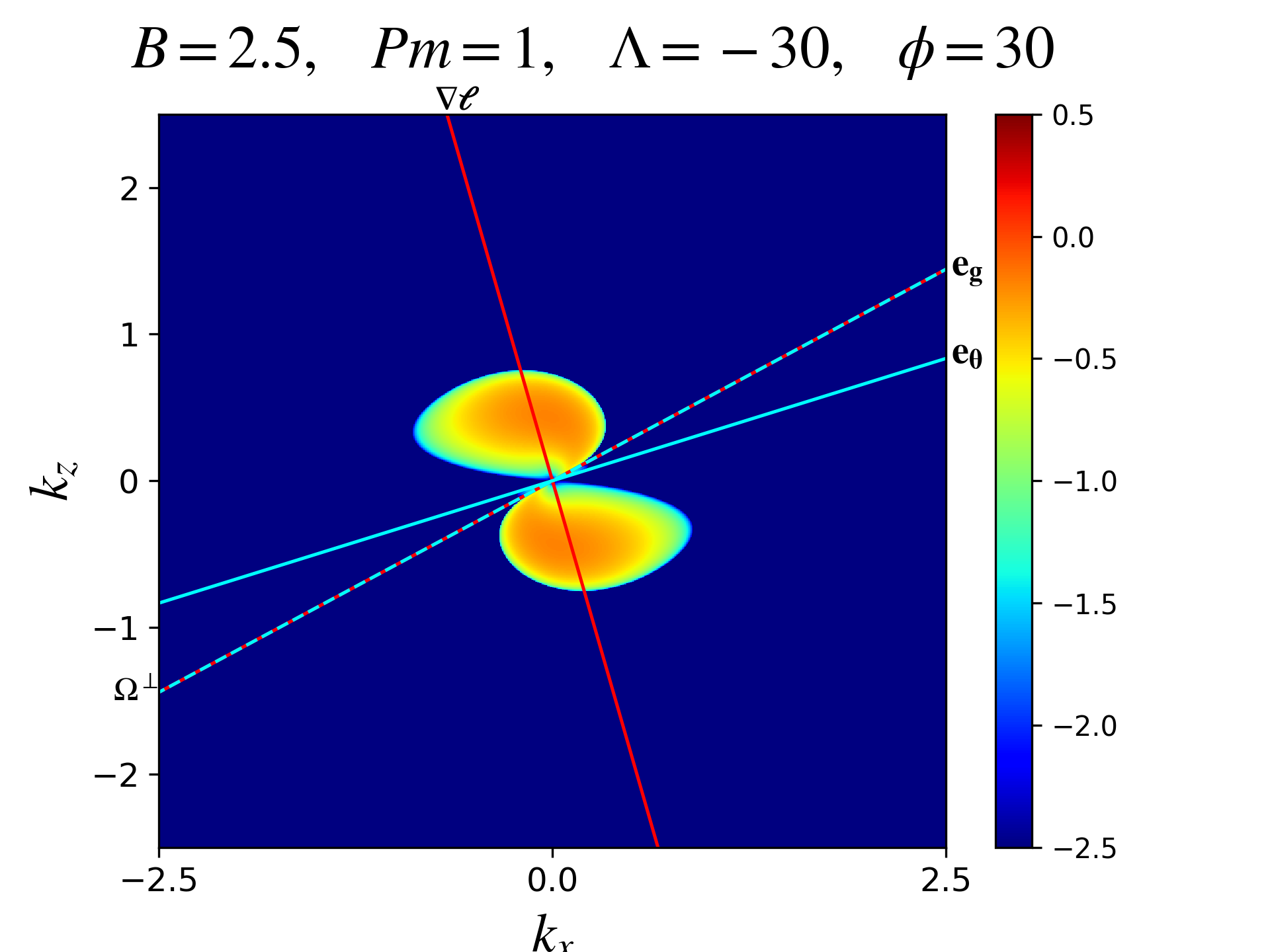}}
    
    \subfigure[]{\includegraphics[trim=0cm 0cm 0cm 0cm,clip=true,
    width=0.33\textwidth]{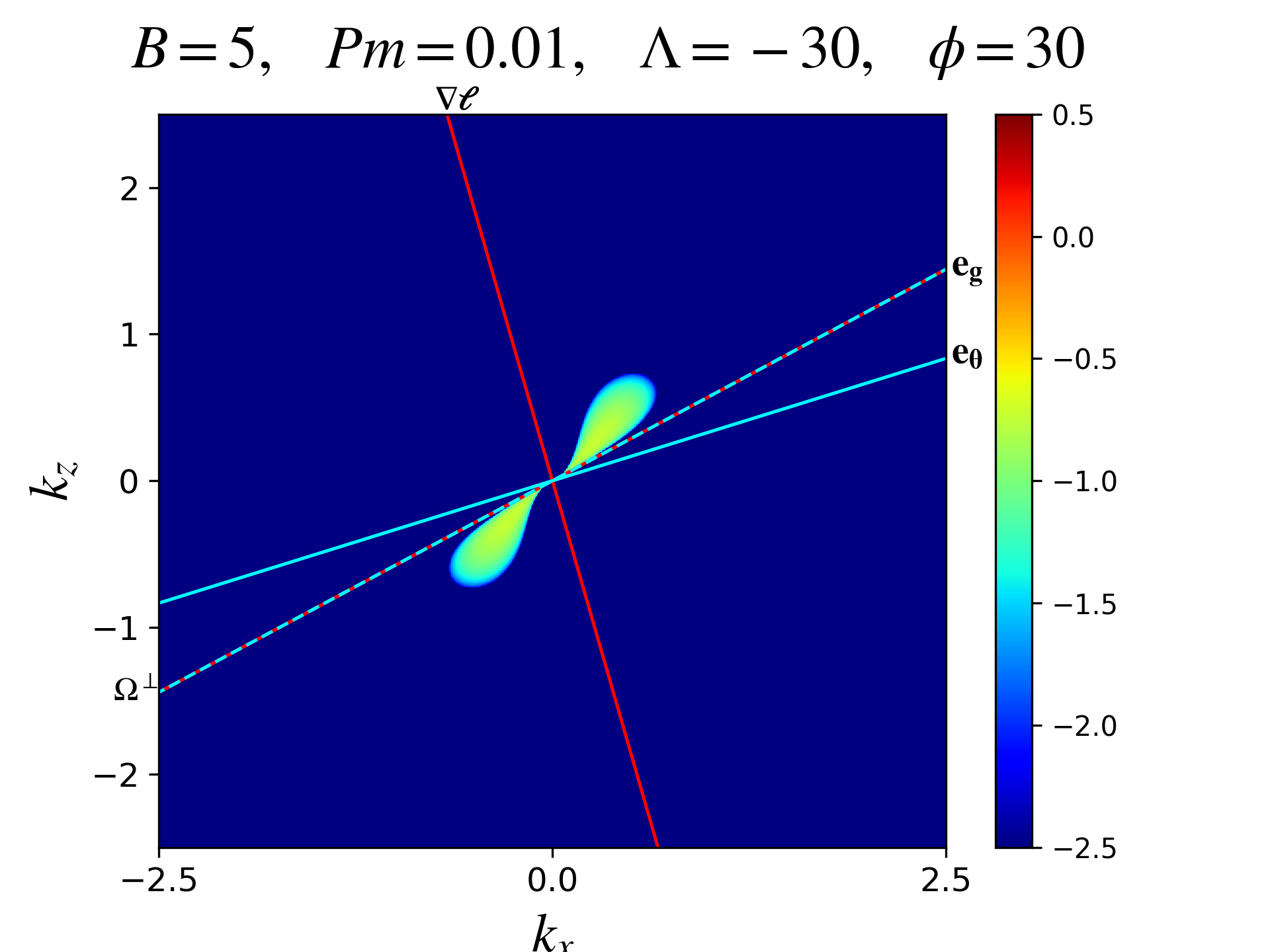}}
    \subfigure[]{\includegraphics[trim=0cm 0cm 0cm 0cm,clip=true,
    width=0.33\textwidth]{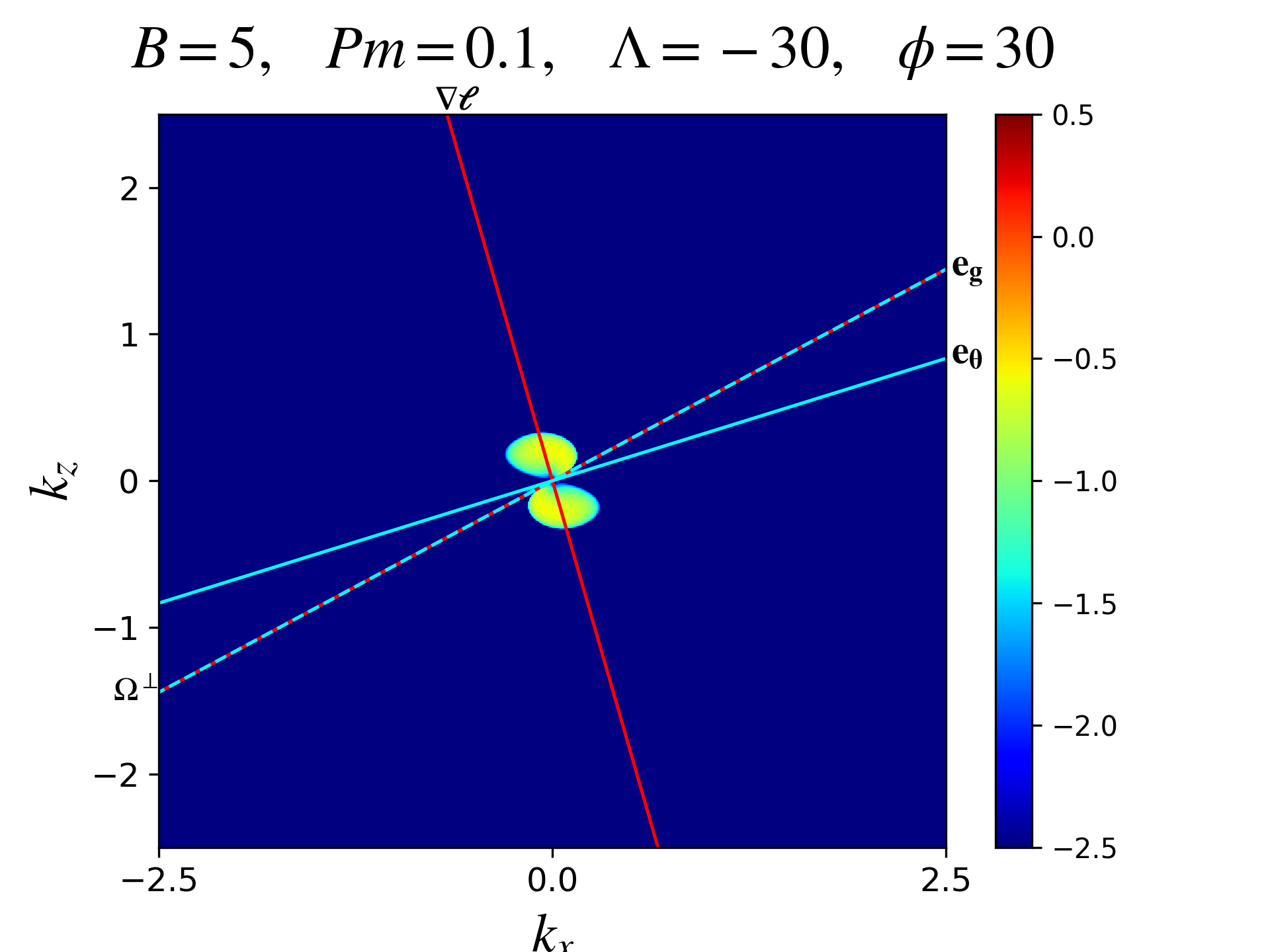}}
    \subfigure[]{\includegraphics[trim=0cm 0cm 0cm 0cm,clip=true,
    width=0.33\textwidth]{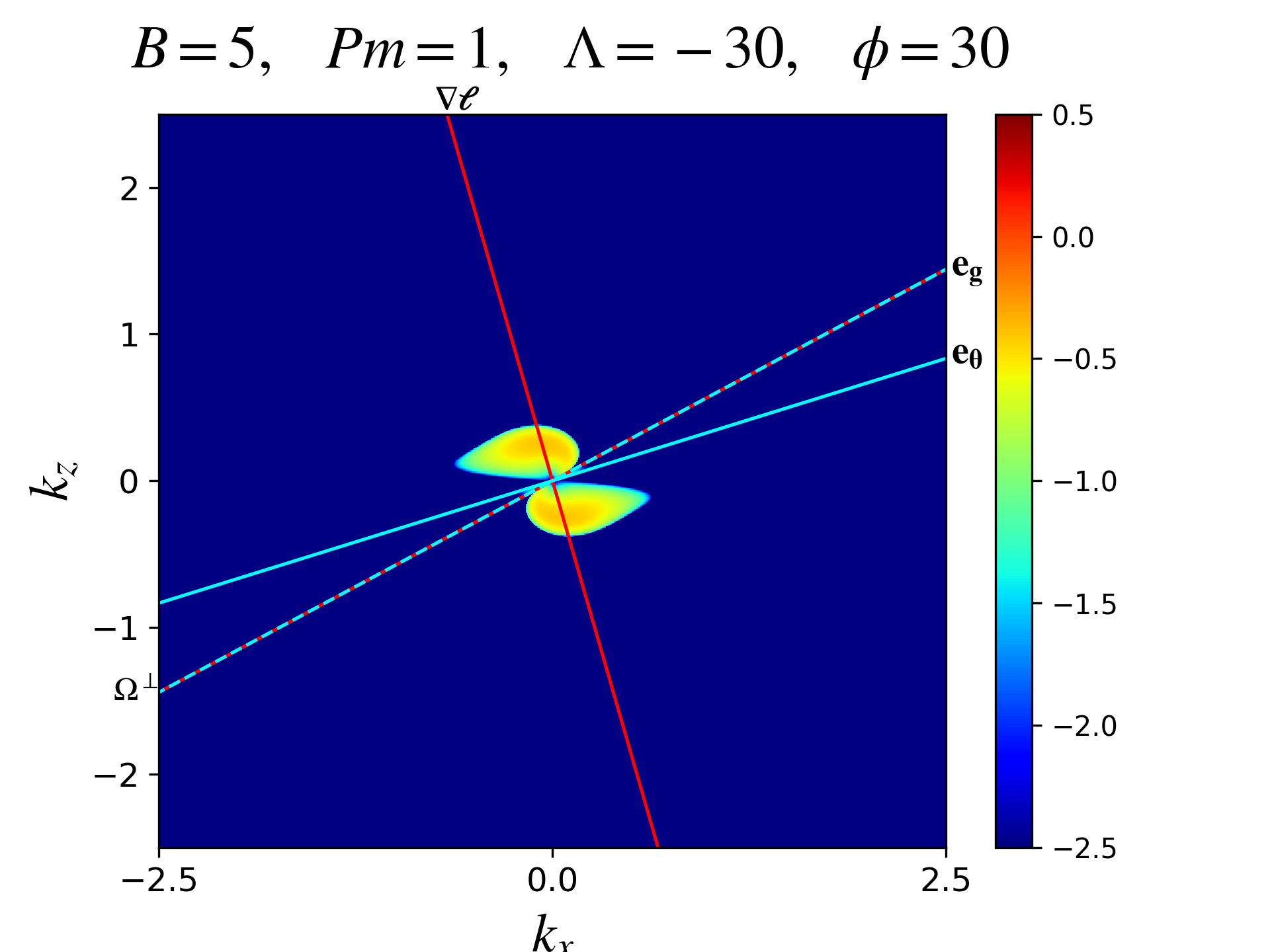}}
    
  \caption{Logarithm of the growth rate $\log_{10} (\sigma/\Omega)$ of axisymmetric perturbations plotted on the $(k_x, k_z)$-plane according to Eq.~\ref{DR}, for various $B_0$ and Pm, with $\phi=30^\circ, \Lambda = -30^\circ$, i.e.~a mixed radial/latitudinal shear at the equator with latitude $\Lambda+\phi=0^\circ$. Parameters are $\mathcal{N}^2/\Omega^2 = 10$, Pr$=10^{-2}$, $\mathcal{S}/\Omega=2$. We vary the strength of the magnetic field from  $B_0=0$ to $B_0=5$ down each column, and vary Pm from $\mathrm{Pm}=0.01$ to $\mathrm{Pm}=1$ along each row.  GSF modes are primarily confined within the wedge bounded by $\hat{\boldsymbol{\Omega}}^\perp$ and $\nabla \ell$ (red lines). However as the field strength increases (downwards) the wavevector orientation is shifted to correspond more with MRI.  Increasing $B_0$ at fixed Pm decreases both the maximum growth rate and the size of the unstable region on the $(k_x,k_z)$-plane. Reducing magnetic diffusivity by increasing Pm on the other hand seems to have the opposite effect and enhances the destabilising effects of the field, leading to a larger region of instability as well as a stronger destabilisation of the dominant mode.}
  \label{Lobes1}
\end{figure*}

\begin{figure*}
    \subfigure[]{\includegraphics[trim=0cm 0cm 0cm 0cm,clip=true,
    width=0.33\textwidth]{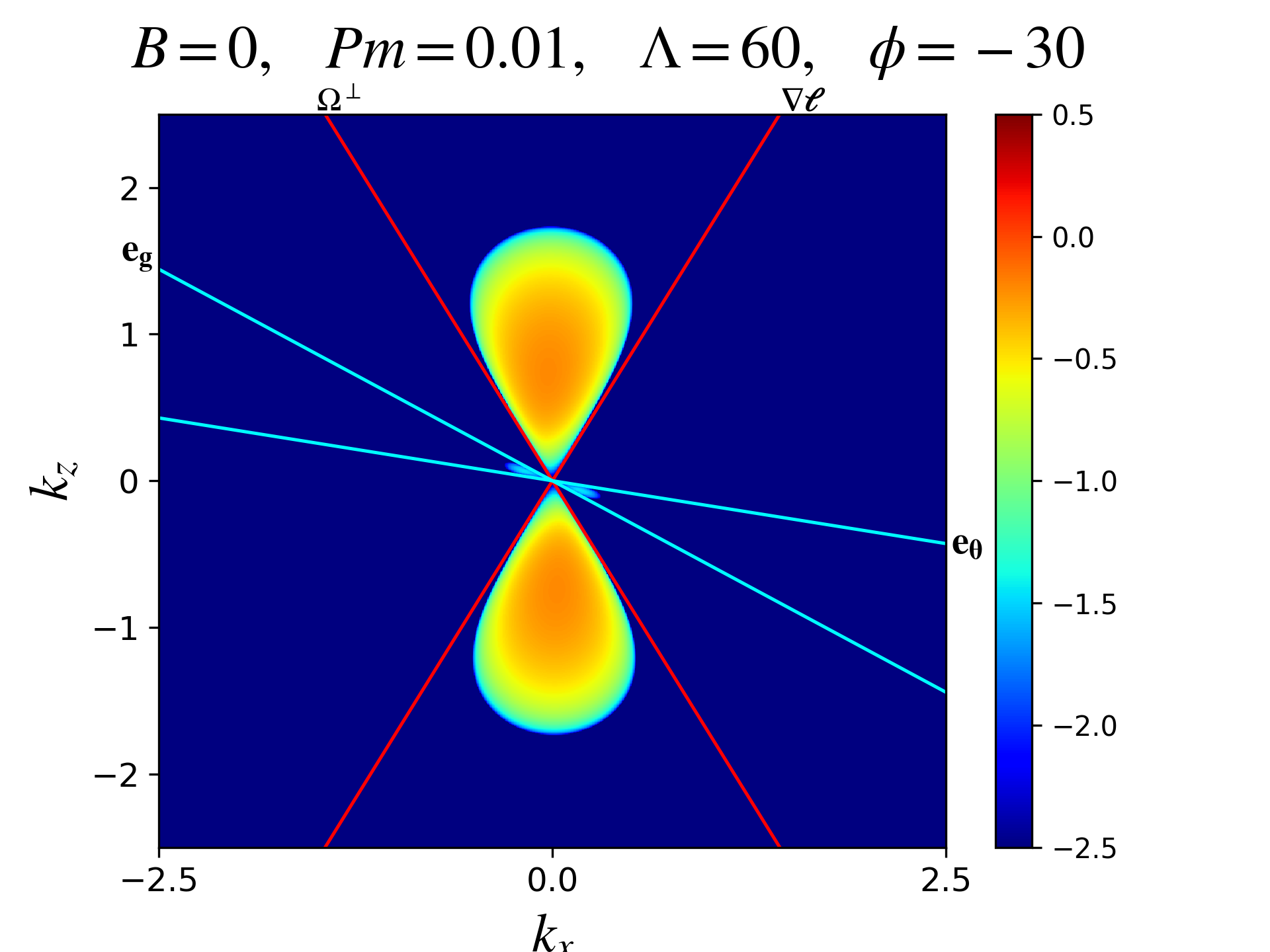}}
    \subfigure[]{\includegraphics[trim=0cm 0cm 0cm 0cm,clip=true,
    width=0.33\textwidth]{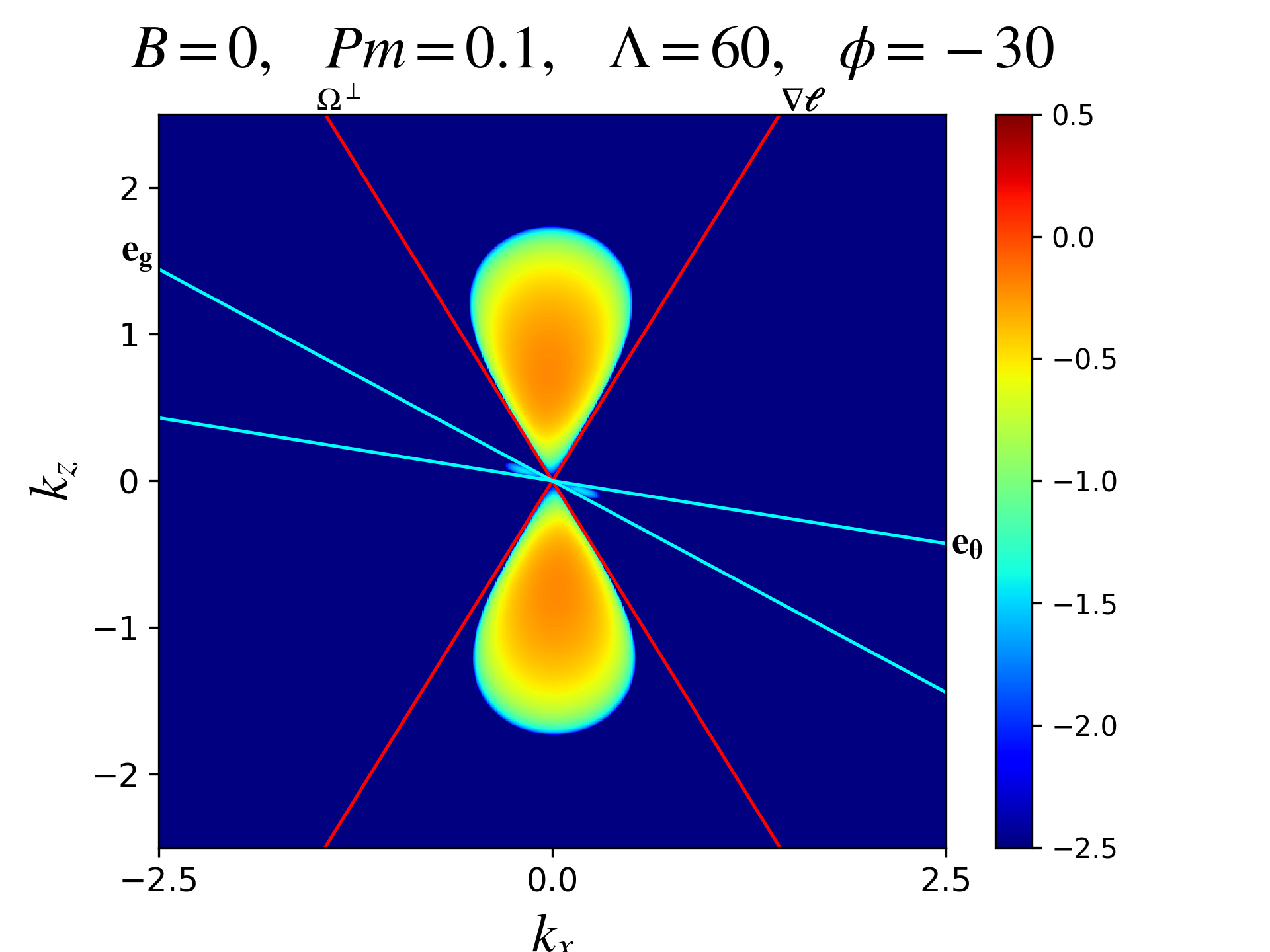}}
    \subfigure[]{\includegraphics[trim=0cm 0cm 0cm 0cm,clip=true,
    width=0.33\textwidth]{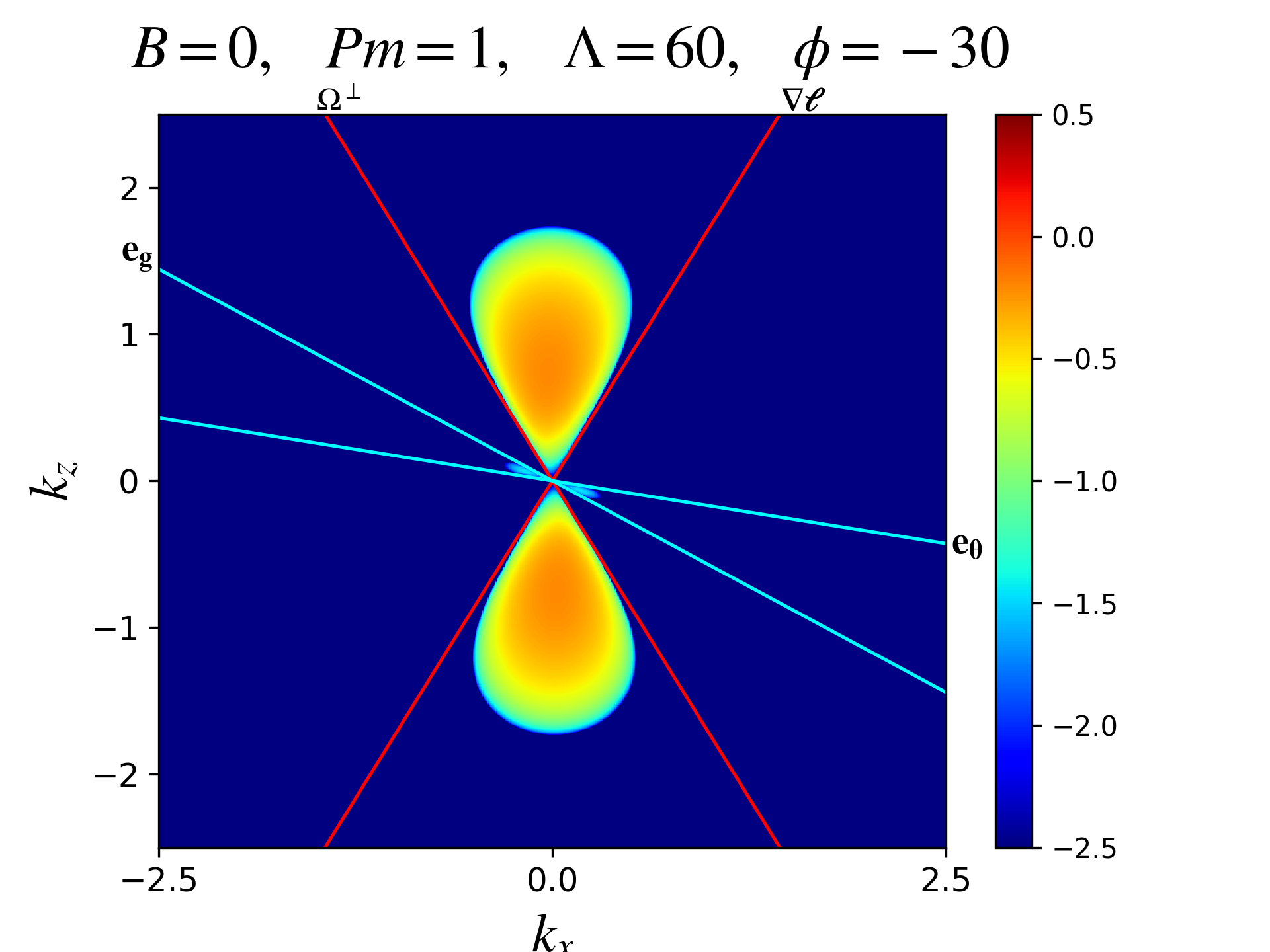}}
    
        \subfigure[]{\includegraphics[trim=0cm 0cm 0cm 0cm,clip=true,
    width=0.33\textwidth]{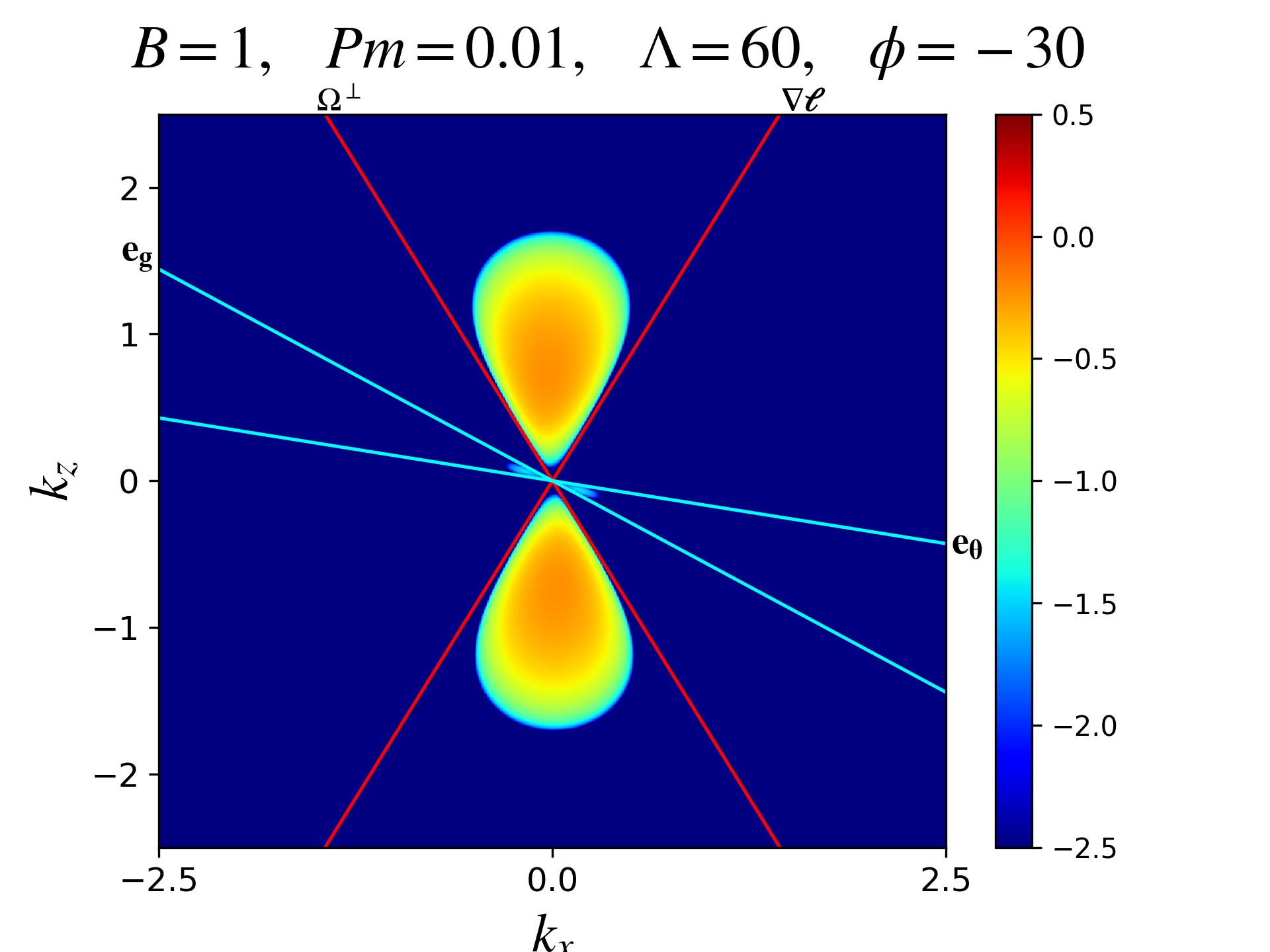}}
    \subfigure[]{\includegraphics[trim=0cm 0cm 0cm 0cm,clip=true,
    width=0.33\textwidth]{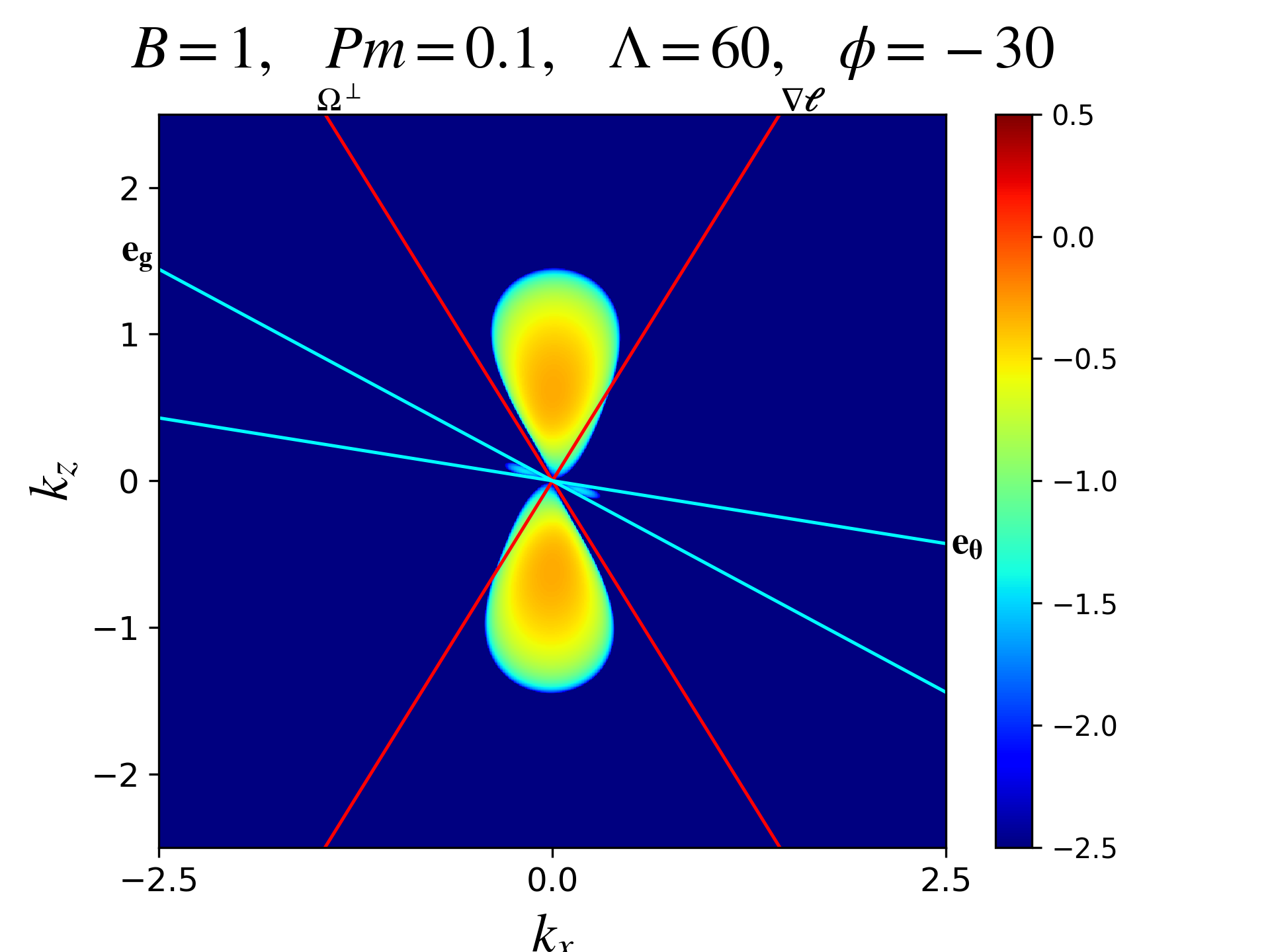}}
    \subfigure[]{\includegraphics[trim=0cm 0cm 0cm 0cm,clip=true,
    width=0.33\textwidth]{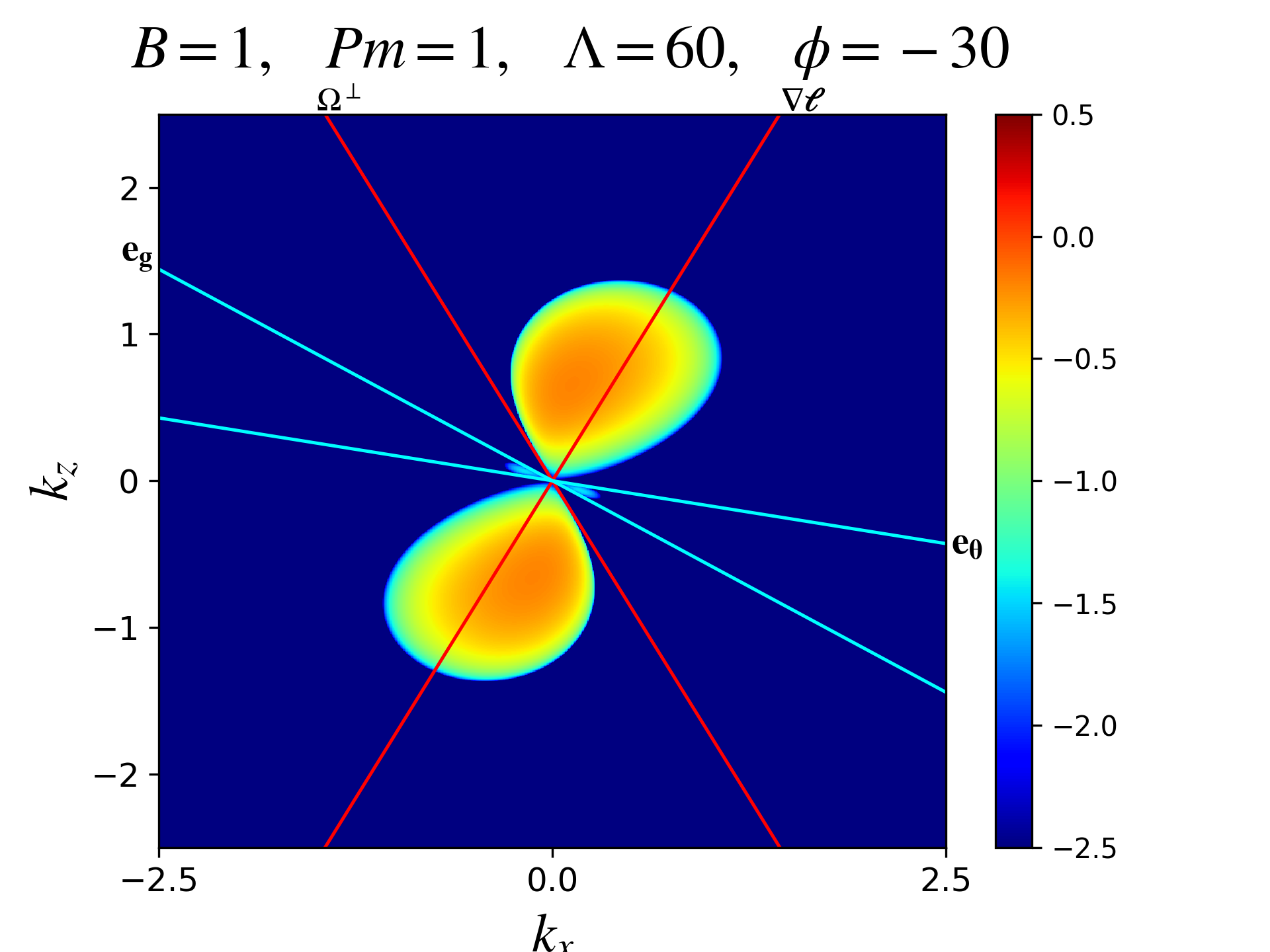}}
    
        \subfigure[]{\includegraphics[trim=0cm 0cm 0cm 0cm,clip=true,
    width=0.33\textwidth]{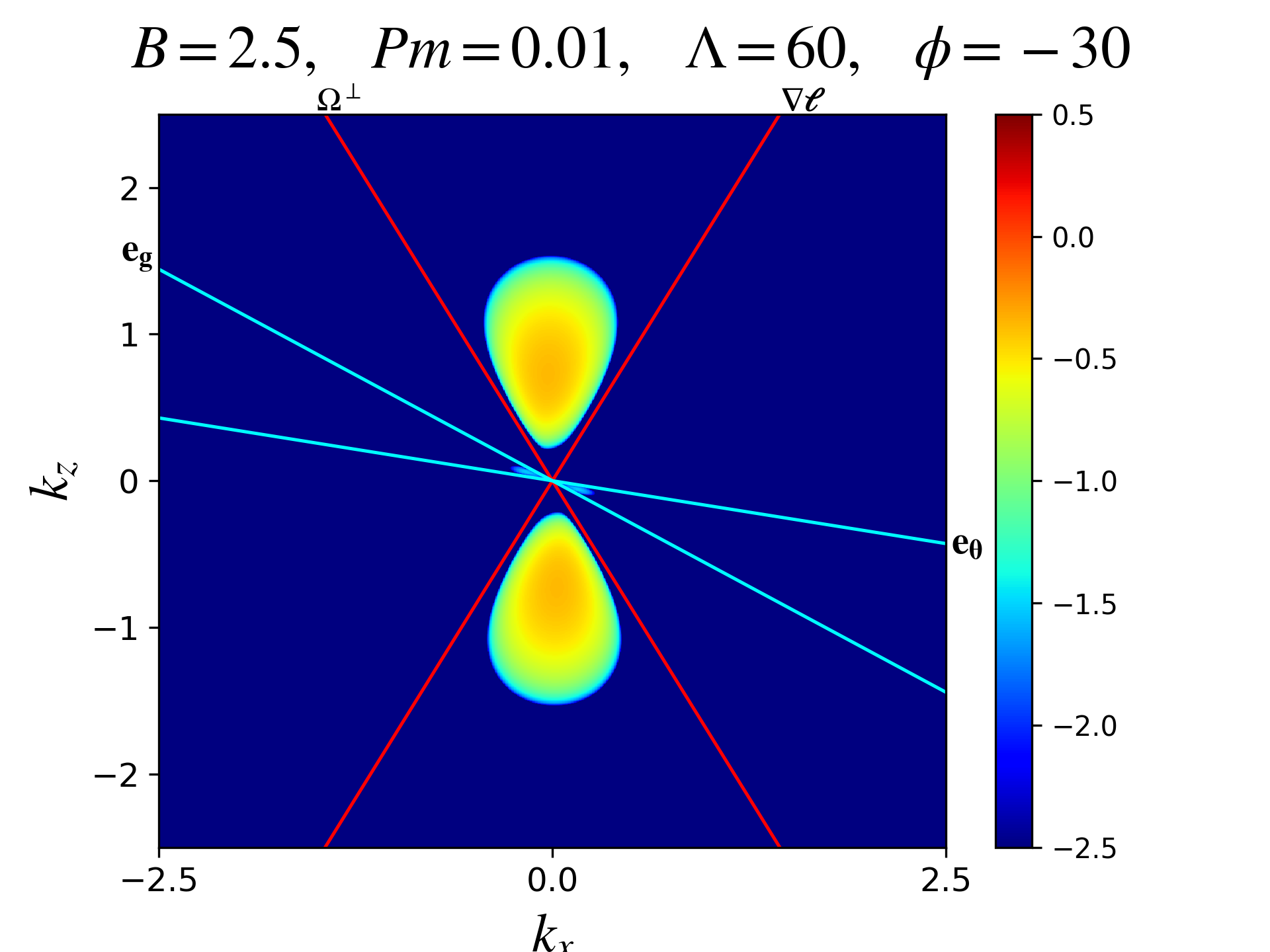}}
    \subfigure[]{\includegraphics[trim=0cm 0cm 0cm 0cm,clip=true,
    width=0.33\textwidth]{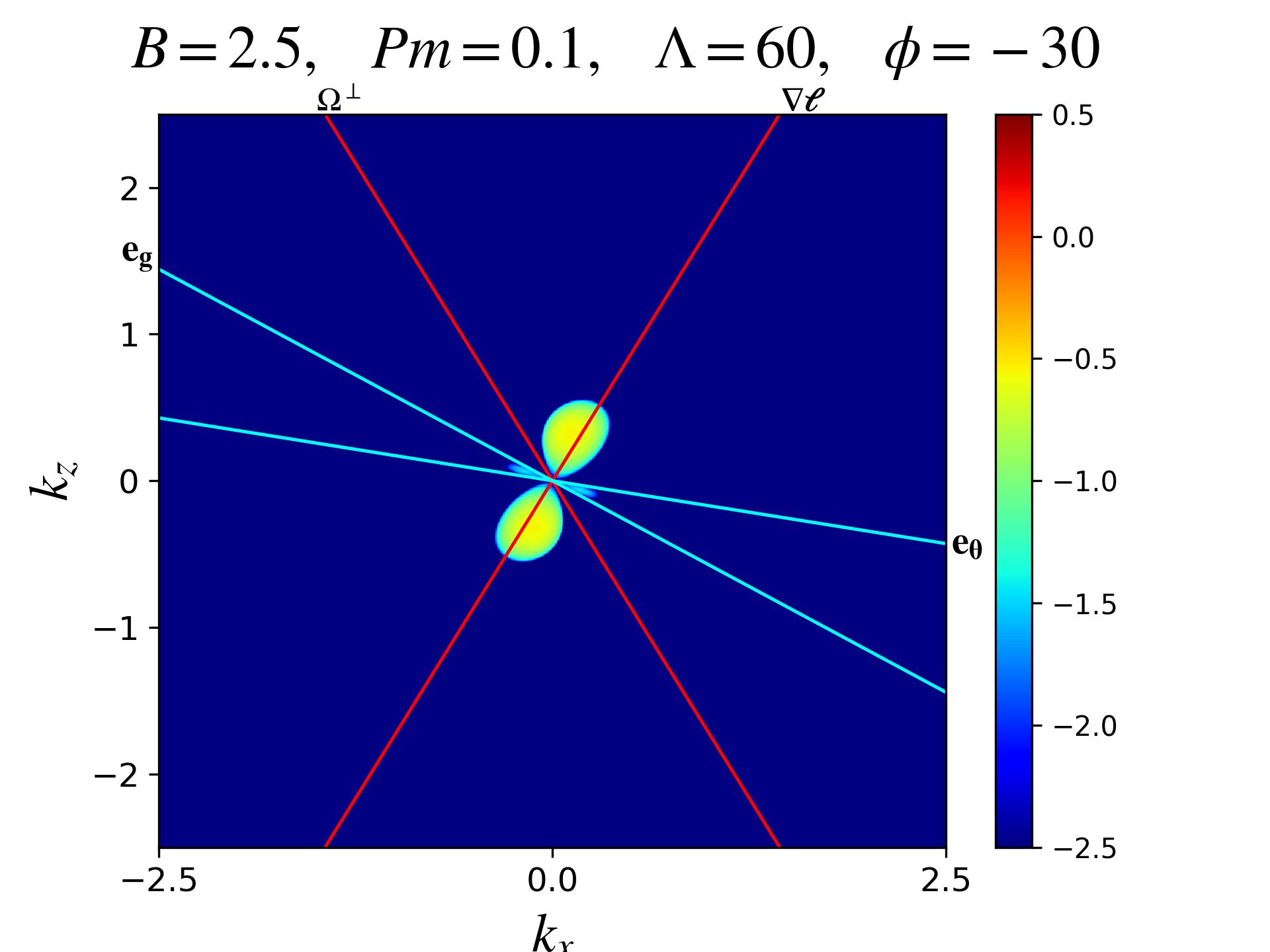}}
    \subfigure[]{\includegraphics[trim=0cm 0cm 0cm 0cm,clip=true,
    width=0.33\textwidth]{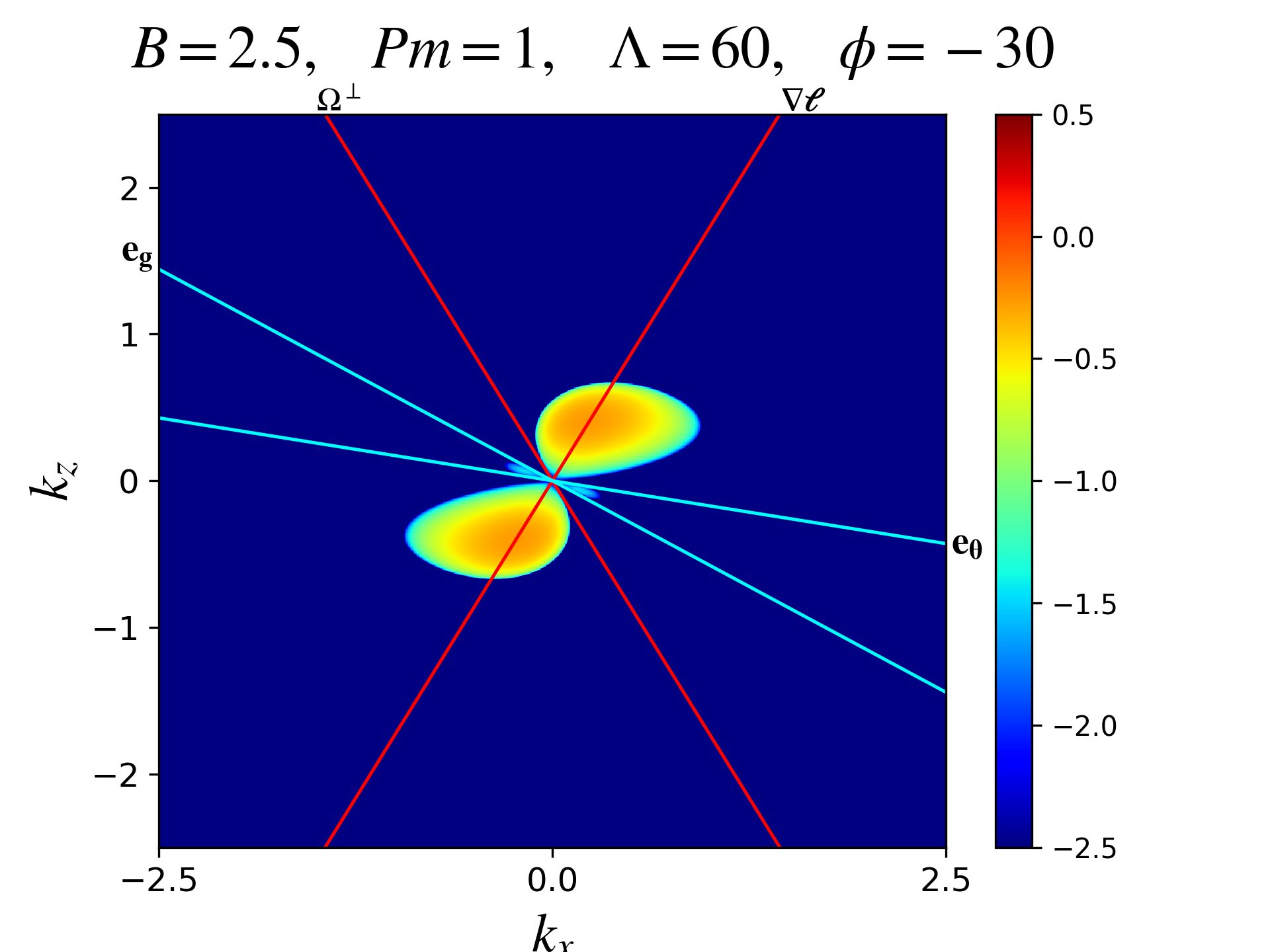}}
    
        \subfigure[]{\includegraphics[trim=0cm 0cm 0cm 0cm,clip=true,
    width=0.33\textwidth]{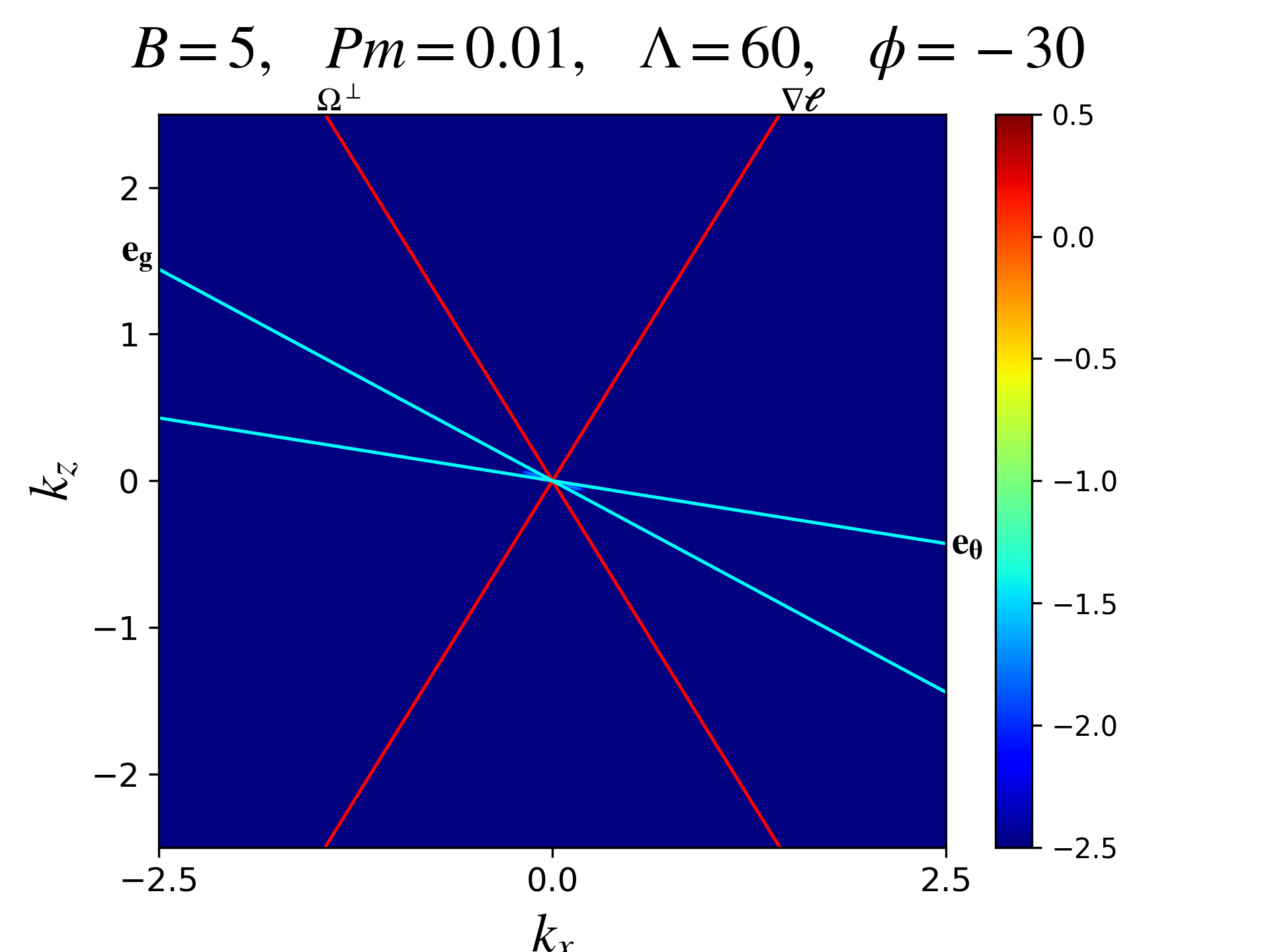}}
    \subfigure[]{\includegraphics[trim=0cm 0cm 0cm 0cm,clip=true,
    width=0.33\textwidth]{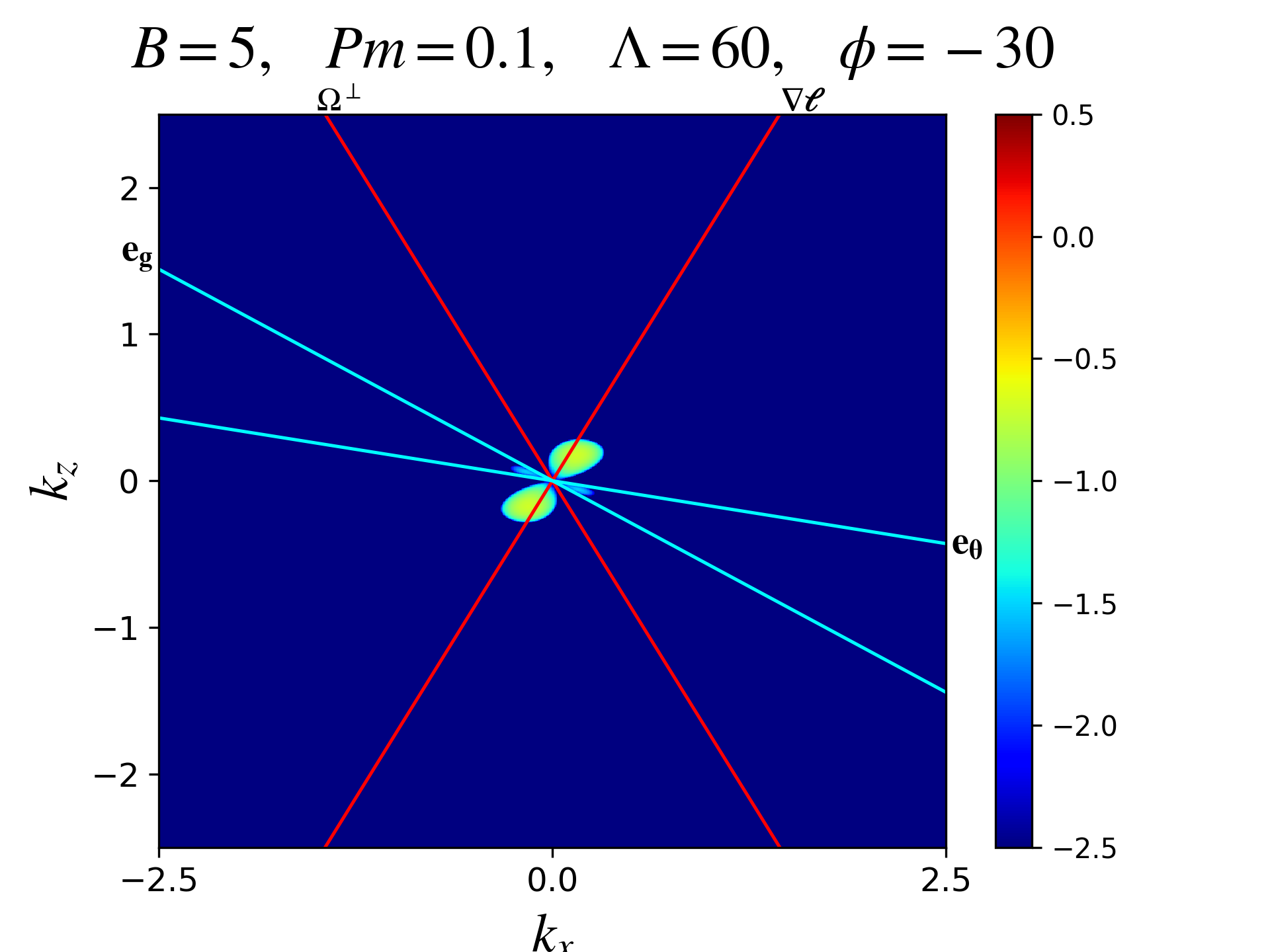}}
    \subfigure[]{\includegraphics[trim=0cm 0cm 0cm 0cm,clip=true,
    width=0.33\textwidth]{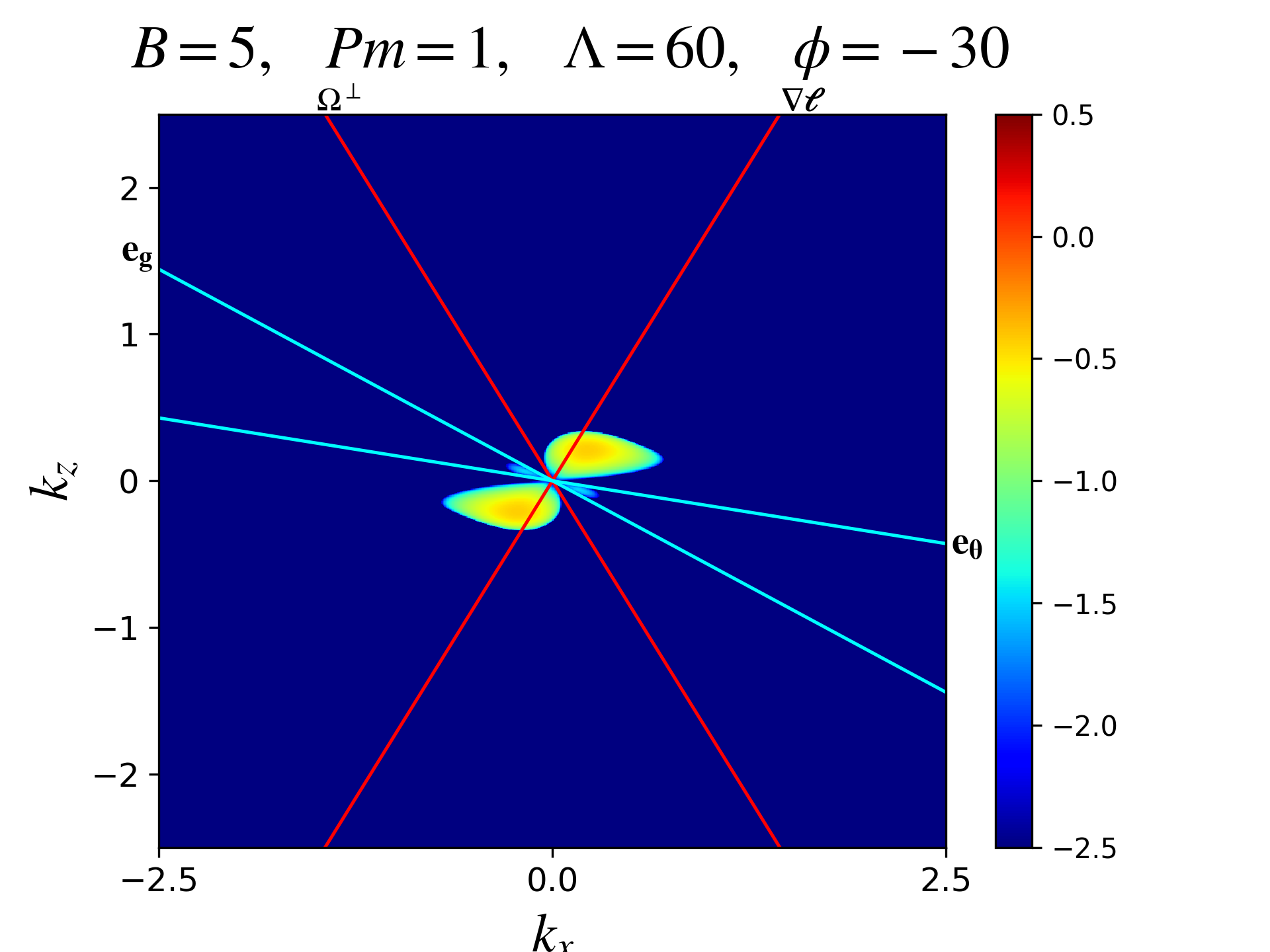}}

  \caption{Logarithm of the growth rate $\log_{10} (\sigma/\Omega)$ of axisymmetric perturbations plotted on the $(k_x, k_z)$-plane according to Eq.~\ref{DR}, for various $B_0$ and Pm, with $\phi=-30^\circ, \Lambda = 60^\circ$, i.e.~a mixed radial/latitudinal shear at latitude $\Lambda+\phi=30^\circ$. Parameters are $\mathcal{N}^2/\Omega^2 = 10$, Pr$=10^{-2}$, $\mathcal{S}/\Omega=2$. We vary the strength of the magnetic field from $B_0=0$ to $B_0=5$ down each column, and vary Pm from $\mathrm{Pm}=0.01$ to $\mathrm{Pm}=1$ along each row. GSF unstable modes are primarily confined to within the wedge bounded by $\hat{\boldsymbol{\Omega}}^\perp$ and $\nabla \ell$ (red lines) for weak fields, but this direction is modified when the  MRI takes over. We observe a secondary set of unstable oscillatory modes, consisting of weakly destabilised magneto-inertial-gravity waves. These grow more weakly than the primary lobes but they are less affected by the stabilising  effects of the magnetic field and even operate (as in the $B_0=5, \mathrm{Pm}=0.01$ case) when the primary lobes have been stabilised by it.}
  \label{Lobes2}
\end{figure*}

\begin{figure*}
    \subfigure[]{\includegraphics[trim=0cm 0cm 0cm 0cm,clip=true,
    width=0.33\textwidth]{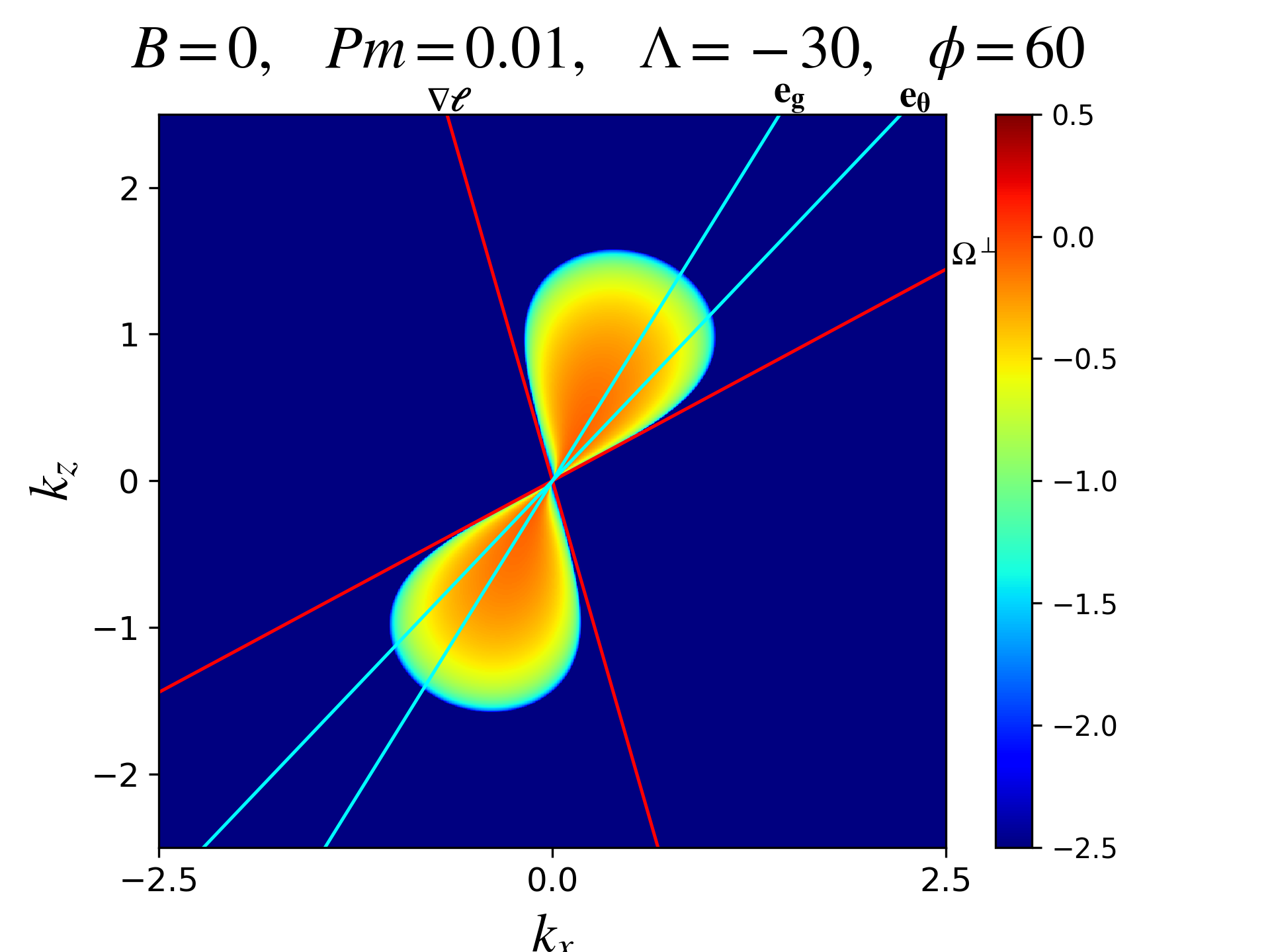}}
    \subfigure[]{\includegraphics[trim=0cm 0cm 0cm 0cm,clip=true,
    width=0.33\textwidth]{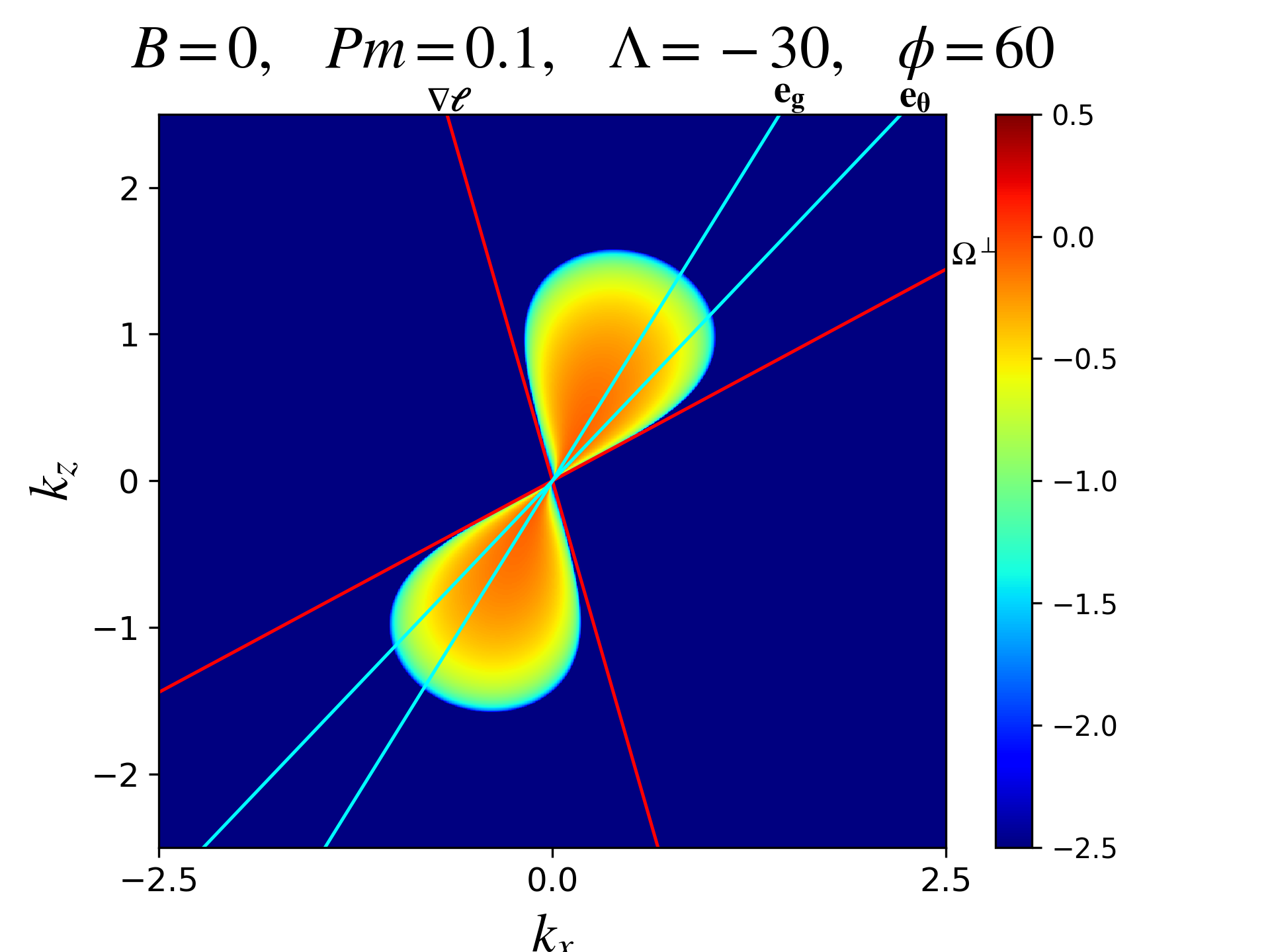}}
    \subfigure[]{\includegraphics[trim=0cm 0cm 0cm 0cm,clip=true,
    width=0.33\textwidth]{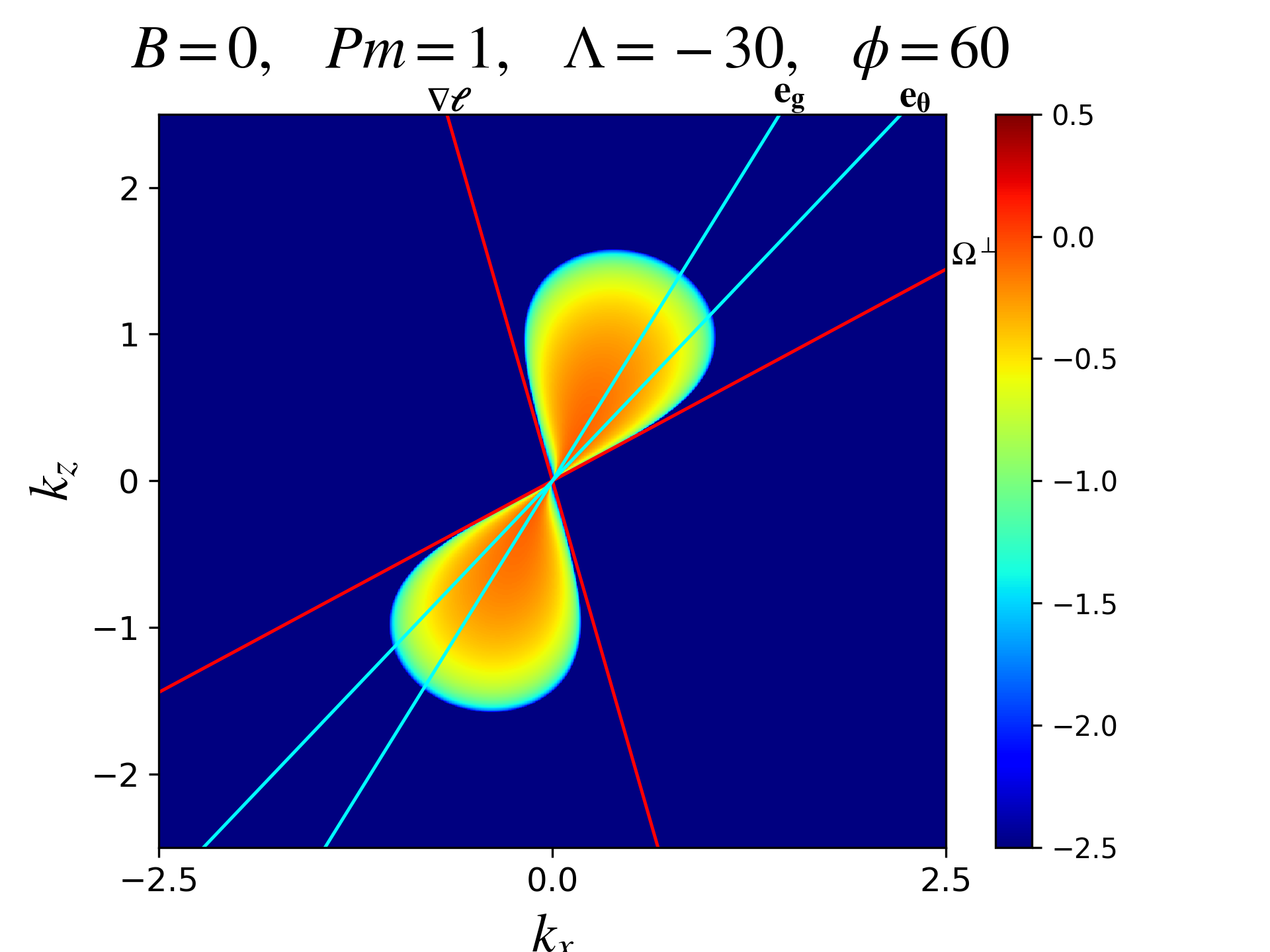}}
  
        \subfigure[]{\includegraphics[trim=0cm 0cm 0cm 0cm,clip=true,
    width=0.33\textwidth]{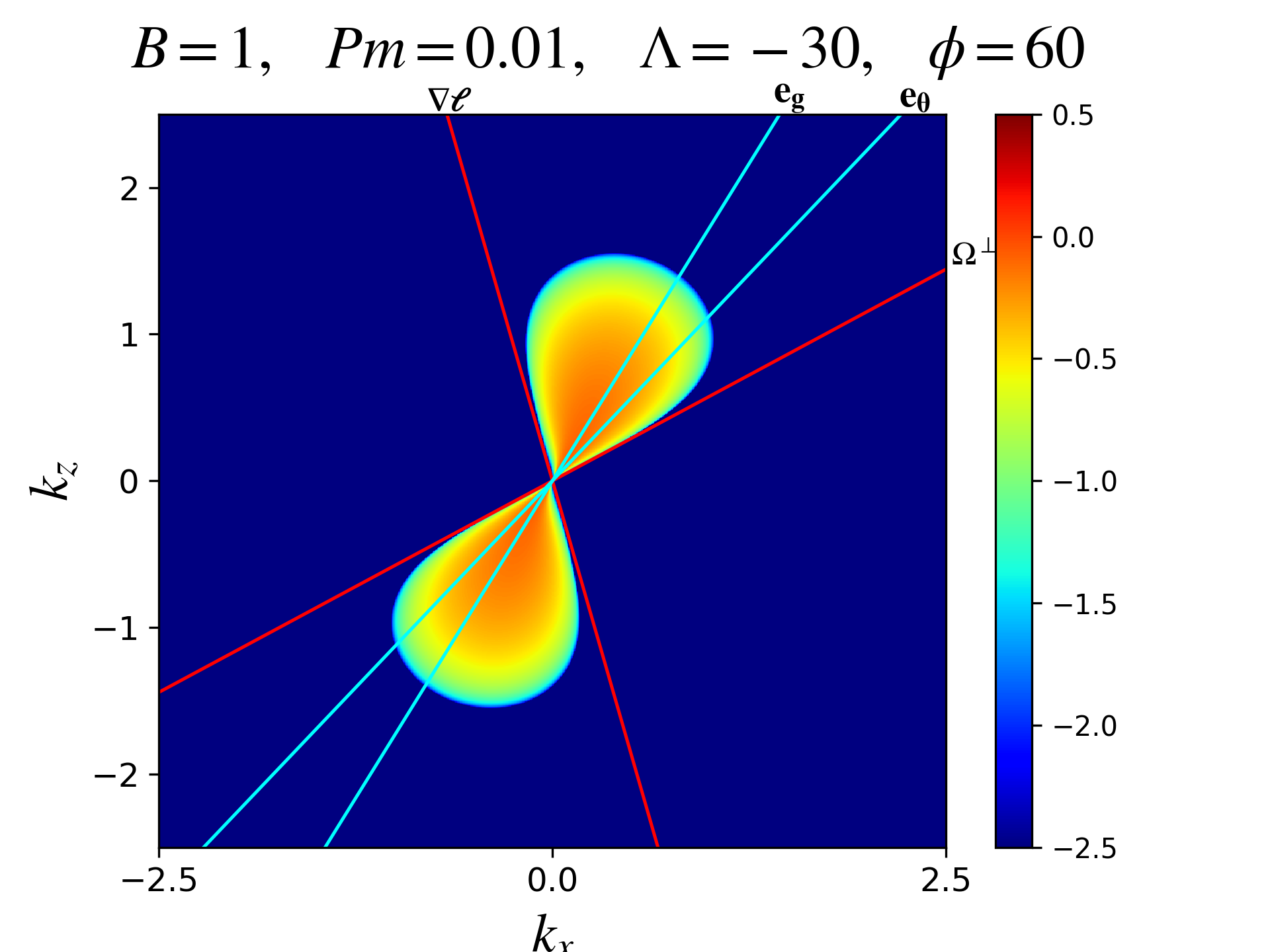}}
    \subfigure[]{\includegraphics[trim=0cm 0cm 0cm 0cm,clip=true,
    width=0.33\textwidth]{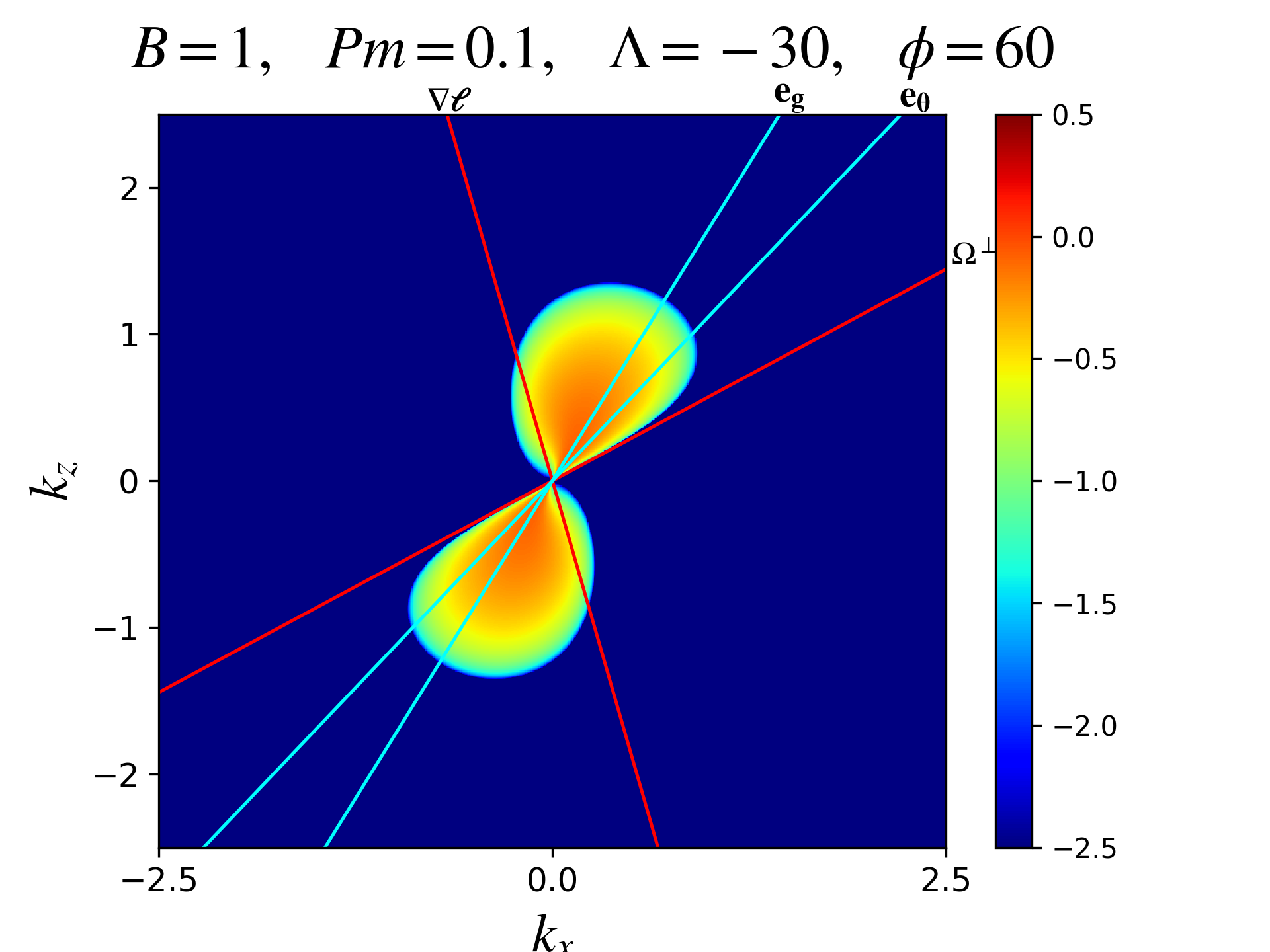}}
    \subfigure[]{\includegraphics[trim=0cm 0cm 0cm 0cm,clip=true,
    width=0.33\textwidth]{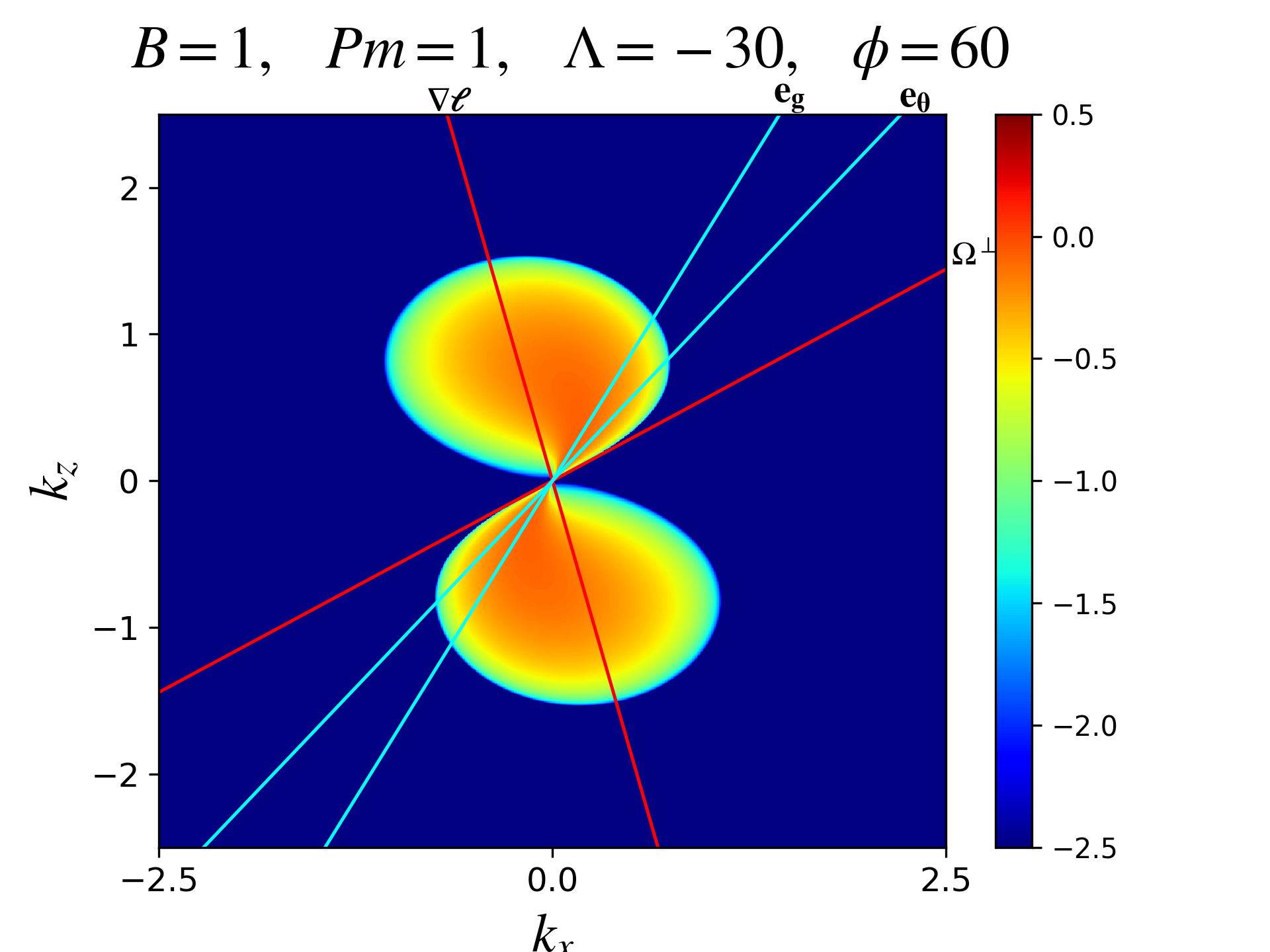}}
    
    \subfigure[]{\includegraphics[trim=0cm 0cm 0cm 0cm,clip=true,
    width=0.33\textwidth]{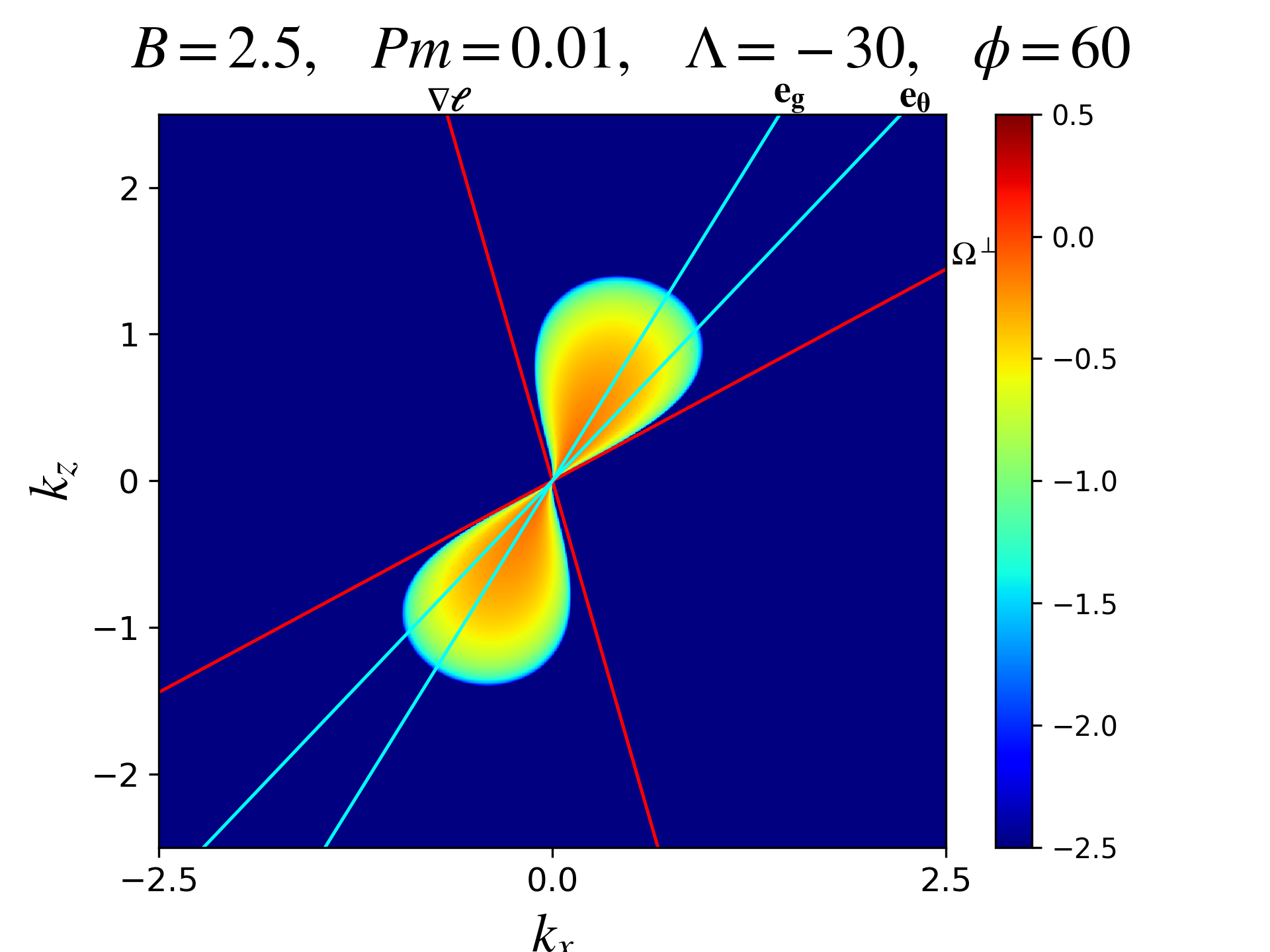}}
    \subfigure[]{\includegraphics[trim=0cm 0cm 0cm 0cm,clip=true,
    width=0.33\textwidth]{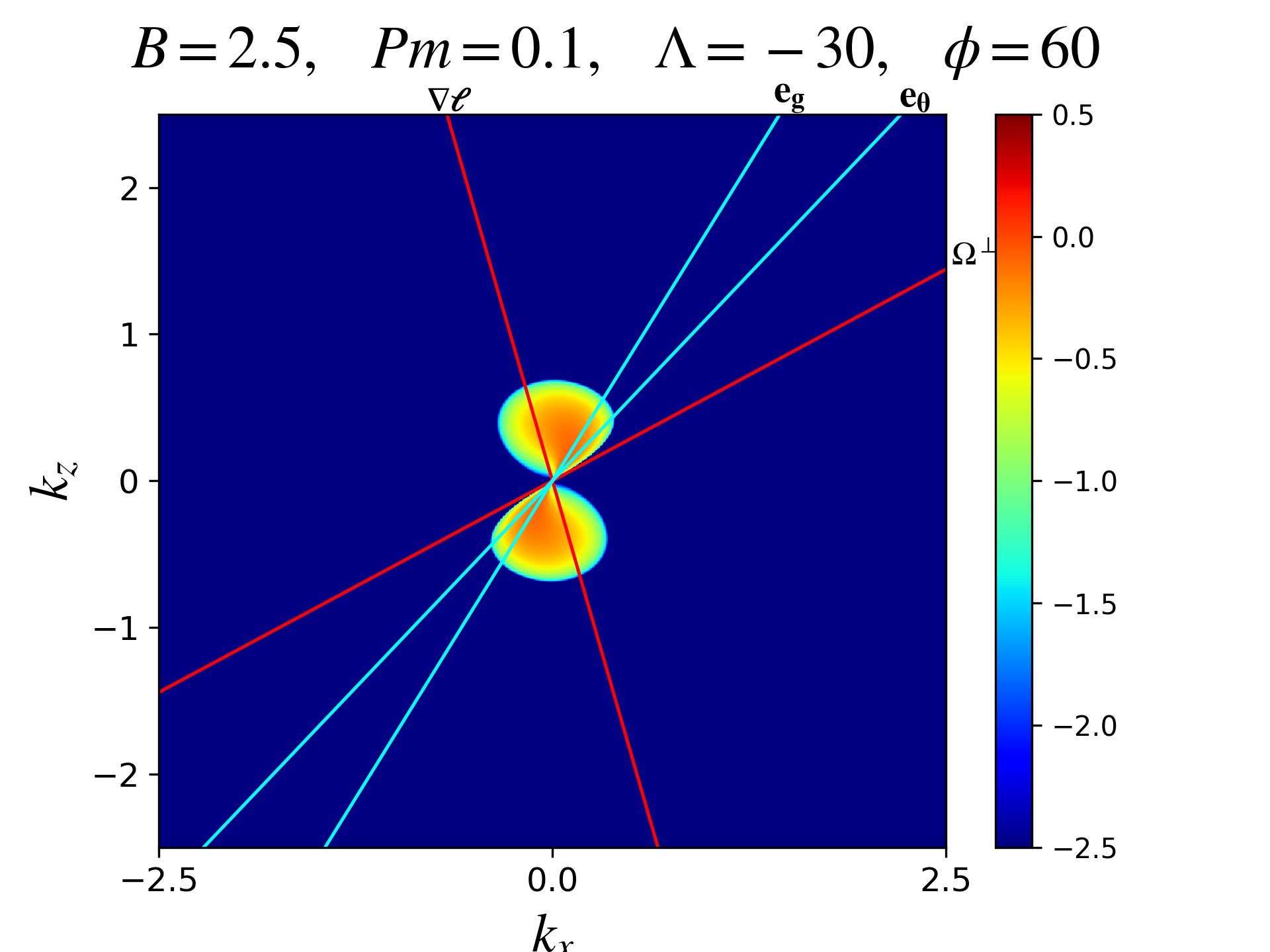}}
    \subfigure[]{\includegraphics[trim=0cm 0cm 0cm 0cm,clip=true,
    width=0.33\textwidth]{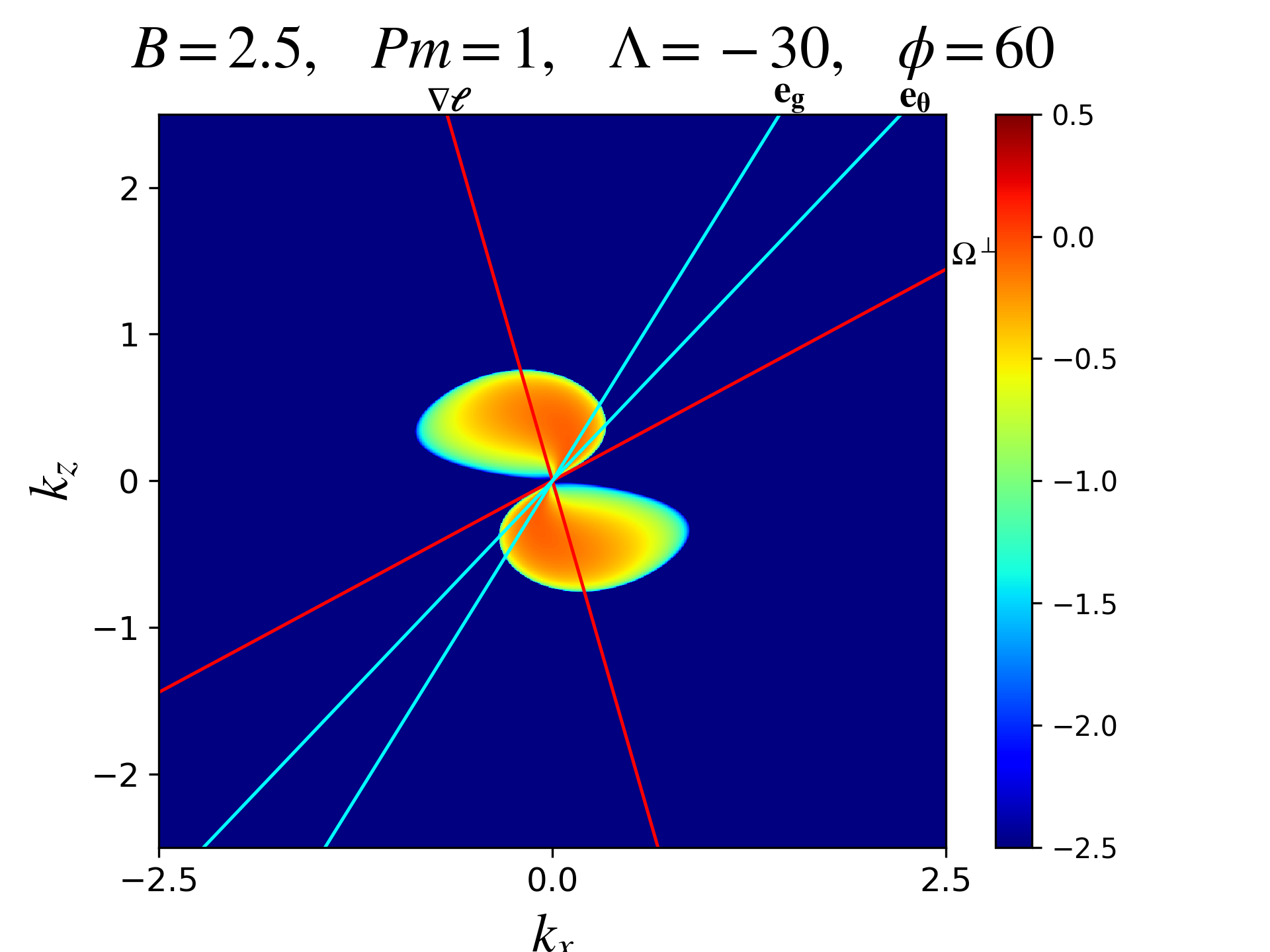}}
    
    \subfigure[]{\includegraphics[trim=0cm 0cm 0cm 0cm,clip=true,
    width=0.33\textwidth]{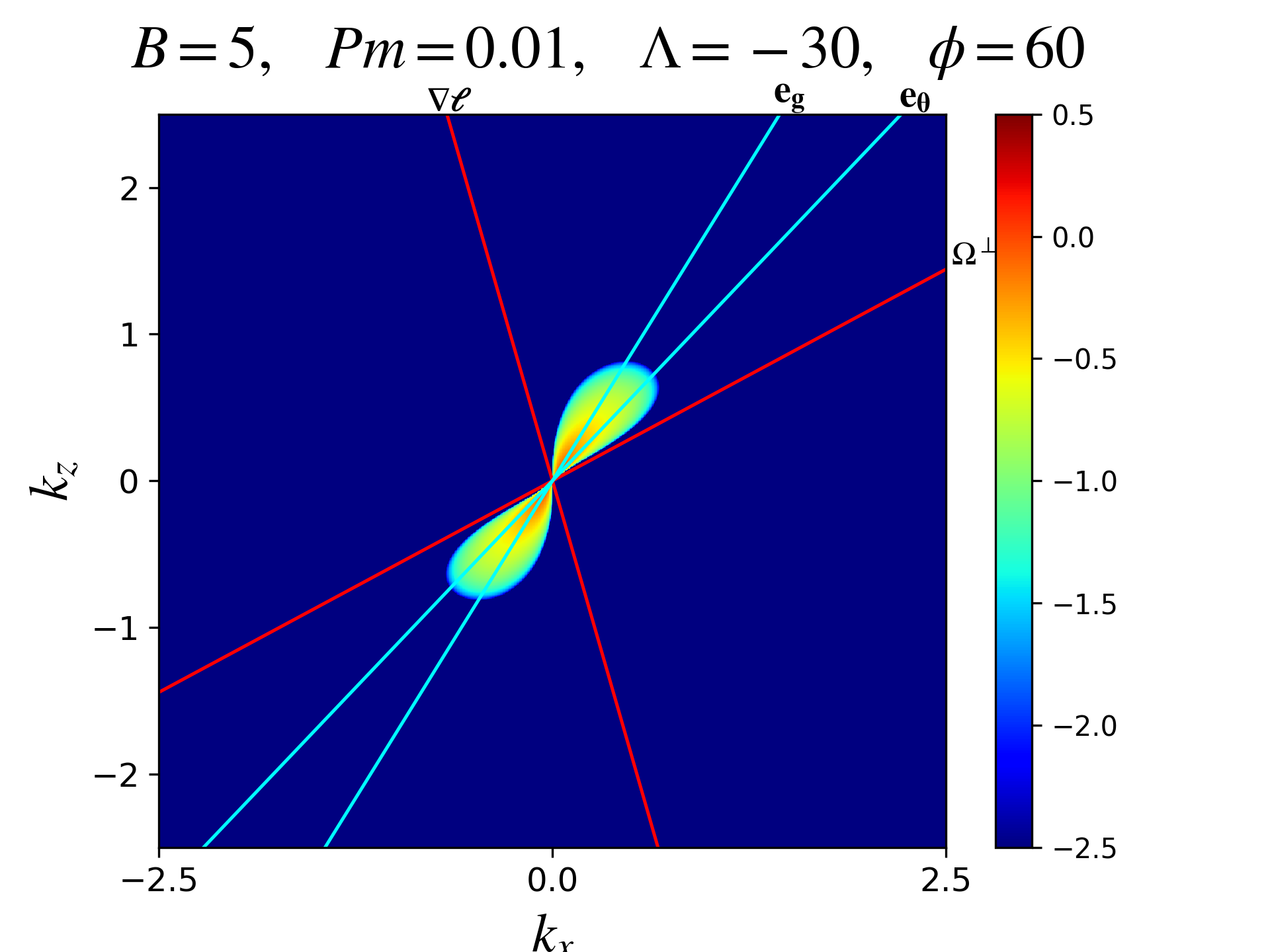}}
    \subfigure[]{\includegraphics[trim=0cm 0cm 0cm 0cm,clip=true,
    width=0.33\textwidth]{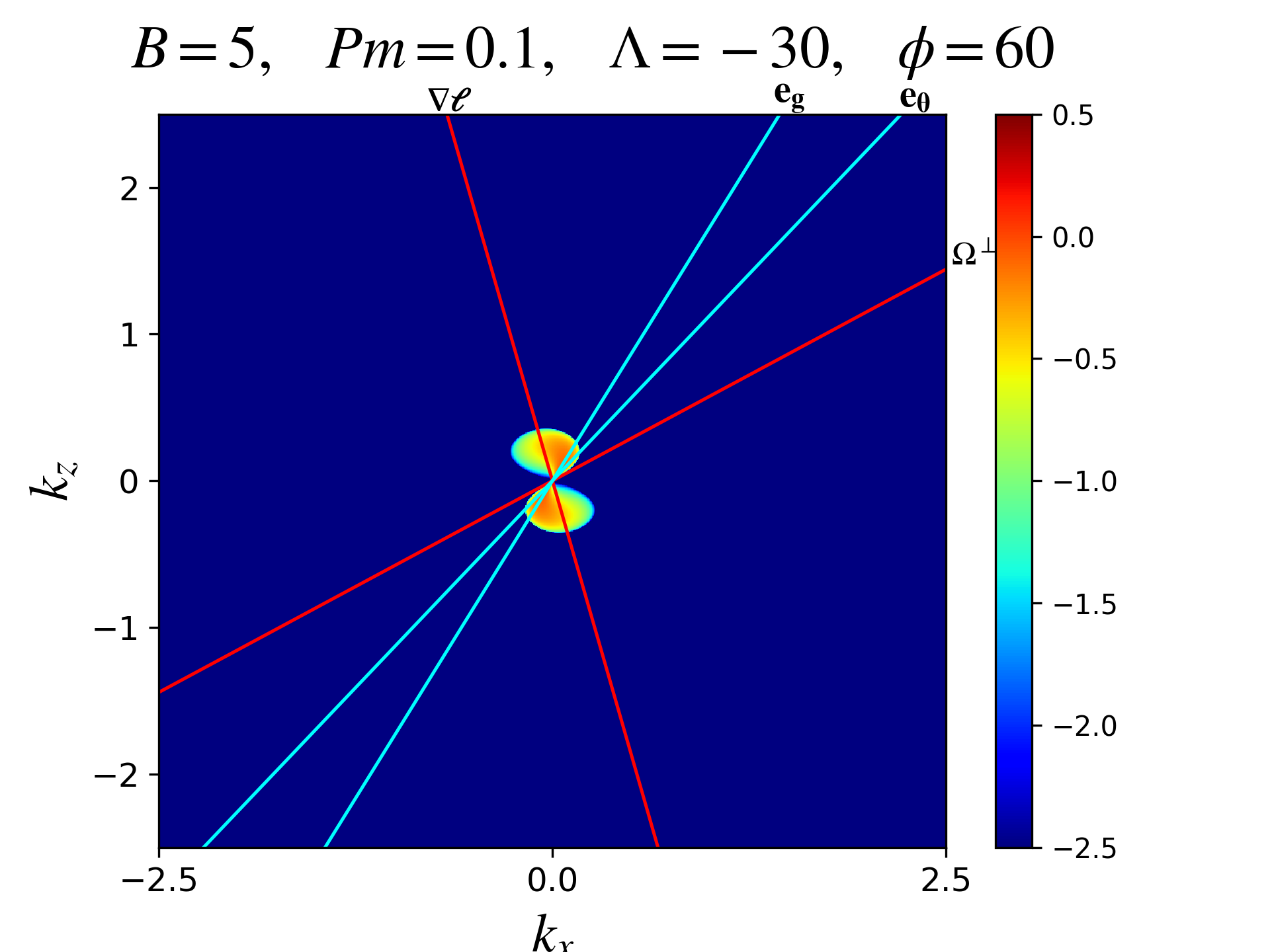}}
    \subfigure[]{\includegraphics[trim=0cm 0cm 0cm 0cm,clip=true,
    width=0.33\textwidth]{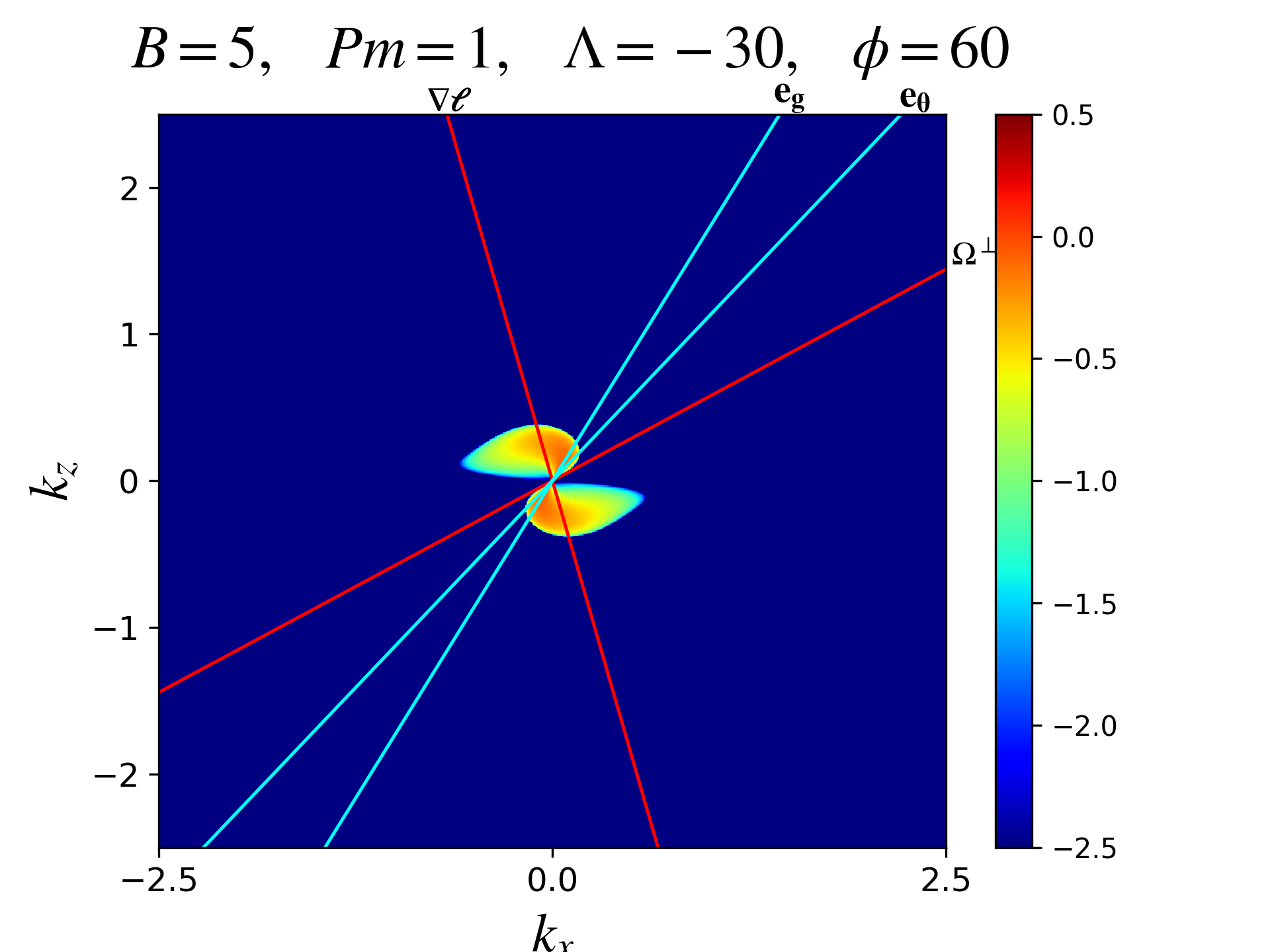}}

  \caption{Logarithm of the growth rate $\log_{10} (\sigma/\Omega)$ of axisymmetric perturbations plotted on the $(k_x, k_z)$-plane according to Eq.~\ref{DR}, for various $B_0$ and Pm, with $\phi=60^\circ, \Lambda = -30^\circ$, i.e.~a mixed radial/latitudinal shear at latitude $\Lambda+\phi=30^\circ$. Parameters are $\mathcal{N}^2/\Omega^2 = 10$, Pr$=10^{-2}$, $\mathcal{S}/\Omega=2$. We vary the strength of the magnetic field from $B_0=0$ to $B_0=5$ down each column, and vary Pm from $\mathrm{Pm}=0.01$ to $\mathrm{Pm}=1$ along each row. When $B_0=0$ the system is  adiabatically unstable since it violates the Solberg-H\o iland criterion. This is visually characterised by a tendency for the fastest growing modes to occur even as $k \to 0$, suggesting that the presence of diffusion leads to the preference of the largest possible wavelengths in this regime. This is in comparison to the GSF and MRI modes cases where the fastest growing modes here have a unique non-zero wavenumber and hence a preferred wavelength in real space.}
  \label{Lobes3}
\end{figure*}

In this section we solve the dispersion relation and graphically analyse the properties of the possible instabilities in our system. In Figures~\ref{Lobes1}--\ref{Lobes3} we probe effects of varying the magnetic field strength $B_0$ and magnetic Prandtl number $\mathrm{Pm}$ for three different configurations with different latitudes $\beta=\Lambda+\phi$ and orientations of the shear with respect to gravity $\phi$. We fix $\mathrm{Pr}=0.01$ small, but motivated by parameters accessible with nonlinear numerical simulations, and $S=2$ \citep[following][]{barker2020,Dymott2023}, since the latter choice would be marginally stable according to Rayleigh's criterion for cylindrical differential rotation. We present pseudocolour plots of the base 10 logarithm of the growth rate of an axisymmetric perturbation in Fourier space ($k_x,k_z$) in these figures. Over-plotted in red are the lines $\hat{\boldsymbol{\Omega}}^\perp$ and $\nabla \ell$, within which the direct GSF instability occurs, and in light blue are the directions of buoyancy (more specifically, the normal to stratification surfaces) and gravity, $\boldsymbol{e}_{\theta}$ and $\boldsymbol{e}_g$, respectively. For comparison the hydrodynamic cases with $B_0=0$ are shown in the top row of each figure. The magnetic field strength $B_0$ is increased within the set $[0,1,2.5,5]$ with each successive row, and Pm is increased within $[0.01,0.1,1]$ with each successive column.

We identify two sets of `lobes' of instability operating in the system. The dominant sets are bounded by $\nabla \ell$ and $\hat{\boldsymbol{\Omega}}^\perp$ in the hydrodynamic case, and they correspond to the dominant direct instability. This is either the double-diffusive GSF or the adiabatic Solberg-H\o iland instability (see \S~3.2 of paper 3 for the explicit conditions required for the latter to operate). The fastest growing modes typically have growth rates $O(1)$, which we note is comparable to $\Omega^{-1}$, given our unit of time, and are initially (in the hydrodynamic case) observed to lie along the line that is approximately half-way between $\nabla \ell$ and $\hat{\boldsymbol{\Omega}}^\perp$. This wedge is perpendicular to the physical wedge within which the unstable mode displacements and velocity perturbations arise due to incompressibility. The second set of smaller lobes, when present, contains oscillatory modes, which are weakly growing internal magneto-inertia-gravity waves that propagate and are destabilised within the wedges bounded by $\boldsymbol{e}_g$ and $\boldsymbol{e}_{\theta}$. 

The introduction of non-zero $\boldsymbol{B}_0$ has observable effects on the orientation, strength and structure of the unstable region in parameter space. Increasing the strength of the field for a fixed Pm has a tendency to force the modes into alignment with the preferred direction for MRI modes, and to shift them to larger scales (smaller $k$ magnitudes), both as predicted in \S~\ref{diffusiveanalysis}. Note that we do not observe the MRI modes to be well aligned with $\boldsymbol{e}_\theta$ here, as explained in \S~\ref{adiab}. This is likely due to the effects of thermal diffusion in eliminating the stabilising effects of buoyancy forces on MRI modes when $\mathrm{Pr}/\mathrm{Pm}=\eta/\kappa$ and $\mathrm{Pr}$ are both small, which is the case here when $\mathrm{Pm}\geq 0.01$ as we fix $\mathrm{Pr}=0.01$. The addition of a field seems to impose a stabilising effect on the hydrodynamically-unstable GSF modes, ultimately resulting from the stabilising effects of magnetic tension \citep[see also][]{LatterPap2018}, and so the majority of cases exhibit a smaller growth rate for the dominant instability as $B_0$ is increased. 

Magnetic diffusion counteracts the effects of magnetic fields, and this is clearest for small Pm in the left panels of Figures~\ref{Lobes1}--\ref{Lobes3}. Cases with $\mathrm{Pm}=1$ have the weakest ohmic diffusion, and cases with $\mathrm{Pm}=0.01$ have much more efficient ohmic than viscous diffusion. Small Pm allows magnetic cases to return to the hydrodynamic limit and larger Pm (closer to unity, in our case) therefore exhibit the strongest magnetic effects for a given $B_0$. When $\mathrm{Pm}=1$, instability is possible outside the hydrodynamic region contained within 
the lines $\hat{\boldsymbol{\Omega}}^{\perp}$ and $\nabla{\ell}$. This can be seen most clearly in the right-most bottom panel of these figures, where magnetic effects are strongest (and magnetic diffusion is weakest). The direction of the preferred modes in that case are better described by the unstratified (due to rapid thermal diffusion) MRI in \S~\ref{diffusiveanalysis}.

The oscillatory modes seem to be only very marginally modified by the magnetic field, as is seen most clearly in Fig.~\ref{Lobes2}. This suggests that the internal inertia-gravity waves observed to be destabilised in paper 3 \citep{Dymott2023} within the wedge between $\boldsymbol{e}_g$ and $\boldsymbol{e}_\theta$ continue to be weakly destabilised magneto-inertial-gravity waves.

\section{Parameter dependence of fastest growing mode}\label{FGMparam}


\begin{figure*}
   \subfigure[$\sigma$]{\includegraphics[trim=3.5cm 9cm 4cm 10cm,clip=true,width=0.325\textwidth]{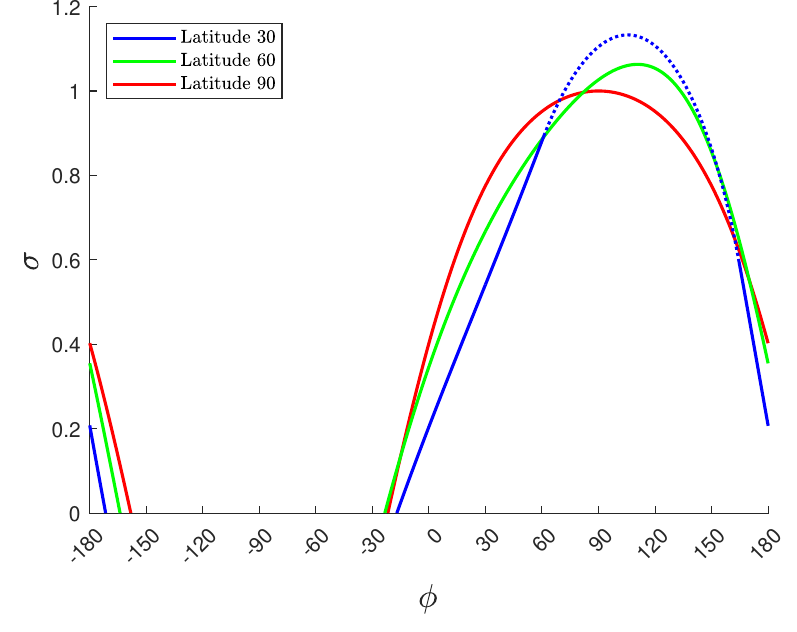}}
   \subfigure[$k$ ($B_0=1$)]{\includegraphics[trim=3.5cm 9cm 4cm 10cm,clip=true,width=0.325\textwidth]{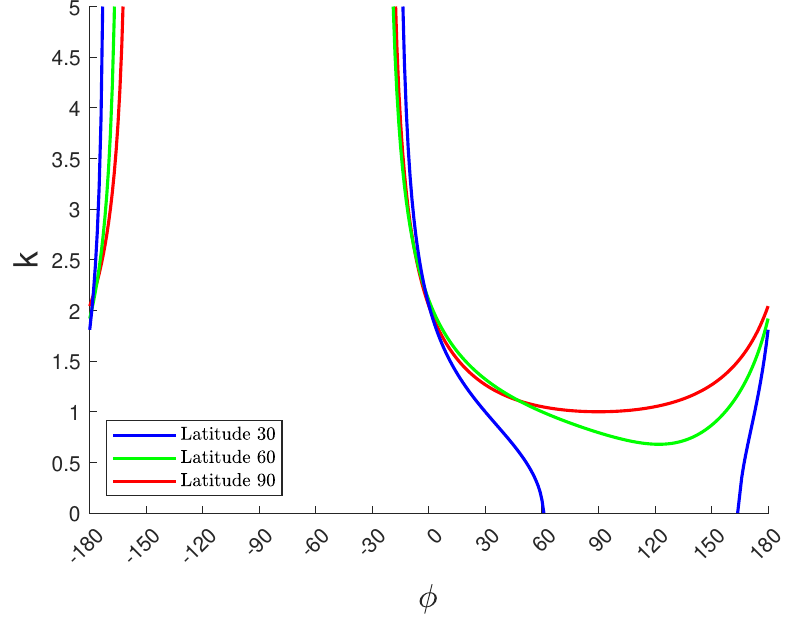}}
   \subfigure[$\theta_k$]{\includegraphics[trim=3.5cm 9cm 4cm 10cm,clip=true,width=0.325\textwidth]{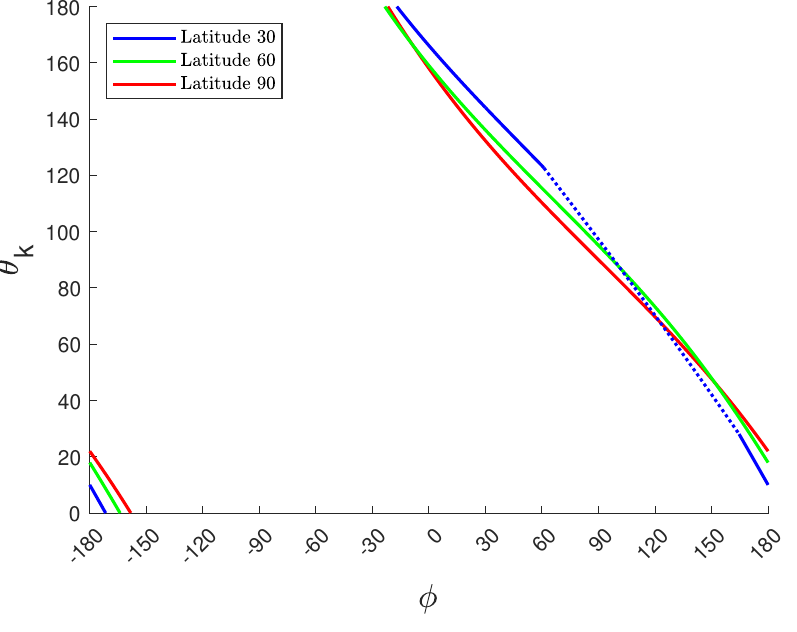}}    
\label{matlabadi}
  \caption{A selection of figures comparing the properties of the fastest growing non-diffusive ($\nu=\kappa=\eta=0$) modes with an imposed  magnetic field for $S=2$ and $N^2=10$ for latitudes $30^\circ$, $60^\circ$, $90^\circ$.  Left panel: maximum growth rate $\sigma$ as a function of $\phi$. The blue dotted curve is where the fastest growing mode is hydrodynamic and the magnetic field plays no role, which prefers modes with $k\to 0$. The solid curves are where the corresponding $k$ is non-zero and magnetic field affects the growth rate. Middle: $k$ when it is finite and there is instability. Right: corresponding wavevector orientation $\theta_k$, which is well-defined for all growing modes.}
  \label{adiMATLAB}
\end{figure*}


\begin{figure*}
    \subfigure[$\sigma$ at $\Lambda + \phi = 0^\circ$]{\includegraphics[trim=3cm 9cm 4cm 9cm,clip=true,
    width=0.32\textwidth]{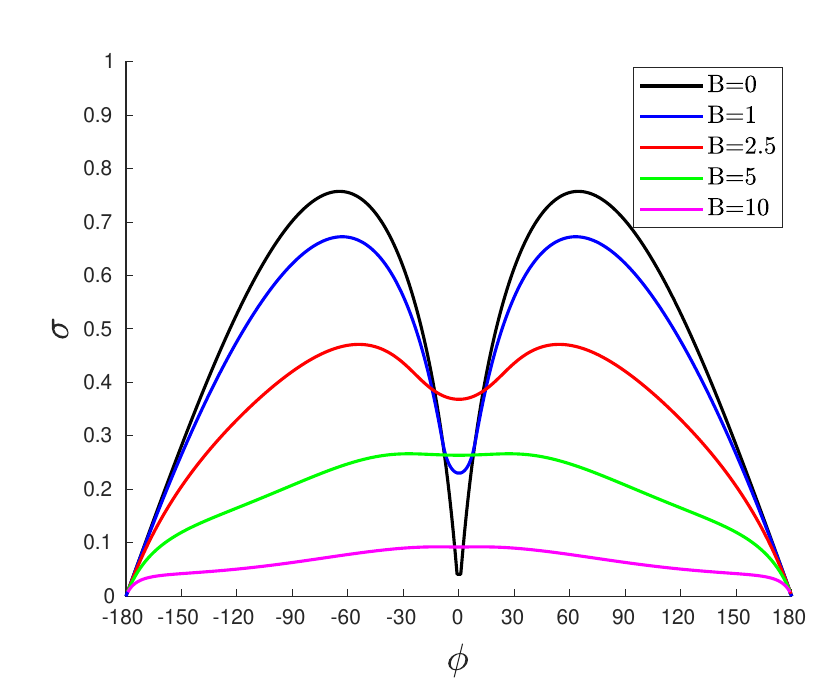}}
    \subfigure[$|k|$ at $\Lambda + \phi = 0^\circ$]{\includegraphics[trim=3cm 9cm 4cm 9cm,clip=true,
    width=0.32\textwidth]{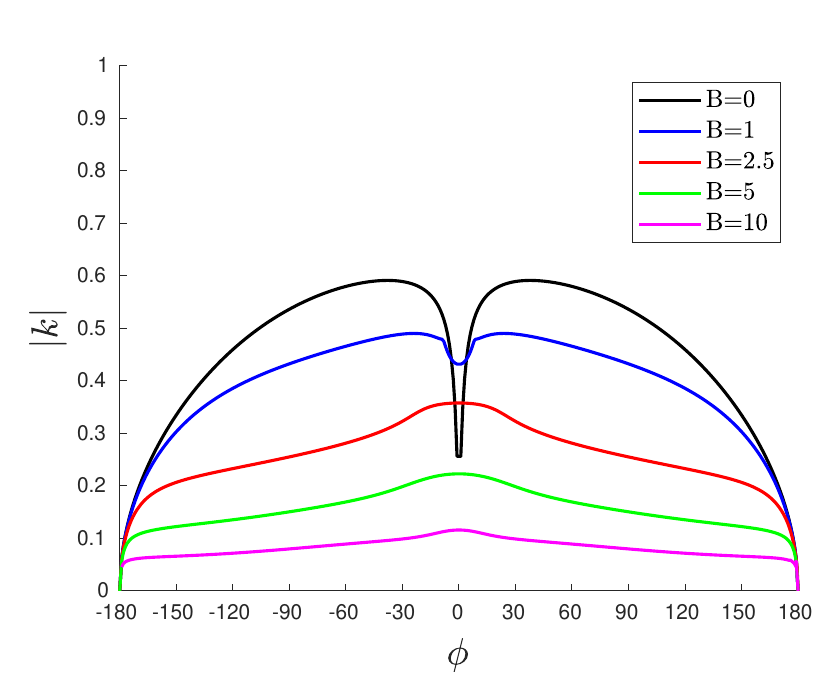}}
    \subfigure[$\theta_k$ at $\Lambda + \phi = 0^\circ$]{\includegraphics[trim=3cm 9cm 4cm 9cm,clip=true,
    width=0.32\textwidth]{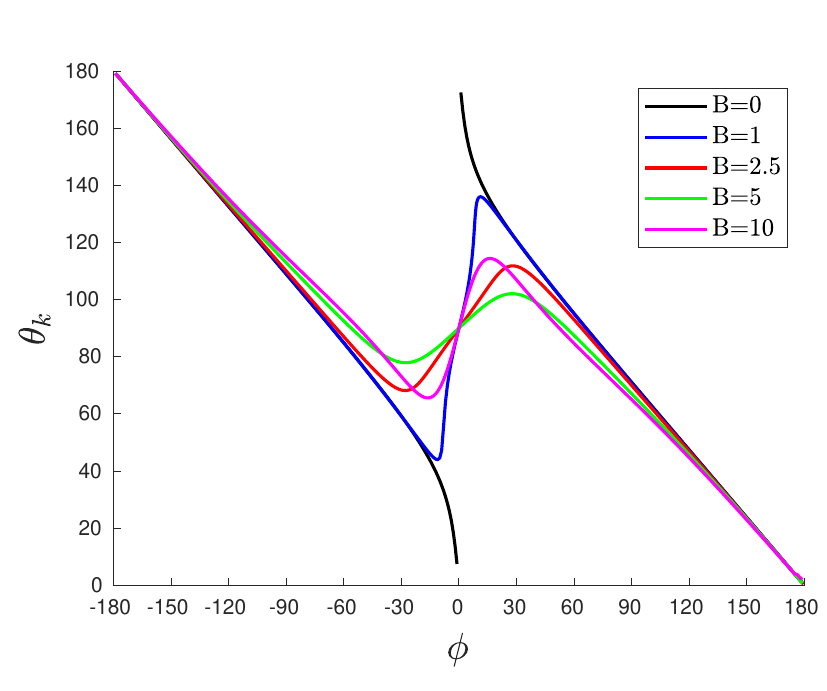}}
  \caption{Properties of the fastest growing modes for various values of the magnetic field $B_0$ with $S=2, \mathrm{Pr}=10^{-2}, N^2=10$, $\mathrm{Pm} =0.1$, for different rotation profiles (values of $\phi$) at the equator. The hydrodynamically stable case of $\phi=0^\circ$, corresponding to cylindrical rotation, is destabilised by the magnetic field. Within close proximity of cylindrical rotation $(-15^\circ \lesssim \phi \lesssim 15^\circ$) increases in field strength of up to roughly $B_0=2.5$ increase the growth rate. This is paired with a decrease in the wavelength of this mode and deviation in orientation from the hydrodynamic case, where $\theta_k$ tends to align itself more so with the orientation of the field. For other $\phi$, the field tends to stabilise the instability over the hydrodynamic case, reducing its maximum growth rate and wavenumber $k$.}
  \label{matlablat0}
\end{figure*}

\begin{figure*}
    \subfigure[$\sigma$ at $\Lambda + \phi = 30^\circ$]{\includegraphics[trim=3cm 9cm 4cm 9cm,clip=true,
    width=0.32\textwidth]{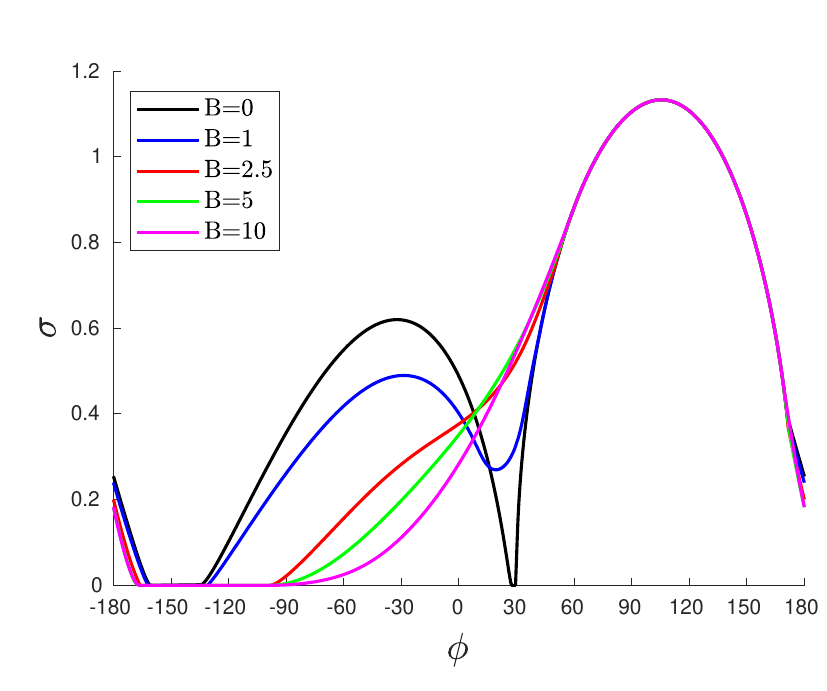}}
    \subfigure[$|k|$ at $\Lambda + \phi = 30^\circ$]{\includegraphics[trim=3cm 9cm 4cm 9cm,clip=true,
    width=0.32\textwidth]{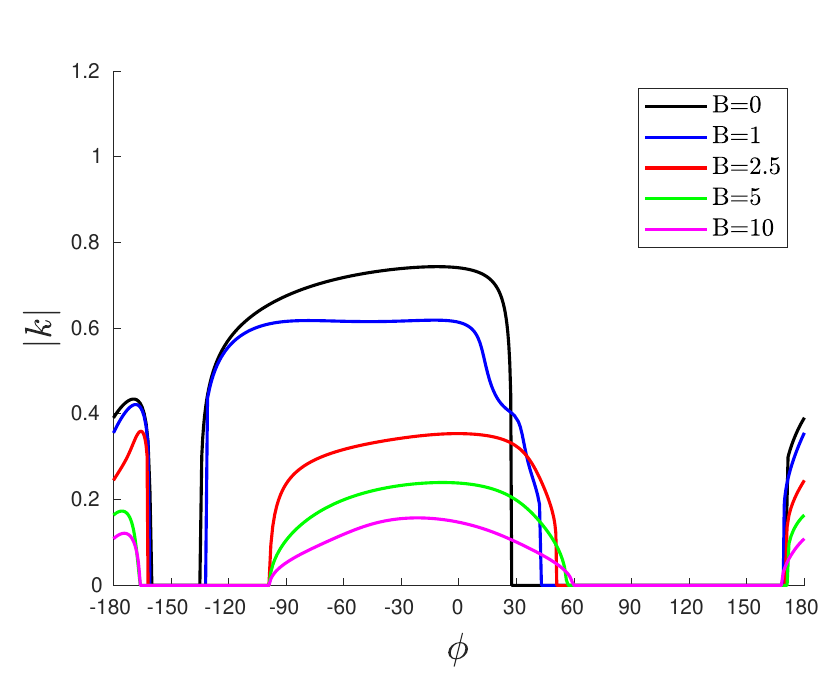}}
    \subfigure[$\theta_k$ at $\Lambda + \phi = 30^\circ$]{\includegraphics[trim=3cm 9cm 4cm 9cm,clip=true,
    width=0.32\textwidth]{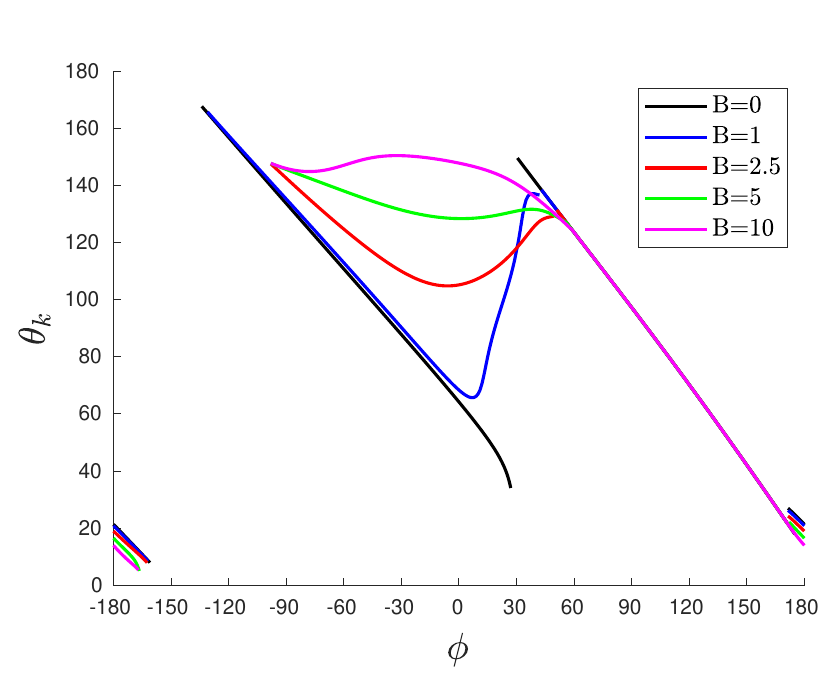}}
  \caption{Properties of the fastest growing modes for various values of the magnetic field $B_0$ with $S=2, \mathrm{Pr}=10^{-2}, N^2=10$, $\mathrm{Pm} =0.1$, for different rotation profiles (values of $\phi$) at a latitude $\beta=\Lambda+\phi=30^\circ$. The addition of a magnetic field significantly alters the linear growth rate of the diffusive modes, and typically acts to reduce both the growth rate $\sigma$ and wavenumber $k$, but it does not affect the adiabatically unstable region for $\phi\in[60^\circ,170]$. The effect of the magnetic field depends on both field strength ($B_0$) and differential rotation profile ($\phi$). Nearly cylindrical differential rotations $\phi\approx 30^\circ$ ($\Lambda=0$) that are hydrodynamically stable are heavily destabilised by the addition of a magnetic field, which corresponds to onset of the MRI.}
  \label{matlablat30}
\end{figure*}

\begin{figure*}
    \subfigure[$\sigma$ at $\Lambda + \phi = 60^\circ$]{\includegraphics[trim=3cm 9cm 4cm 9cm,clip=true,
    width=0.32\textwidth]{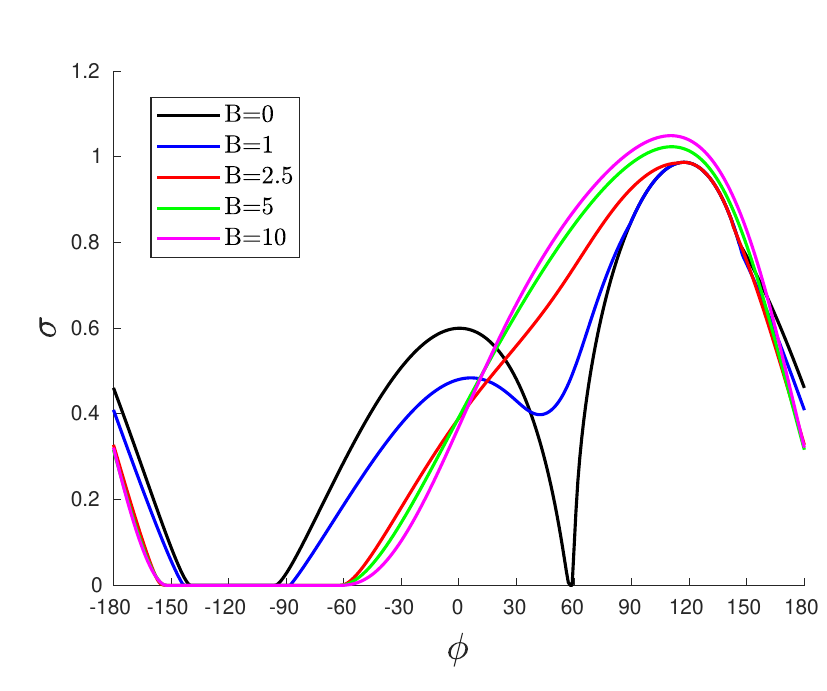}}
    \subfigure[$|k|$ at $\Lambda + \phi = 60^\circ$]{\includegraphics[trim=3cm 9cm 4cm 9cm,clip=true,
    width=0.32\textwidth]{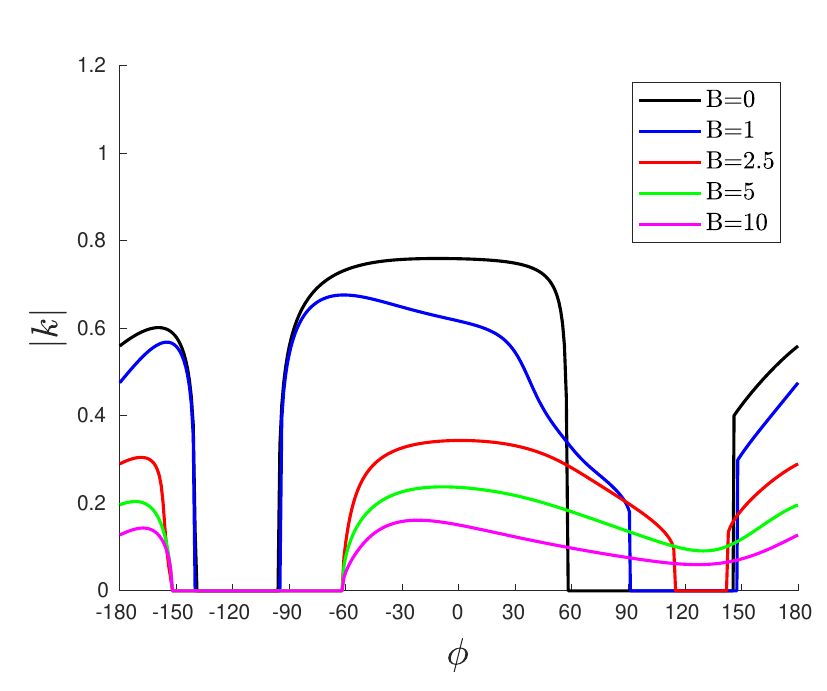}}
    \subfigure[$\theta_k$ at $\Lambda + \phi = 60^\circ$]{\includegraphics[trim=3cm 9cm 4cm 9cm,clip=true,
    width=0.32\textwidth]{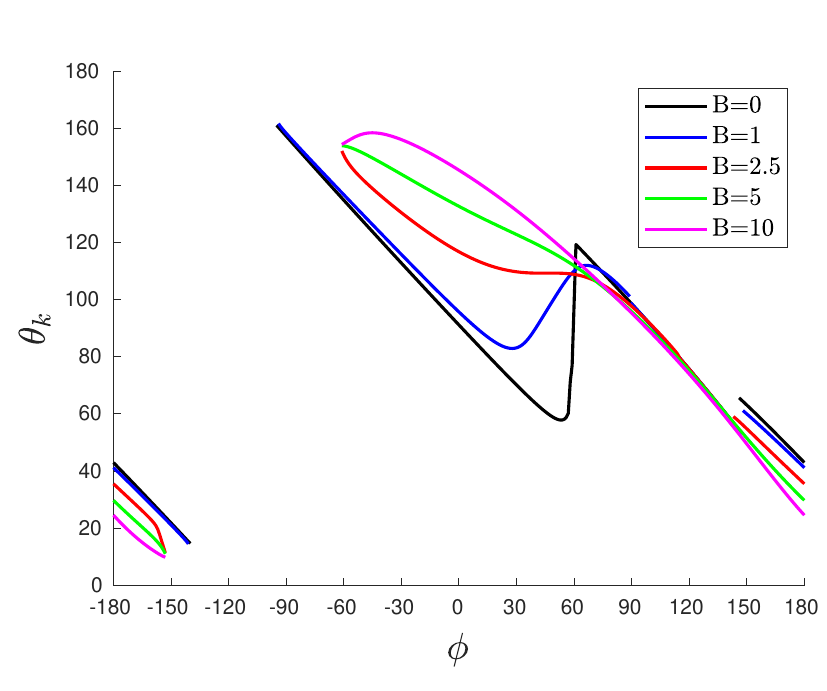}}
  \caption{Properties of the fastest growing modes for various values of the magnetic field $B_0$ with $S=2, \mathrm{Pr}=10^{-2}, N^2=10$, $\mathrm{Pm} =0.1$, for different rotation profiles (values of $\phi$) at a latitude $\beta=\Lambda+\phi=60^\circ$.}
  \label{matlablat60}
\end{figure*}

\begin{figure*}
    \subfigure[$\sigma$ at $\Lambda + \phi = 90^\circ$]{\includegraphics[trim=3cm 9cm 4cm 9cm,clip=true,
    width=0.32\textwidth]{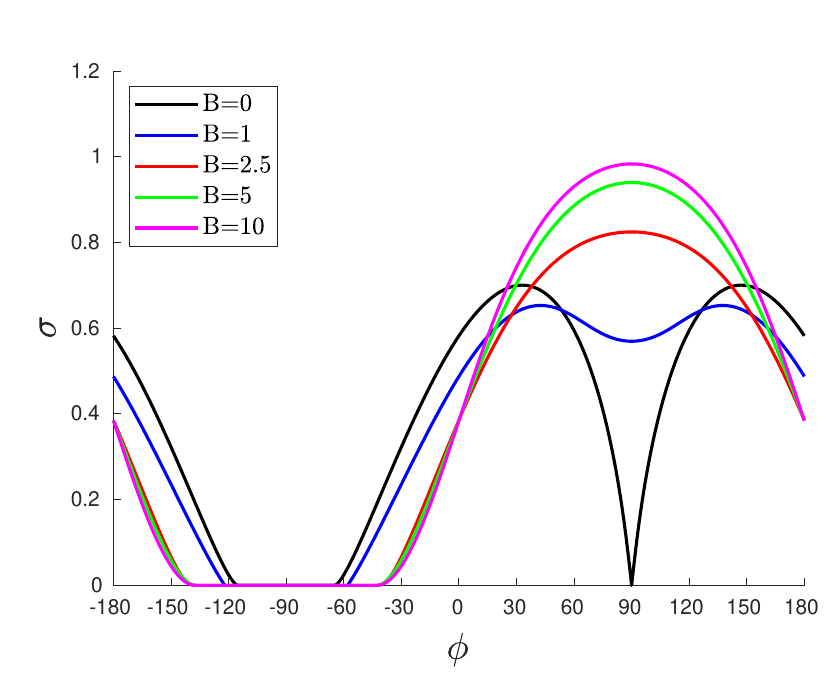}}
    \subfigure[$|k|$ at $\Lambda + \phi = 90^\circ$]{\includegraphics[trim=3cm 9cm 4cm 9cm,clip=true,
    width=0.32\textwidth]{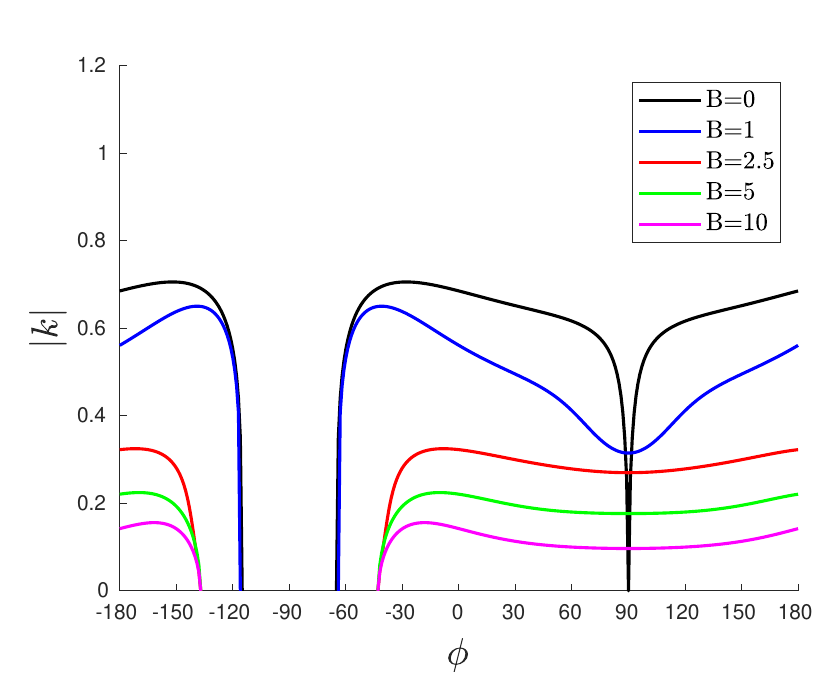}}
    \subfigure[$\theta_k$ at $\Lambda + \phi = 90^\circ$]{\includegraphics[trim=3cm 9cm 4cm 9cm,clip=true,
    width=0.32\textwidth]{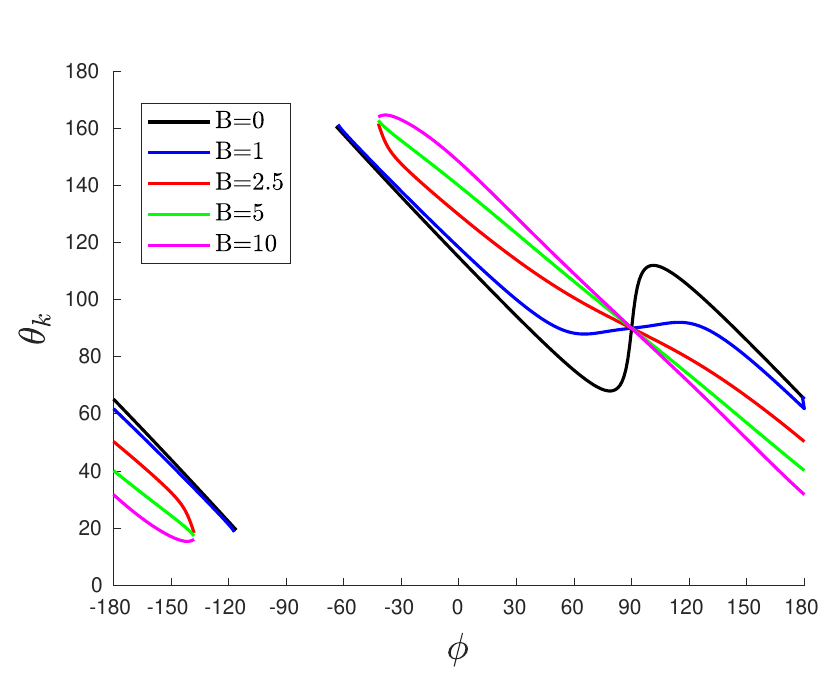}}
  \caption{Properties of the fastest growing modes for various values of the magnetic field $B_0$ with $S=2, \mathrm{Pr}=10^{-2}, N^2=10$, $\mathrm{Pm} =0.1$, for different rotation profiles (values of $\phi$) at a latitude $\beta=\Lambda+\phi=90^\circ$.}
  \label{matlablat90}
\end{figure*}

\begin{figure*}
    \subfigure[$\sigma$ at $\Lambda + \phi = 30^\circ$]{\includegraphics[trim=0cm 0cm 0cm 0.65cm,clip=true,width=0.32\textwidth]{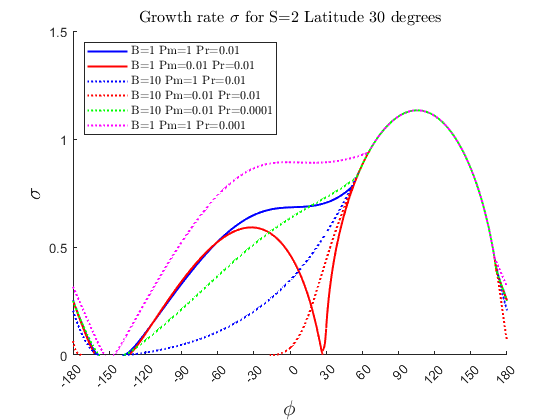}}
    \subfigure[$|k|$ at $\Lambda + \phi = 30^\circ$]{\includegraphics[trim=0cm 0cm 0cm 0.65cm,clip=true,width=0.32\textwidth]{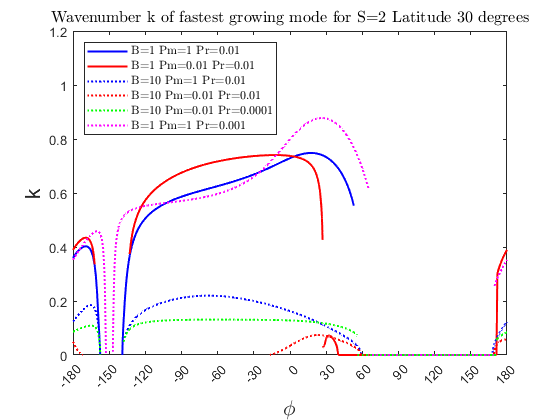}}
    \subfigure[$\theta_k$ at $\Lambda + \phi = 30^\circ$]{\includegraphics[trim=0cm 0cm 0cm 0.65cm,clip=true,width=0.32\textwidth]{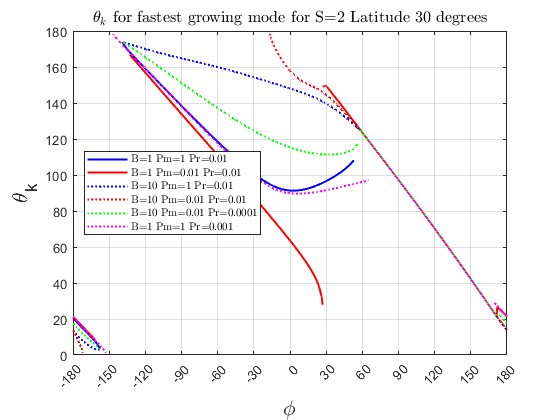}}
  \caption{Properties of the fastest growing modes for various values of the magnetic field $B_0$, $\mathrm{Pm}$ and $\mathrm{Pr}$ with $S=2, \mathrm{Pr}=10^{-2}, N^2=10$, for different rotation profiles (values of $\phi$) at a latitude $\beta=\Lambda+\phi=30^\circ$.}
  \label{matlablat30Pm}
\end{figure*}

After displaying the properties of all unstable axisymmetric modes on the $(k_x,k_z)$-plane in our system as $B_0$ and Pm are varied, we now turn to explore the variation in the fastest growing mode optimised over $k_x$ and $k_z$ as the parameters are varied. We primarily consider $S=2$, $N^2=10$ at 4 latitudes   ($\beta=\phi + \Lambda = 0^\circ, 30^\circ, 60^\circ$ and $90^\circ$).

\subsection{Non-diffusive instabilities}

We first explore non-diffusive (stratified) instabilities by solving the quartic dispersion relation in Eq.~\ref{eq_adiab} numerically (using \verb|fminsearch| on $-\Re[s]$ in Matlab), to determine the properties of the fastest growing mode, which we present in Fig.~\ref{adiMATLAB} as a function of the direction of the differential rotation $\phi$, for three different latitudes $\beta=30^\circ$, $60^\circ$ and $90^\circ$. The left panel shows the growth rate $\sigma$ of the fastest growing mode, the middle panel the corresponding wavenumber magnitude $k$, and the right panel the wavevector orientation $\theta_k=\tan^{-1}(-k_z/k_x)$. Note that the magnetic field only appears in Eq.~\ref{eq_adiab} through powers of $\omega_A$, therefore the results in the left and right panels of Fig.~\ref{adiMATLAB} are independent of the magnetic field, whereas the middle panels show results for $B_0=1$ but $k$ can be straightforwardly scaled to consider any $B_0$ since the $y$-axis can be interpreted as $B_0 k$ (and $k_z$ can be obtained using the corresponding $\theta_k$). This means that for strong fields, instability prefers small $k$.

The blue lines in Fig.~\ref{adiMATLAB} are results at a latitude of $30^\circ$. Blue dashed lines are where the non-diffusive hydrodynamic Solberg-H\o iland instability operates, and blue lines where the magnetic field modifies the instability over the hydrodynamic case \citep[shown in Fig.~4(b) in][]{Dymott2023}. Values of $\phi$ for which there are no blue or red line are non-diffusively stable. When the non-diffusive hydrodynamic Solberg-H\o iland modes are unstable in red, between approximately $\phi\in[60^\circ,150^\circ]$, there is no preferred $k$, only a preferred wavevector orientation. In this limit, the blue lines suggest the wavevector $k\to 0$ for this range of $\phi$. The magnetic field widens the unstable region to below $\phi=0$ from $\phi\approx 30^\circ$.

For latitudes $60^\circ$ (green lines) and $90^\circ$ (red lines), there are no purely hydrodynamically unstable non-diffusive modes, but it can be seen by comparison with Fig.~4 in paper 3 that the field widens the unstable range of $\phi$. The growth rate has a similar maximum value to the hydrodynamic case, with $\sigma\sim 1$ for the maximal $\phi$. $\phi\approx -30^\circ$ to $180^\circ$ are typically the most unstable configurations, and they also have the largest wavelength (smallest $k$) instabilities, whereas $\phi$ approximately between $-150^\circ$ and $-30^\circ$ are typically non-diffusively stable. Note that we have obtained a new region of instability near to $\phi\approx -180^\circ$ that we omitted from Fig.~4 in paper 3 but is also present in the hydrodynamic case.

\subsection{Diffusive instabilities}

In stellar radiation zones, rapid thermal diffusion means that $\mathrm{Pr}\ll 1$, $\mathrm{Pm}\ll 1$ and $\mathrm{Pr}/\mathrm{Pm} =\eta/\kappa \ll 1$. Hence, the stratified non-diffusive instabilities we have just analysed are likely to be modified by thermal diffusion. To explore this, we solve the full triply-diffusive dispersion relation Eq.~\ref{DR} with $\mathrm{Pr}=10^{-2}$ and $\mathrm{Pm}=0.1$, for which $\mathrm{Pr}/\mathrm{Pm}=0.1$ and is therefore small. Though these are small values they are not in the regime of stellar interiors --- howver these values are accessible for nonlinear calculations. We show the growth rate (left panels), wavevector magnitudes (middle panels) and orientations $\theta_k$ (right panels) as a function of $\phi$ for various field strengths strengths $B_0\in[0,1,2.5,5,10]$ in Figs.~\ref{matlablat0}, \ref{matlablat30}, \ref{matlablat60}, \ref{matlablat90} for latitudes $0^\circ, 30^\circ, 60^\circ$ and $90^\circ$, respectively. These demonstrate the effects of a 
magnetic field on linear growth rates over the complete range of differential rotation configurations (value of $\phi$) with $S=2$.
 
The equatorial case in Fig.~\ref{matlablat0} is symmetric about $\phi =0$ and is adiabatically (non-diffusively) stable for $S=2$ for any $\phi$ and $B_0$. The hydrodynamic $B_0=0$ case is stable when $\phi =0$, corresponding to cylindrical rotation at the equator, but it becomes destabilised by even a weak magnetic field. This destabilisation is seen at all latitudes and is a result of a change in the stability criteria governing instability here. In the hydrodynamic case we require a violation of Rayleigh's criterion, which requires angular momentum to decrease outwards on isobars for instability, whereas in the magnetic case this criteria can -- for certain field strengths -- correspond to an MRI mode that requires angular velocity to decrease along isobars instead, which is much easier to satisfy. Within close proximity to the cylindrically-rotating profile $(-15^\circ \lesssim \phi \lesssim 15^\circ)$, the magnetic instability operating is significantly more unstable than the hydrodynamic case. Increases in field strength of up to $B_0\approx 2.5$ increase linear growth rates of the dominant modes, paired with a decrease in their wavelengths (increase in $k$) and a significant deviation in orientation from the hydrodynamic case there (towards $\theta_k\sim 90^\circ$, implying $\boldsymbol{k}$ is along $z$). 
Outside of this region (particularly for $\phi$ outside of $-60^\circ \lesssim \phi \lesssim 60^\circ$), the field inhibits growth of the hydrodynamic GSF modes and increases their wavelength (reduces its $k$). For these parameters it is clear that the magnetic field typically has a stabilising effect except for close to cylindrical rotation profiles.

Fig.~$\ref{matlablat30}$ shows the same results for latitude $\Lambda+\phi=30^\circ$. Here the symmetry about $\phi=0$ seen at the equator is broken and varying $\phi$ has more complicated effects. Cylindrical differential rotation corresponds here with $\phi=30^\circ$ (since then $\Lambda=0^\circ$), and we observe that it this stable when $B_0=0$ but is destabilised by the addition of a magnetic field, with  more magnetised cases becoming more unstable until the growth rate becomes independent of $B_0$ for $B_0\gtrsim 2.5$. After $\phi \approx 60^\circ$ there is very good agreement between all cases. This is when the non-diffusive hydrodynamic Solberg-H\o iland instability operates (as seen in Fig.~\ref{adiMATLAB}), which prefers $k\to 0$, and magnetic fields have little effect on it.

There is a notable change in the $\phi$ range of non-zero $|k|$ values as $B_0$ is increased, which goes from $169^\circ \lesssim \phi \lesssim -157^\circ \cup -135^\circ \lesssim \phi \lesssim 28^\circ$ in the hydrodynamic case (note that the boundary between $180^\circ$ and $-180^\circ$ is continuous due to symmetry) to $168^\circ\lesssim \phi \lesssim -160^\circ \cup -100^\circ \lesssim \phi \lesssim 60^\circ$ in the strongest $B_0=10$ case. Note that in any region where more than one $B_0$ has well-defined $|k|$ values, the smaller $B_0$ always has the shorter wavelength. In regions where the magnetic instability operates the field again acts to force $\theta_k$ into alignment with the preferred direction for unstratified MRI modes discussed in \S~\ref{diffusiveanalysis} for these parameters. Regions where the dominant mode switches from one form of instability to another can also be seen by a discontinuity in $\theta_k$, as seen in panel (c).

At latitude $\Lambda+\phi=60^\circ$, shown in Fig.~\ref{matlablat60}, the effects of the field are in many ways similar to Fig.~\ref{matlablat30}. The range of unstable $\phi$ values does however decrease with increasing $B_0$, with the smallest range of instability being $-145^\circ \lesssim \phi \lesssim -100^\circ$ in the hydrodynamic case up to $-155^\circ \lesssim \phi \lesssim -59^\circ$ at $B_0=10$. However, the field is destabilising, leading to larger growth rates between $15^\circ \lesssim \phi \lesssim 160^\circ$, where the largest $B_0$ is the most unstable. 

Similar results are found at latitude $\Lambda+\phi=90^\circ$ in Fig.~\ref{matlablat90}. Modes with $20^\circ \lesssim \phi \lesssim 150^\circ$ are destabilised by the field, with nearly cylindrical rotation profiles ($\Lambda\sim 0^\circ$) being most strongly destabilised by the field. 
Cylindrical rotation is marginally stable in the hydrodynamic case but is the most unstable configuration for any $B_0>1$ plotted here, and grows faster than any hydrodynamic case in this figure. 

\subsubsection{Variation of $B_0$, Pm and Pr}

In Fig.~\ref{matlablat30Pm} we explore the dependence of our results on $\mathrm{Pm}$, $\mathrm{Pr}$ and $B_0$ at a latitude of $30^\circ$, using the same parameters otherwise as in Fig.~\ref{matlablat30}. Our prior results indicate that increasing magnetic diffusion can mitigate the effects of the field and bring predictions closer to the hydrodynamic case when GSF-unstable. We observe in Fig.~\ref{matlablat30Pm} that increasing $B_0$ inhibits instability for $\phi<0$ in the northern hemisphere, but that smaller $\mathrm{Pm}/\mathrm{Pr}$ (more efficient magnetic diffusion) can mitigate this, as seen by comparing the solid and dashed blue lines with the green dashed line. Fig.~\ref{matlablat30Pm} indicates that there does not appear to be a single parameter neatly describing the competing influences of magnetic field and diffusion on instability. For example, increasing $B_0$ tends to enhance instability around $\phi \sim 0$ when comparing the solid red and dashed red lines (with the same Pr and Pm) but it has the opposite effect for the solid blue and dashed blue lines. Decreasing Pr, thereby reducing viscosity relative to thermal diffusion, enhances instability for $\phi<0$ (compare the pink dashed and solid blue lines). Varying Pm can have different effects depending on other parameters, but we see reducing Pm tends to enhance instability for $\phi<0$ when comparing the green and blue dashed lines (with $B_0=10$), but inhibit it when comparing the solid blue and red lines. (Fixing $\mathrm{Pr}/\mathrm{Pm}=\eta/\kappa$ is also clearly not the sole parameter of importance.)

\subsubsection{Critical value of $S$}
\begin{figure}
    \subfigure[$S_{\mathrm{crit}}$]{\includegraphics[trim=0cm 0cm 0cm 0cm,clip=true,
    width=0.45\textwidth]{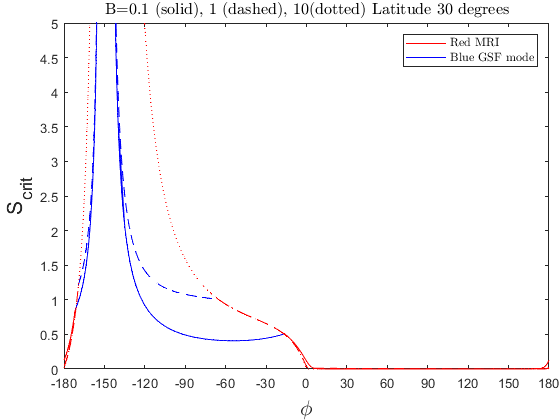}}
    \subfigure[$|k|$]{\includegraphics[trim=0cm 0cm 0cm 0cm,clip=true,
    width=0.45\textwidth]{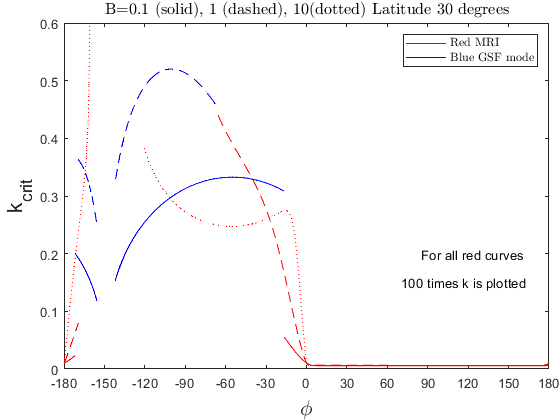}}
    \subfigure[$\theta_k$]{\includegraphics[trim=0cm 0cm 0cm 0cm,clip=true,
    width=0.45\textwidth]{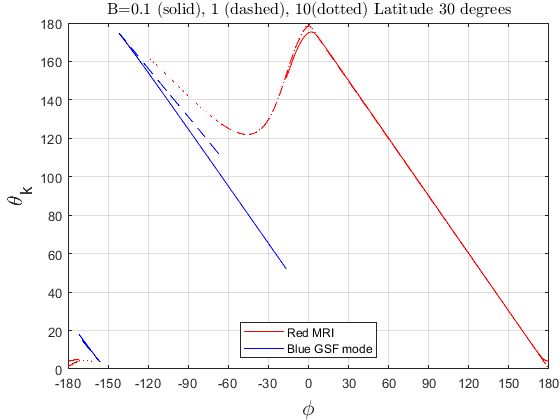}}
  \caption{Critical value of $S$ for instability (top), and the corresponding wavevector magnitude (middle) and orientation (bottom), for $B_0=0.1$, $1$ and $10$, with $S=2, \mathrm{Pr}=10^{-2}, N^2=10$, $\mathrm{Pm} =0.1$, for different rotation profiles (values of $\phi$) at latitude $\Lambda+\phi=30^\circ$. For $\phi>0$ there is instability for any $S>0$ for $B_0\ne 0$, consistent with results obtained in \S~\ref{LinearTheory}, due to the operation of the MRI (shown in red). This instability prefers arbitrarily small wavenumbers. For $\phi<0$, the instability is similar to the hydrodynamic GSF instability (shown in blue; cf. Fig.~6 in paper 3), and exhibits a preferred $k\sim d$. The magnetic field weakens operation of the GSF instability for $\phi<0$. }
  \label{Scrit}
\end{figure}

Finally, we determine numerically the critical value of $S$ for instability ($S_\mathrm{crit}$), once again by optimising over $k_x$ and $k_z$. We show results in Fig.~\ref{Scrit} for $S_\mathrm{crit}$ as a function of $\phi$ for $B_0=0.1$ (solid lines), $B_0=1$ (dashed lines) and $B_0=10$ (dotted lines) at a latitude $30^\circ$, along with the corresponding wavevector magnitude $k$ and orientation $\theta_k$. Our numerical results confirm the arguments presented in \S~\ref{diffusiveanalysis2}. In complete contrast to the hydrodynamic case, we find MRI occurs for any $S>0$ for $\phi>0$, such that $S_\mathrm{crit}=0$ for such differential rotations. Note that if we had chosen a negative latitude value, $S_\mathrm{crit}=0$ would have occurred for $\phi < 0$. The corresponding wavenumber $k$ also becomes arbitrarily small, implying arbitrarily large wavelength instabilities according to our local model. When $\phi<0$, the dominant instability is primarily the hydrodynamic GSF instability, weakly modified by magnetic fields. This has a preferred $k=O(1)$ (in units of $d^{-1}$) when it operates, and it is weakly inhibited by the presence of the magnetic field. We expect to find similar results -- in terms of the modification of the hydrodynamic results shown in Fig.~6 of paper 3 -- for different latitudes and field strengths. Stronger fields would widen operation of the MRI and inhibit the GSF modes further. However, GSF modes may still be the dominant instability for differential rotations with $\phi<0$.

Overall, the addition of a magnetic field tends to inhibit diffusive rotational instabilities by reducing $\sigma$ for $\phi\lesssim 0$, and to promote (increase $\sigma$) instability for $\phi\gtrsim 0$, particularly for nearly cylindrical differential rotations ($\Lambda\sim 0$, where $\phi$ equals the latitude). The wavelength of the dominant instability is typically affected by the strength of the field, with stronger fields generally exciting larger wavelengths (smaller $k$'s). The orientation of the mode also differs from the hydrodynamic prediction for strong enough fields. The effects of magnetic fields on diffusive rotational instabilities are therefore complex, but in nearly all cases the field strongly modifies the growth rate or wavenumber of the dominant mode. We may thus expect magnetic fields to substantially modify turbulent transport in stellar radiative regions.

\section{Energetics of the instabilities}
\label{energy}
\subsection{Derivation of the energy equations and evaluation for linear modes}
In this section we analyse the energetics of the instabilities in our model. This helps identify the physical mechanisms and energy sources that drive the various instabilities and quantify the role of magnetic fields. In order to derive the total energy equation we must first calculate equations that govern the different types of energy in our system, namely, kinetic, thermal/potential and magnetic. We start with the equations governing the evolution of perturbations given by Eqs.~\ref{eq1a}--\ref{eq6a}. To obtain volume-averaged energy equations we take the product of the relevant equation and quantity there (scalar product of Eq.~\ref{eq1a} with $\boldsymbol{u}$ and \ref{eq3a} with $\boldsymbol{B}$ for kinetic and magnetic energies, and multiplication of \ref{eq2a} by $\theta$ for thermal energy) and volume average. We denote volume averages by $\langle \cdot\rangle$ where $\langle \cdot\rangle = \frac{1}{V}\iiint \cdot \, \mathrm{d}V$, where $V$ is the volume of our box, which for linear modes is taken to be a single wavelength of the dominant mode. We define the kinetic, magnetic and thermal energies of our perturbations according to (assuming $\mathcal{N}^2>0)$
\begin{align}
    \mathcal{K} = \frac{1}{2}\langle |\boldsymbol{u}|^2\rangle, \qquad
    \mathcal{M} = \frac{1}{2}\langle |\boldsymbol{B}|^2\rangle, \qquad
    \mathcal{P} = \frac{1}{2}\langle \frac{|\theta|^2}{\mathcal{N}^2}\rangle.
\end{align}

For the kinetic energy equation we obtain
\begin{align}
\nonumber
        \nonumber \partial_t {\mathcal{K}} = -\langle\boldsymbol{u}\cdot (\boldsymbol{u}\cdot \nabla) \boldsymbol{U_0}\rangle &+\langle\theta \boldsymbol{u} \cdot \boldsymbol{e}_g\rangle 
        + \nu\langle\boldsymbol{u}\cdot \nabla ^2 \boldsymbol{u}\rangle
        \\&+ \langle \boldsymbol{u} \cdot (\boldsymbol{B}_0\cdot \nabla)\boldsymbol{B}\rangle.
\end{align}
Note that $\langle\boldsymbol{u}\cdot(\boldsymbol{u}\cdot \nabla) \boldsymbol{u}\rangle=\langle\boldsymbol{u}\cdot(\boldsymbol{U}_0\cdot \nabla) \boldsymbol{u}\rangle=\langle\boldsymbol{u}\cdot\nabla p\rangle=0$ using the chain rule, incompressibility and the divergence theorem (applying periodic boundary conditions), and noting that the Coriolis force does no work ($\boldsymbol{u}\cdot(2\boldsymbol{\Omega}\times\boldsymbol{u})=0$).
We then substitute $\boldsymbol{U}_0=-\mathcal{S}x\boldsymbol{e}_y$ to yield
\begin{equation}
    \partial_t {\mathcal{K}} = \mathcal{S}\langle u_{x}u_{y}\rangle +\langle\theta \boldsymbol{u} \cdot \boldsymbol{e}_g\rangle + \langle \boldsymbol{u} \cdot (\boldsymbol{B}_0\cdot \nabla)\boldsymbol{B}\rangle+\nu\langle\boldsymbol{u}\cdot \nabla^2 \boldsymbol{u}\rangle.
    \label{KEeqn}
\end{equation}
This indicates that the kinetic energy of perturbations can grow by extracting kinetic energy from the shear flow/differential rotation (first term), from conversions of thermal to kinetic energy (second term), from conversions of magnetic to kinetic energy (third term), and that it is dissipated by viscosity (fourth term, which can be shown to be negative definite through an integration by parts).

In a similar manner we can obtain the magnetic energy equation, noting that  $\langle\boldsymbol{B}\cdot (\boldsymbol{u}\cdot \nabla)\boldsymbol{B}\rangle = \langle\boldsymbol{B}\cdot (\boldsymbol{U_0}\cdot \nabla) \boldsymbol{B}\rangle=0$, giving 
\begin{equation}
    \partial_t \mathcal{M} = \langle\boldsymbol{B}\cdot (\boldsymbol{B}_0\cdot \nabla) \boldsymbol{u}\rangle - \mathcal{S}\langle B_xB_y\rangle +\eta \langle\boldsymbol{B} \cdot \nabla^2 \boldsymbol{B}\rangle.
    \label{MEeqn}
\end{equation}
This indicates that magnetic energy of perturbations can grow from conversion of kinetic to magnetic energy (first term), from extracting kinetic energy from the background shear flow/differential rotation (second term), and that it is dissipated ohmically (third term; negative definite). Note that the term $\langle\boldsymbol{B}\cdot (\boldsymbol{B}_0\cdot \nabla) \boldsymbol{u}\rangle$ in Eq.~\ref{MEeqn} can be shown to be equivalent with $-\langle \boldsymbol{u} \cdot (\boldsymbol{B}_0\cdot \nabla)\boldsymbol{B}\rangle$ in Eq.~\ref{KEeqn} using integration by parts, which indicates that these just convert kinetic to magnetic energy and vice versa, and do not inject total energy into the system.

The final energy equation that we need to consider is the one governing thermal energy. We obtain
\begin{equation}
    \partial_t \mathcal{P} = - \langle\theta \boldsymbol{u}\cdot \boldsymbol{e}_{\theta}\rangle + \kappa\langle\frac{\theta \nabla^2 {\theta}}{\mathcal{N}^2}\rangle,
\end{equation}
using the results $\langle\theta (\boldsymbol{u}\cdot \boldsymbol{\nabla})\theta\rangle = \langle\theta (\boldsymbol{U}_0\cdot \nabla)\theta\rangle = 0$. This shows that thermal energy grows through conversion from kinetic to thermal energy (first term) and that it is dissipated by thermal diffusion (second term; negative definite). Hence, the equation for the total energy $\mathcal{E} = \mathcal{K} + \mathcal{M} + \mathcal{P}$, is
\begin{align}
      \nonumber  \partial_t \mathcal{E} = &\mathcal{S}\left(\langle u_x u_y\rangle - \langle B_xB_y\rangle\right) + \langle\theta \boldsymbol{u} \cdot (\boldsymbol{e}_g-\boldsymbol{e}_{\theta})\rangle \\
        & + \nu\langle\boldsymbol{u}\cdot \nabla ^2  \boldsymbol{u}\rangle+\eta \langle\boldsymbol{B} \cdot \nabla^2 \boldsymbol{B}\rangle + \kappa\langle\frac{\theta \nabla^2 {\theta}}{\mathcal{N}^2}\rangle.
        \label{totalE}
\end{align}
This indicates that the total energy of perturbations can grow only via extraction of kinetic energy from the shear flow/differential rotation into perturbation kinetic or magnetic energies (first two terms), or via the baroclinic term that extracts potential energy from the basic state into kinetic and thermal energies (last term on first the line), if and only if these contributions exceed the sum of the viscous, ohmic and thermal dissipations (terms on the bottom line).

We can use these results to analyse the energy sources contributing to the instabilities described by Eq.~\ref{DR}. To do this, we calculate each of the terms in these energy equations for a single axisymmetric Fourier mode with a wavevector $\boldsymbol{k}=(k_x,0,k_z$). This can be used to understand better both the driving forces of the instability and the momentum transporting properties of the instability. We first express $\langle u_{x}u_{y}\rangle$ for a single mode with $u_x = \Re\left[\hat{u}_x\exp{\left(\mathrm{i} \boldsymbol{k}\cdot \boldsymbol{x} + st\right)}\right]$ and $u_y = \Re\left[\hat{u}_y\exp{\left(\mathrm{i} \boldsymbol{k}\cdot \boldsymbol{x} + st\right)}\right]$. Using the properties of complex numbers, $\Re(A)\Re(B) = \frac{1}{2}\Re(AB+AB^*)$, where $*$ denotes the complex conjugate, this can be written
\begin{align}
\nonumber
   \langle u_{x}u_{y}\rangle&=\frac{1}{2}\langle\Re\left(\hat{u}_x\hat{u}_y\exp{(2st+2\mathrm{i}\boldsymbol{k}\cdot \boldsymbol{x}}) + \exp(2\Re[s]t)\hat{u}_x\hat{u}^{*}_{y}\right)\rangle \\
   &=\frac{1}{2}\exp(2\Re[s]t)\Re(\hat{u}_x\hat{u}^{*}_{y}),
   \label{uxuyhat}
\end{align}
upon applying the periodic boundary conditions (thereby eliminating the first term on the top line).

For a single linear mode we can use Eqs.~\ref{eq1a}--\ref{eq6a} to relate $\hat{u}_y$ to $\hat{u}_x$ (and similarly for all other variables) to obtain
\begin{equation}
   \hat{u}_y = \frac{s_{\eta}}{s_{\nu}s_\eta + \omega_A^2}\left(\mathcal{S}\left(1+\frac{\omega_A^2}{s^2_{\eta}}\right)-2\Omega\left(c_{\Lambda}+\frac{k_x}{k_z}s_{\Lambda}\right)\right)\hat{u}_x,
\end{equation}
and we also note that $\hat{u}_z=-(k_x/k_z) \hat{u}_x$.
Substituting this into Eq.~\ref{uxuyhat}, and using $\Re[\hat{u}_x\hat{u}_y^*] = \Re[\hat{u}_x^*\hat{u}_y]$, gives the concise form for the $xy$-component of the Reynolds stress
\begin{equation}
    \langle u_x u_y\rangle = \frac{|\hat{u}_x|^2}{2} \Re\left[ \frac{s_{\eta} \left(\mathcal{S}(1+\frac{\omega_A^2}{s^2_{\eta}})-2\Omega(c_{\Lambda}+\frac{k_x}{k_z}s_{\Lambda})\right)}{s_{\nu}s_\eta + \omega_A^2}\right].
    \label{RS}
\end{equation}
We also have
\begin{align}
\nonumber
\hat{B}_y &= \frac{\mathrm{i}\omega_A}{s_\eta}\left[\hat{u}_y-\frac{\mathcal{S}}{s_\eta}\hat{u}_x\right]\\
&= \frac{\mathrm{i}\omega_A \hat{u}_x}{s_\eta}\left[-\frac{\mathcal{S}}{s_\eta}+\frac{\mathcal{S}\left(1+\frac{\omega_A^2}{s_\eta^2}\right)-2\Omega\left(c_\Lambda+s_\Lambda \frac{k_x}{k_z}\right)}{s_\nu+\frac{\omega_A^2}{s_\eta}}\right],
\end{align}
along with $\hat{B}_z=-(k_x/k_z) \hat{B}_x$, which allows the $xy$ component of the Maxwell stress for a single mode to be written
\begin{align}
    &\langle B_{x}B_{y}\rangle = \frac{1}{2} \Re[\hat{B}_x\hat{B}^{*}_{y}]\exp(2\Re[s]t) \\
   &= \Re\left[\frac{\omega_A^2|\hat{u}_x|^2}{2s_\eta^2}\left(-\frac{\mathcal{S}}{s_\eta}+\frac{\mathcal{S}\left(1+\frac{\omega_A^2}{s_\eta^2}\right)-2\Omega\left(c_\Lambda+s_\Lambda \frac{k_x}{k_z}\right)}{s_\nu+\frac{\omega_A^2}{s_\eta}}\right)\right].
    \label{MS}
\end{align}

The third and final term that can inject energy into the system is 
\begin{equation}
    \langle\theta \boldsymbol{u} \cdot (\boldsymbol{e}_g-\boldsymbol{e}_{\theta})\rangle = \langle\theta [{u_x}(c_{\phi}-c_{\Gamma})+ {u_z}(s_{\phi}-s_{\Gamma})]\rangle,
\end{equation}
which is essentially a measure of the extent that baroclinicity (i.e.~non-coincidence of constant density and pressure surfaces) drives the instability. We use 
\begin{align}
\hat{\theta} = -\frac{\mathcal{N}^2}{s_{\kappa}}\left(c_{\Gamma} - \frac{k_x}{k_z}s_{\Gamma}\right)\hat{u}_{x},
\end{align}
along with incompressibility to write:
\begin{align}
 \label{baro_energy}
    &\langle\theta \boldsymbol{u} \cdot (\boldsymbol{e}_g-\boldsymbol{e}_{\theta})\rangle\\
    & = \frac{1}{2} \Re\left[\frac{\mathcal{N}^2}{s_{\kappa}}\left(\frac{k_x}{k_z}s_{\Gamma}-c_{\Gamma}\right)\left((c_{\phi}-c_{\Gamma})-\frac{k_x}{k_z}(s_{\phi}-s_{\Gamma}) \right)\right]|\hat{u}_x|^2.
   \nonumber
\end{align}
Note that this term vanishes when $\boldsymbol{e}_g=\boldsymbol{e}_\theta$, and is thus unimportant when $s_\Lambda=0$ (no differential rotation along the rotation axis), and it is small compared to the other terms in the strongly stratified limit for which $\mathcal{N}^2\gg 2\Omega \mathcal{S} s_\Lambda$.

The kinetic energy is
\begin{align}
    \mathcal{K}=\frac{1}{2}\langle |\boldsymbol{u}|^2\rangle = \frac{1}{4}[|\hat{u}_x|^2 + |\hat{u}_y|^2 + |\hat{u}_z|^2] \exp({2\Re[s]t}),
\end{align}
which can be expressed in terms of $|\hat{u}_x|^2$ using the above results. The magnetic energy is
\begin{align}
\mathcal{M}=\frac{1}{2}\langle |\boldsymbol{B}|^2\rangle = \frac{1}{4}\left[|\hat{B}_x|^2 + |\hat{B}_y|^2 + |\hat{B}_z|^2\right] \exp{2\Re[s]t},
\end{align}
which can also be expressed in terms of $|\hat{u}_x|^2$ using the following expressions:
\begin{align}
 |\hat{B}_x|^2 &= \frac{\omega_A^2}{|s_\eta|^2} |\hat{u}_x|^2, \\
    |\hat{B}_y|^2 &=-\frac{\omega^{2}_A}{|s_\eta|^2}\left|-\frac{\mathcal{S}}{s_\eta}+\frac{\mathcal{S}\left(1+\frac{\omega_A^2}{s_\eta^2}\right)-2\Omega\left(c_\Lambda+s_\Lambda \frac{k_x}{k_z}\right)}{s_\nu+\frac{\omega_A^2}{s_\eta}}\right|^2 |\hat{u}_x|^2, \\
   |\hat{B}_z|^2 &= \frac{\omega_A^2}{|s_\eta|^2} \left(\frac{k_x}{k_z}\right)^2|\hat{u}_x|^2.
\end{align}
The thermal energy for a single mode can be expressed as
\begin{equation}
    \langle \frac{|\theta|^2}{2\mathcal{N}^2} \rangle =
     \frac{\mathcal{N}^2}{4|s_{\kappa}|^2}\left(c_{\Gamma}-\frac{k_x}{k_z}s_{\Gamma}\right)^2 |\hat{u}_x |^2.
    \label{total_E_theta}
\end{equation}

Using these expressions we can determine the energetic contributions to the growth rate for a single Fourier mode by noting that 
\begin{align}
    2\Re[s] = \partial_t \ln\mathcal{E}=\frac{1}{\mathcal{E}}\partial_t \mathcal{E},
    \label{Ereason}
\end{align}
where the right hand side contains all six terms in Eq.~\ref{totalE} and is independent of the mode amplitude since $|\hat{u}_x|^2$ cancels in both the numerator and denominator. This can also be used as a check of our codes by ensuring that the growth rate $\Re[s]$ is predicted to close to machine precision by using the linear relations between the components that we have just derived. Once this has been confirmed we can compute the  contribution of each of the first three possible driving terms on the right hand side of Eq.~\ref{totalE} to the growth rate to determine whether a given instability is driven by Reynolds stresses, Maxwell stresses or the baroclinic term.


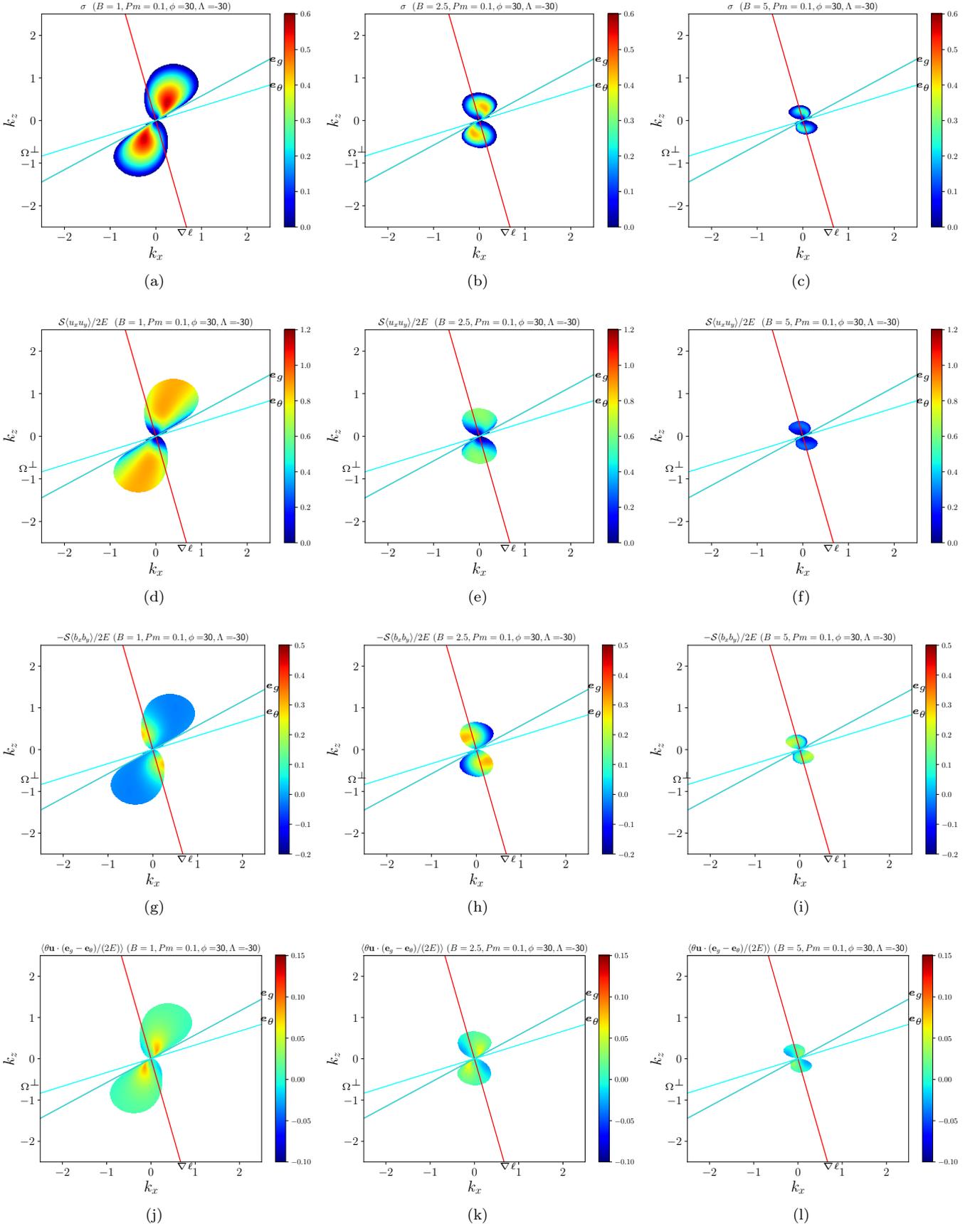
\begin{figure*}
    \subfigure[]{\includegraphics[trim=0cm 0cm 0cm 0cm,clip=true,width=0.33\textwidth]{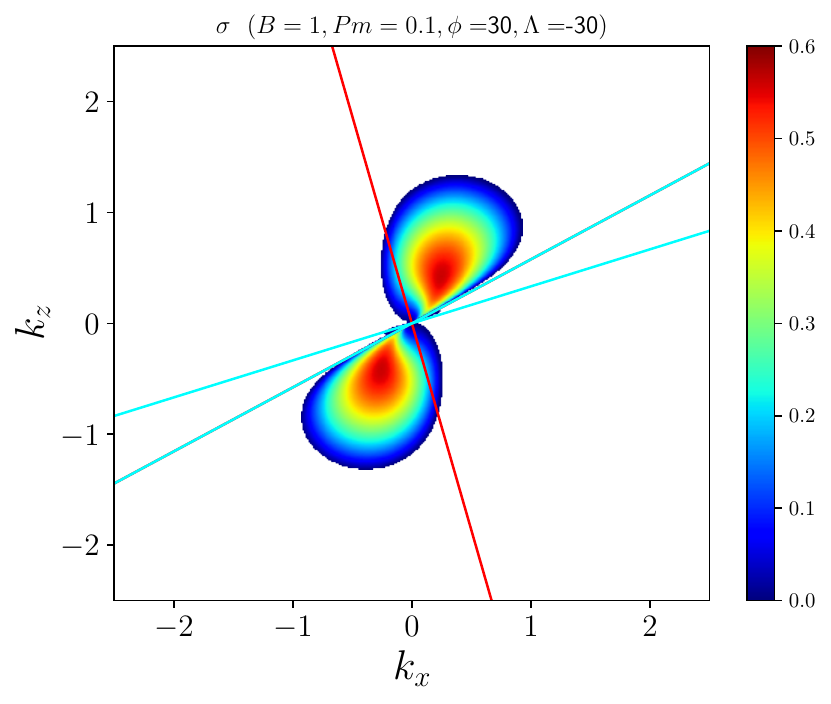}}
    \subfigure[]{\includegraphics[trim=0cm 0cm 0cm 0cm,clip=true,
    width=0.33\textwidth]{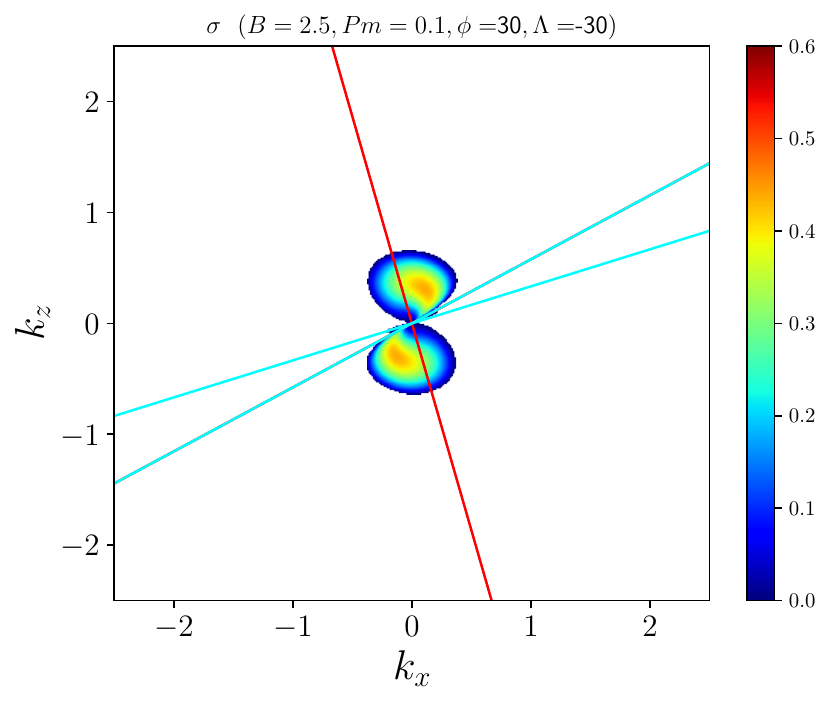}}
    \subfigure[]{\includegraphics[trim=0cm 0cm 0cm 0cm,clip=true,
    width=0.33\textwidth]{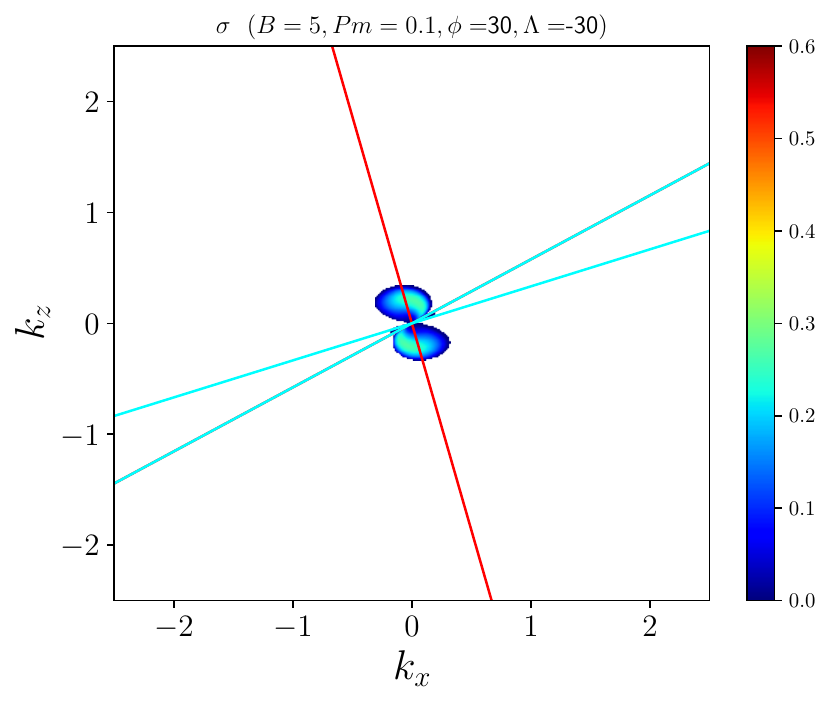}}
    
    \begin{picture}(0,0)
        \put(-242,76){\tiny${\Omega^{\perp}}$}
        \put(-70,76){\tiny$\Omega^{\perp}$}
        \put(95,76){\tiny$\Omega^{\perp}$}
        \put(-160,34){\tiny$\nabla{\ell}$}
        \put(10,34){\tiny${\nabla{\ell}}$}
        \put(180,34){\tiny$\nabla{\ell}$}
        \put(-111.2,125.5){\tiny$\boldsymbol{e}_g$}
        \put(60,125.5){\tiny$\boldsymbol{e}_g$}
        \put(229,125.5){\tiny$\boldsymbol{e}_g$}
        \put(-111.2,112){\tiny$\boldsymbol{e}_{\theta}$}
        \put(60,112){\tiny$\boldsymbol{e}_{\theta}$}
        \put(229,112){\tiny$\boldsymbol{e}_{\theta}$}
    \end{picture}

\subfigure[]{\includegraphics[trim=0cm 0cm 0cm 0cm,clip=true,width=0.33\textwidth]{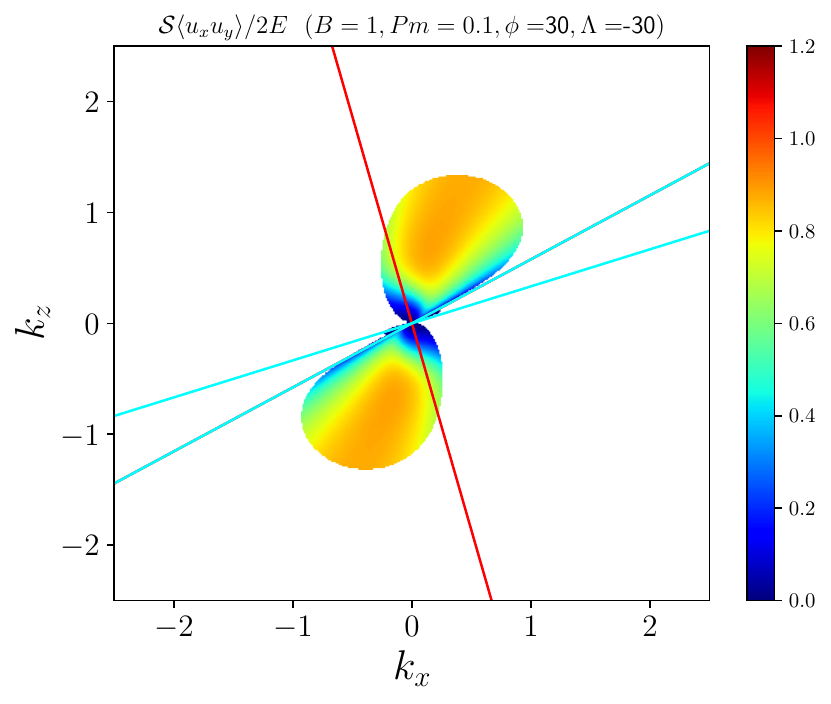}}
\subfigure[]{\includegraphics[trim=0cm 0cm 0cm 0cm,clip=true,
width=0.33\textwidth]{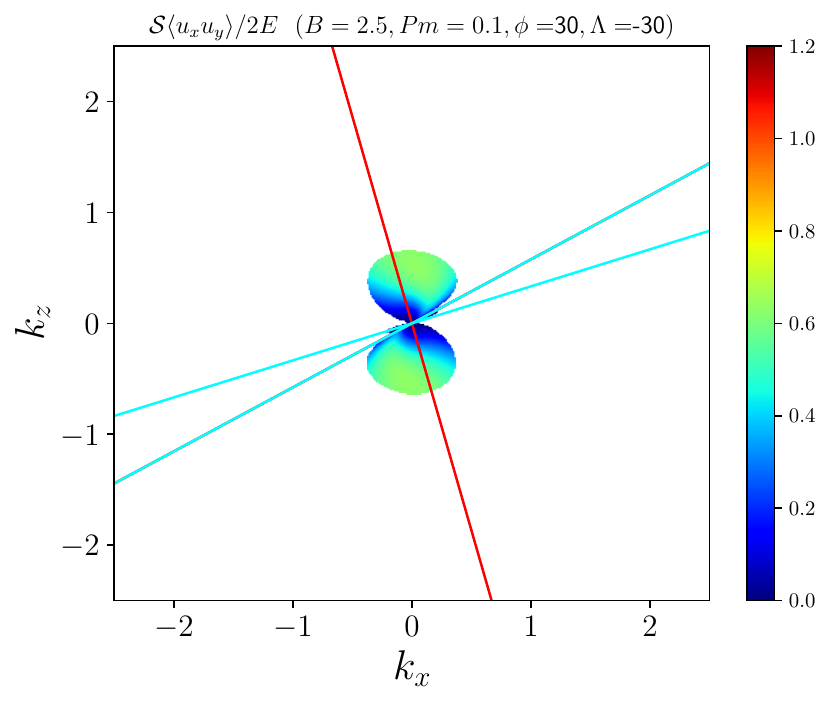}}
\subfigure[]{\includegraphics[trim=0cm 0cm 0cm 0cm,clip=true,
width=0.33\textwidth]{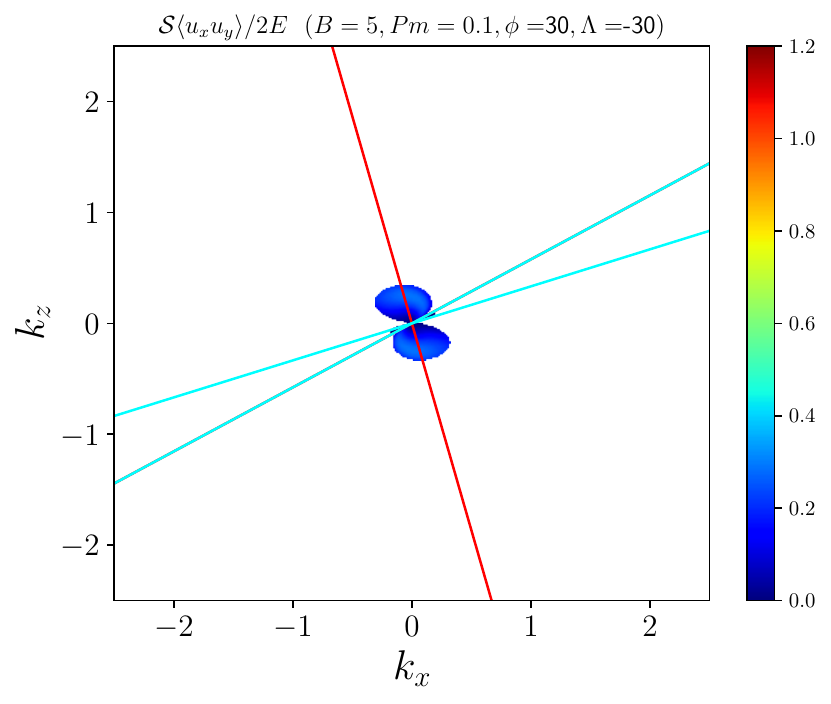}}
    
    \begin{picture}(0,0)
        \put(-243,76){\tiny${\Omega^{\perp}}$}
        \put(-73,76){\tiny$\Omega^{\perp}$}
        \put(97,76){\tiny$\Omega^{\perp}$}
        \put(-160,34){\tiny$\nabla{\ell}$}
        \put(10,34){\tiny${\nabla{\ell}}$}
        \put(180,34){\tiny$\nabla{\ell}$}
        \put(-111.2,125.5){\tiny$\boldsymbol{e}_g$}
        \put(60,125.5){\tiny$\boldsymbol{e}_g$}
        \put(229,125.5){\tiny$\boldsymbol{e}_g$}
        \put(-111.2,112){\tiny$\boldsymbol{e}_{\theta}$}
        \put(60,112){\tiny$\boldsymbol{e}_{\theta}$}
        \put(229,112){\tiny$\boldsymbol{e}_{\theta}$}
    \end{picture}

 \subfigure[]{\includegraphics[trim=0cm 0cm 0cm 0cm,clip=true,width=0.33\textwidth]{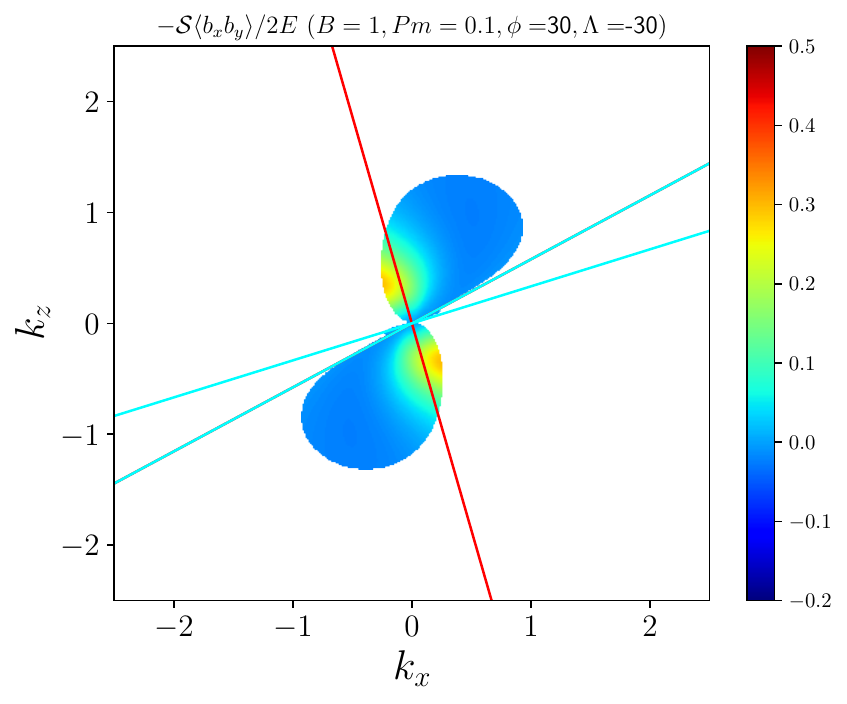}}
\subfigure[]{\includegraphics[trim=0cm 0cm 0cm 0cm,clip=true,
width=0.33\textwidth]{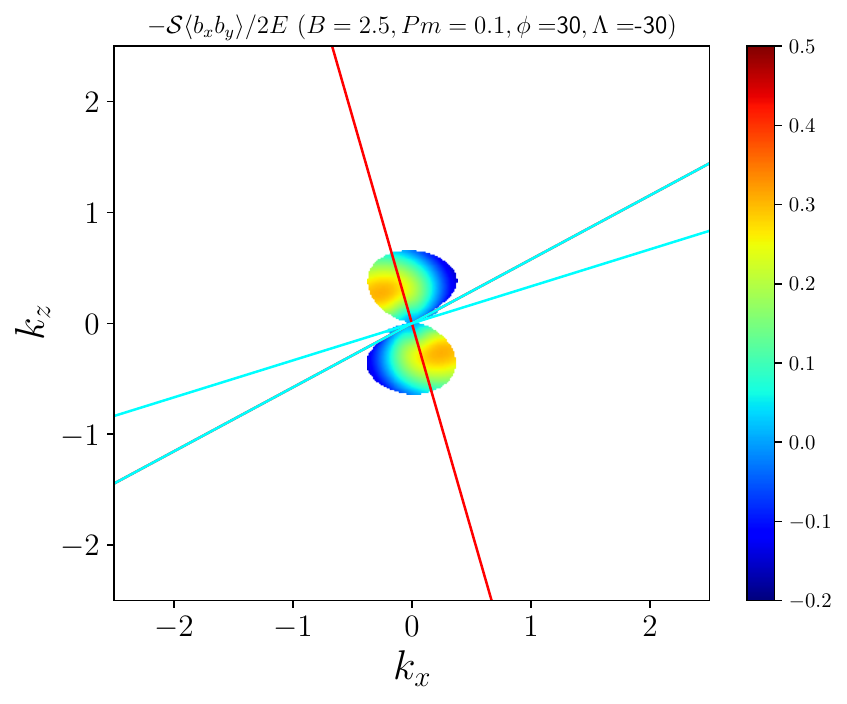}}
\subfigure[]{\includegraphics[trim=0cm 0cm 0cm 0cm,clip=true,
 width=0.33\textwidth]{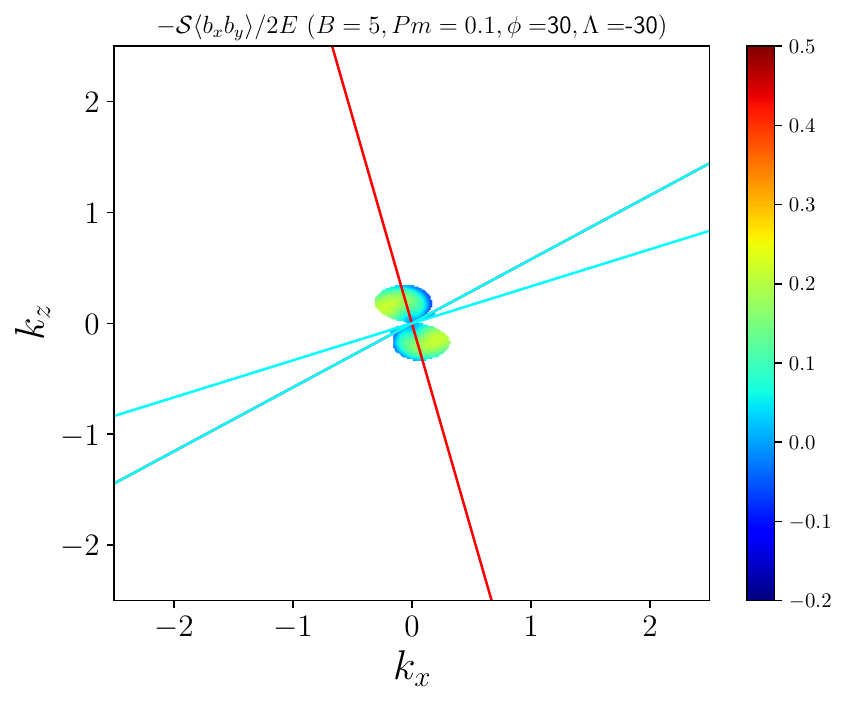}}
    
    \begin{picture}(0,0)
        \put(-242,76){\tiny${\Omega^{\perp}}$}
        \put(-70,76){\tiny$\Omega^{\perp}$}
        \put(101,76){\tiny$\Omega^{\perp}$}
        \put(-160,34){\tiny$\nabla{\ell}$}
        \put(10,34){\tiny${\nabla{\ell}}$}
        \put(180,34){\tiny$\nabla{\ell}$}
        \put(-113,125.5){\tiny$\boldsymbol{e}_g$}
        \put(56.2,125.5){\tiny$\boldsymbol{e}_g$}
        \put(227,125.5){\tiny$\boldsymbol{e}_g$}
        \put(-113,112){\tiny$\boldsymbol{e}_{\theta}$}
        \put(56.2,112){\tiny$\boldsymbol{e}_{\theta}$}
        \put(227,112){\tiny$\boldsymbol{e}_{\theta}$}
    \end{picture}

 \subfigure[]{\includegraphics[trim=0cm 0cm 0cm 0cm,clip=true,width=0.33\textwidth]{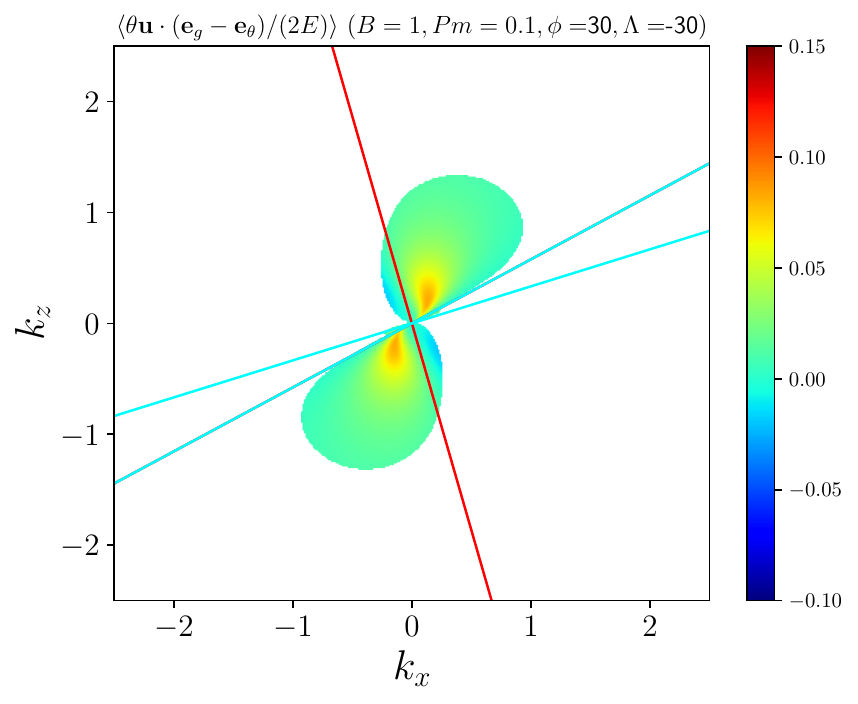}}
\subfigure[]{\includegraphics[trim=0cm 0cm 0cm 0cm,clip=true,
width=0.33\textwidth]{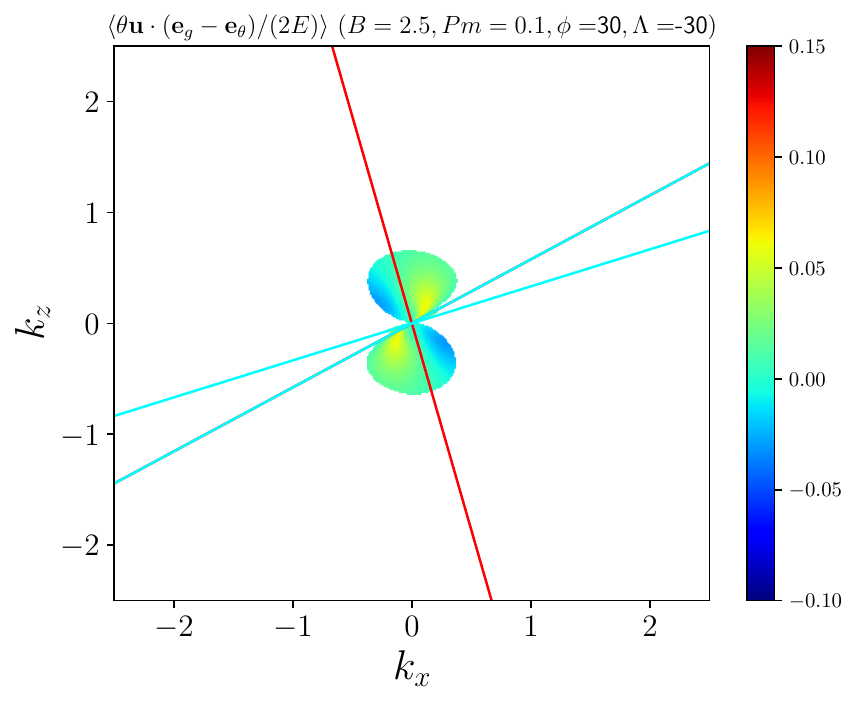}}
\subfigure[]{\includegraphics[trim=0cm 0cm 0cm 0cm,clip=true,
 width=0.33\textwidth]{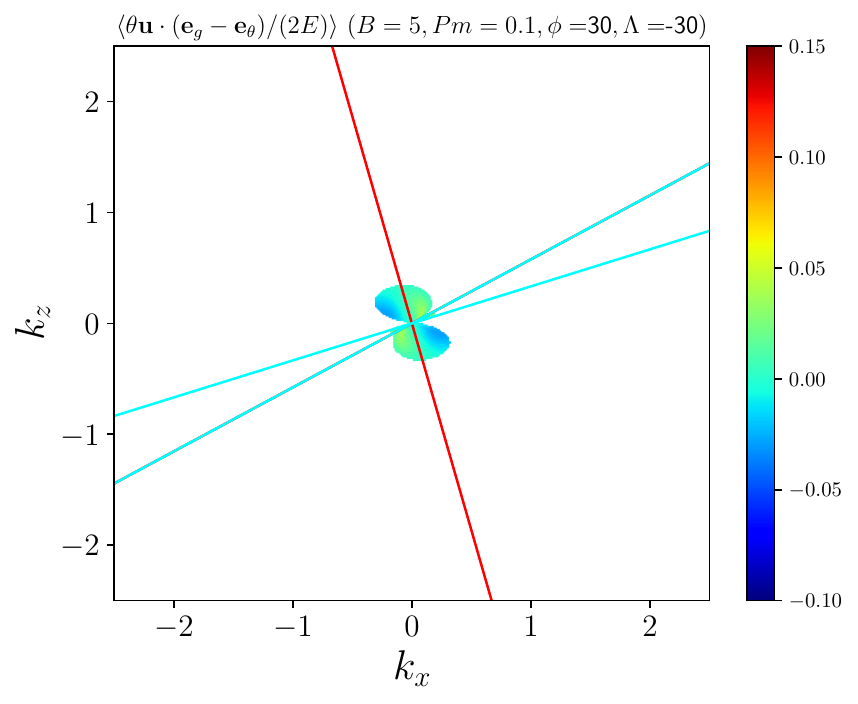}}
    
    \begin{picture}(0,0)
        \put(-242,76){\tiny${\Omega^{\perp}}$}
        \put(-73,76){\tiny$\Omega^{\perp}$}
        \put(97,76){\tiny$\Omega^{\perp}$}
        \put(-160,34){\tiny$\nabla{\ell}$}
        \put(10,34){\tiny${\nabla{\ell}}$}
        \put(180,34){\tiny$\nabla{\ell}$}
        \put(-116,125.5){\tiny$\boldsymbol{e}_g$}
        \put(55,125.5){\tiny$\boldsymbol{e}_g$}
        \put(225,125.5){\tiny$\boldsymbol{e}_g$}
        \put(-116,112){\tiny$\boldsymbol{e}_{\theta}$}
        \put(55,112){\tiny$\boldsymbol{e}_{\theta}$}
        \put(225,112){\tiny$\boldsymbol{e}_{\theta}$}
    \end{picture}
 
\caption{Energetic contributions to instability on the ($k_x,k_z$)-plane for $\Lambda=-30^\circ$ and $\phi=30^\circ$, $B_0=1,2.5$ and $5$ (increasing in columns as we go from left to right) all with $S=2$, $N^2=10$, $\mathrm{Pr}=0.01$ and $\mathrm{Pm}=0.1$. Top row: growth rate. Second row: Reynolds stress contribution. Third row: Maxwell stress contribution. Fourth row: baroclinic contribution. Stable modes with $\Re[s]\leq 0$ are indicated in white for clarity.}
\label{Energy_figs_Lam-30phi30}
\end{figure*}


\begin{figure*}

    \subfigure[]{\includegraphics[trim=0cm 0cm 0cm 0cm,clip=true,width=0.33\textwidth]{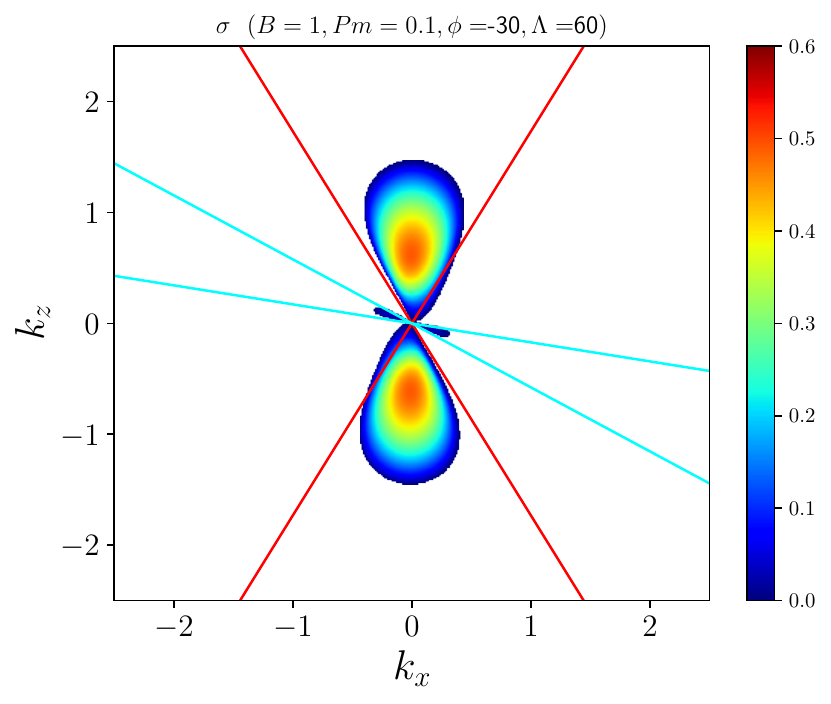}}
    \subfigure[]{\includegraphics[trim=0cm 0cm 0cm 0cm,clip=true,
    width=0.33\textwidth]{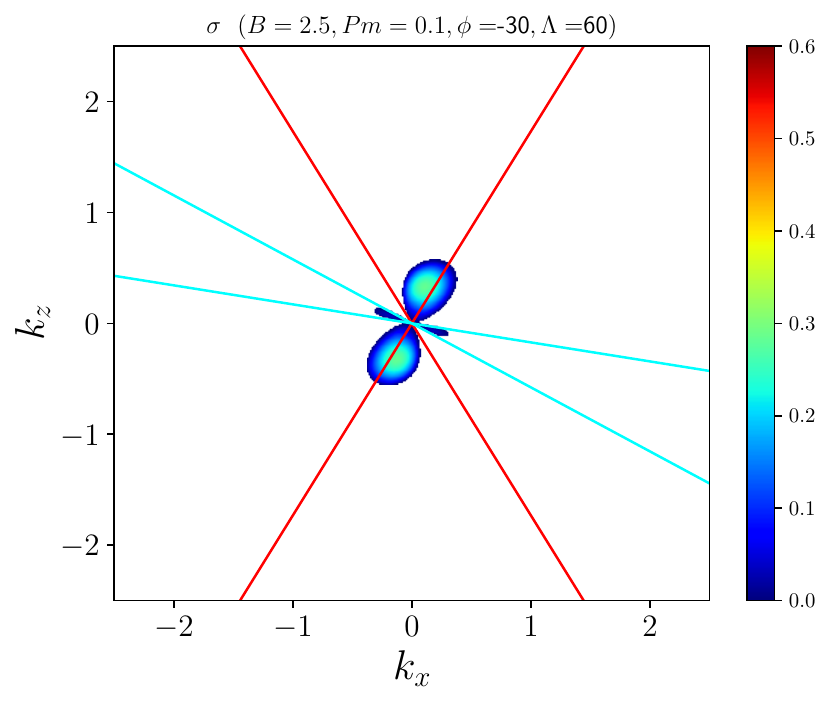}}
    \subfigure[]{\includegraphics[trim=0cm 0cm 0cm 0cm,clip=true,
    width=0.33\textwidth]{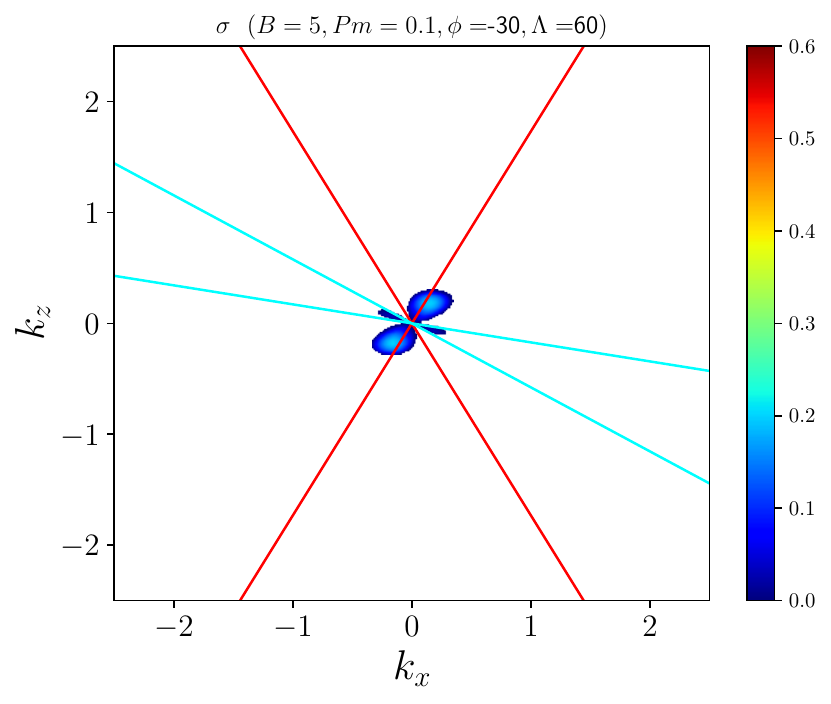}}
    
    \begin{picture}(0,0)
        \put(-140,34){\tiny${\Omega^{\perp}}$}
        \put(30,34){\tiny$\Omega^{\perp}$}
        \put(200,34){\tiny$\Omega^{\perp}$}
        \put(-210,34){\tiny$\nabla{\ell}$}
        \put(-40,34){\tiny${\nabla{\ell}}$}
        \put(130,34){\tiny$\nabla{\ell}$}
        \put(-108.,62){\tiny$\boldsymbol{e}_g$}
        \put(58.7,62){\tiny$\boldsymbol{e}_g$}
        \put(229.5,60){\tiny$\boldsymbol{e}_g$}
        \put(-108.,83){\tiny$\boldsymbol{e}_{\theta}$}
        \put(58.7,83){\tiny$\boldsymbol{e}_{\theta}$}
        \put(229.5,82){\tiny$\boldsymbol{e}_{\theta}$}
    \end{picture}

\subfigure[]{\includegraphics[trim=0cm 0cm 0cm 0cm,clip=true,width=0.33\textwidth]{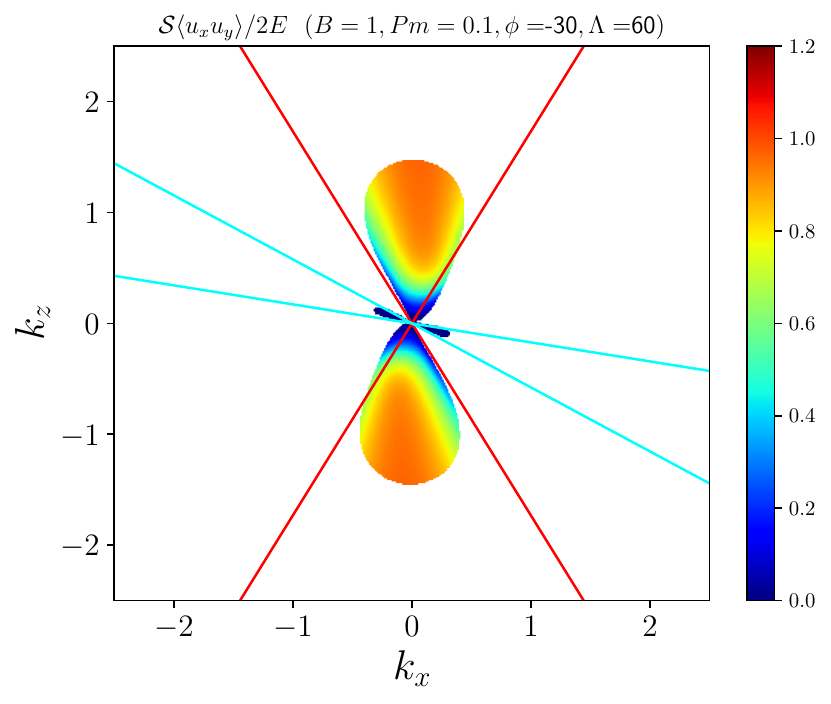}}
    \subfigure[]{\includegraphics[trim=0cm 0cm 0cm 0cm,clip=true,
    width=0.33\textwidth]{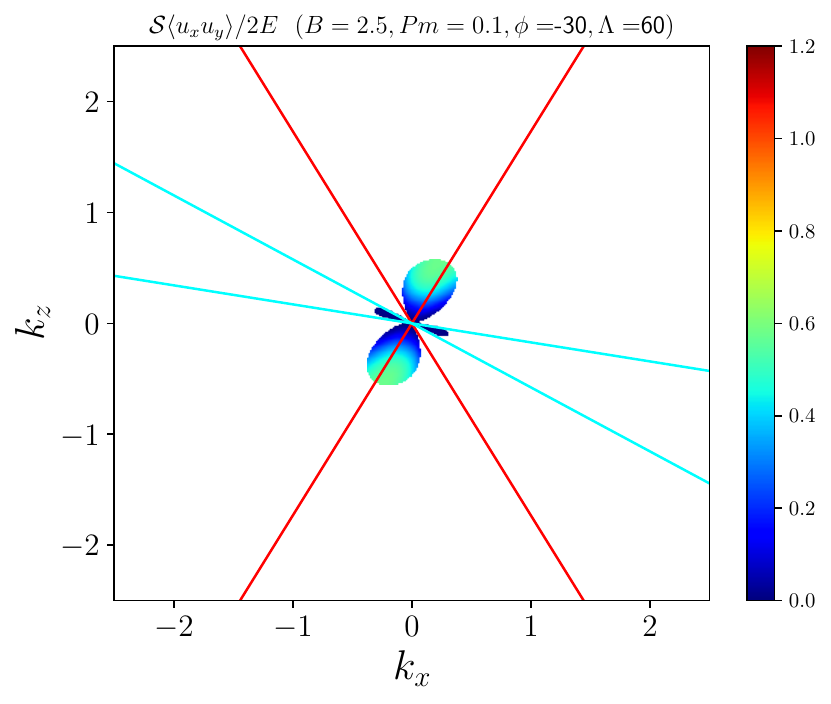}}
    \subfigure[]{\includegraphics[trim=0cm 0cm 0cm 0cm,clip=true,
    width=0.33\textwidth]{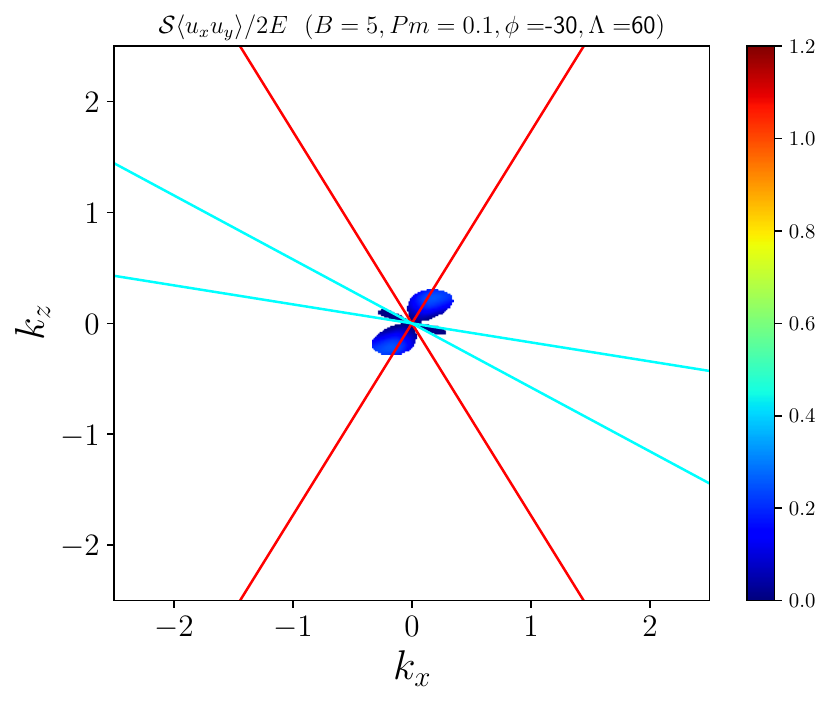}}
    
    \begin{picture}(0,0)
        \put(-140,34){\tiny${\Omega^{\perp}}$}
        \put(30,34){\tiny$\Omega^{\perp}$}
        \put(200,34){\tiny$\Omega^{\perp}$}
        \put(-210,34){\tiny$\nabla{\ell}$}
        \put(-40,34){\tiny${\nabla{\ell}}$}
        \put(130,34){\tiny$\nabla{\ell}$}
        \put(-111.2,62){\tiny$\boldsymbol{e}_g$}
        \put(60,62){\tiny$\boldsymbol{e}_g$}
        \put(228.5,60){\tiny$\boldsymbol{e}_g$}
        \put(-111.2,83){\tiny$\boldsymbol{e}_{\theta}$}
        \put(60,83){\tiny$\boldsymbol{e}_{\theta}$}
        \put(228.5,82){\tiny$\boldsymbol{e}_{\theta}$}
    \end{picture}
    
    \subfigure[]{\includegraphics[trim=0cm 0cm 0cm 0cm,clip=true,width=0.33\textwidth]{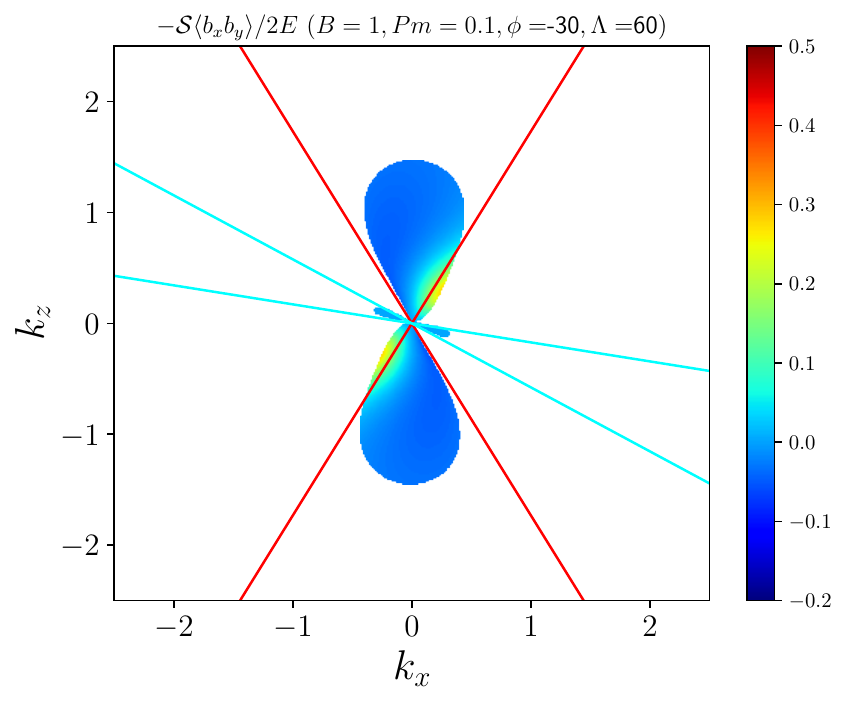}}
    \subfigure[]{\includegraphics[trim=0cm 0cm 0cm 0cm,clip=true,
    width=0.33\textwidth]{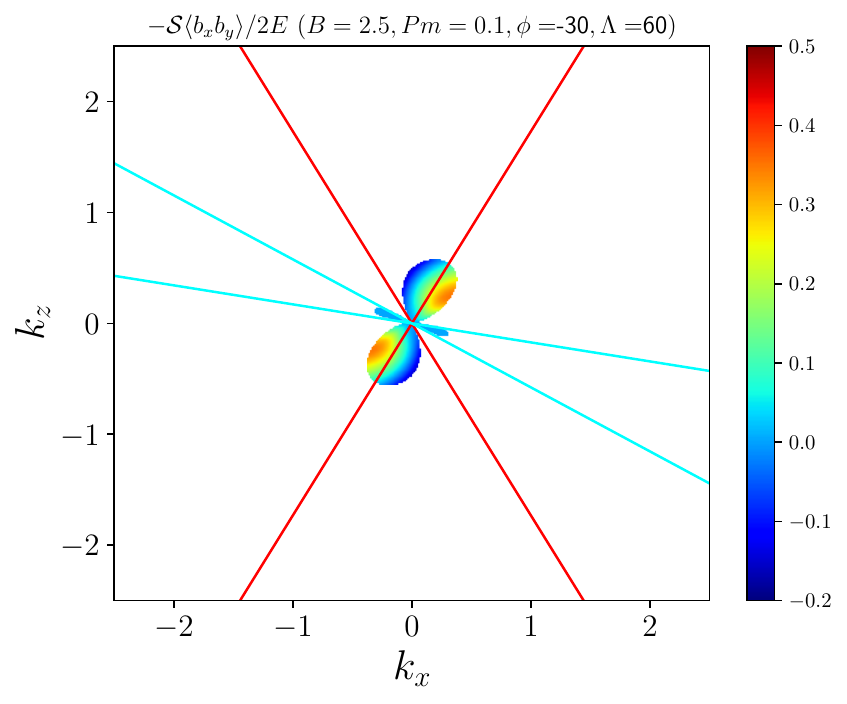}}
    \subfigure[]{\includegraphics[trim=0cm 0cm 0cm 0cm,clip=true,
    width=0.33\textwidth]{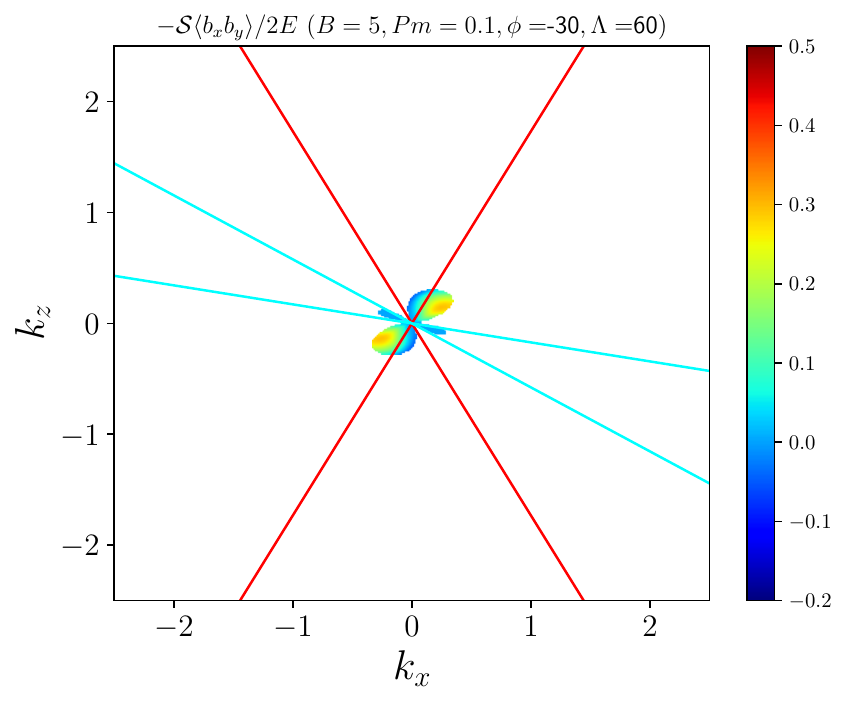}}
    
    \begin{picture}(0,0)
        \put(-140,34){\tiny${\Omega^{\perp}}$}
        \put(30,34){\tiny$\Omega^{\perp}$}
        \put(200,34){\tiny$\Omega^{\perp}$}
        \put(-210,34){\tiny$\nabla{\ell}$}
        \put(-40,34){\tiny${\nabla{\ell}}$}
        \put(130,34){\tiny$\nabla{\ell}$}
        \put(-112.8,62){\tiny$\boldsymbol{e}_g$}
        \put(57,62){\tiny$\boldsymbol{e}_g$}
        \put(226.5,60){\tiny$\boldsymbol{e}_g$}
        \put(-112.8,83){\tiny$\boldsymbol{e}_{\theta}$}
        \put(57,83){\tiny$\boldsymbol{e}_{\theta}$}
        \put(226.5,82){\tiny$\boldsymbol{e}_{\theta}$}
    \end{picture}

    \subfigure[]{\includegraphics[trim=0cm 0cm 0cm 0cm,clip=true,width=0.33\textwidth]{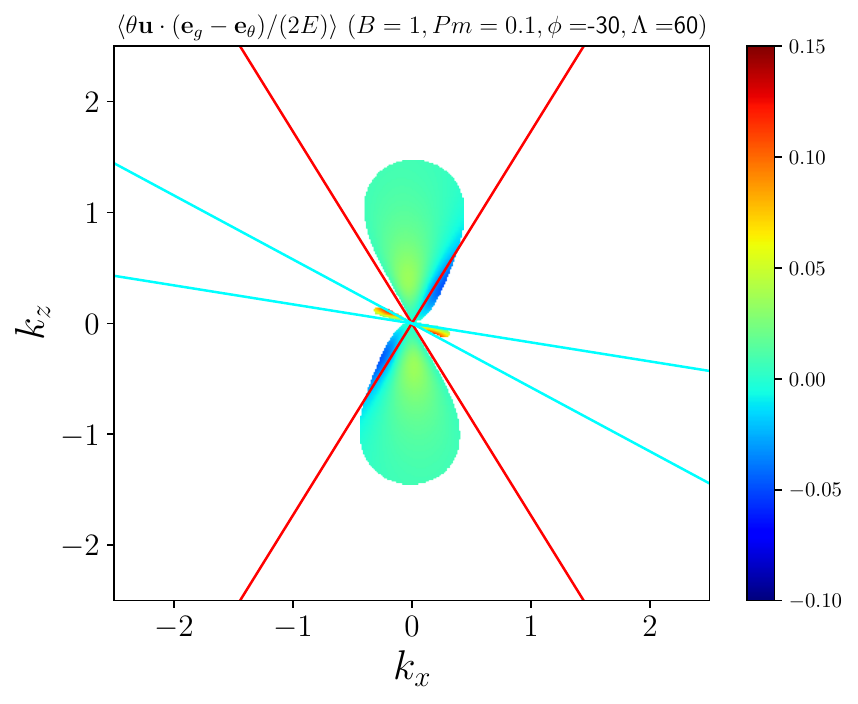}}
    \subfigure[]{\includegraphics[trim=0cm 0cm 0cm 0cm,clip=true,
    width=0.33\textwidth]{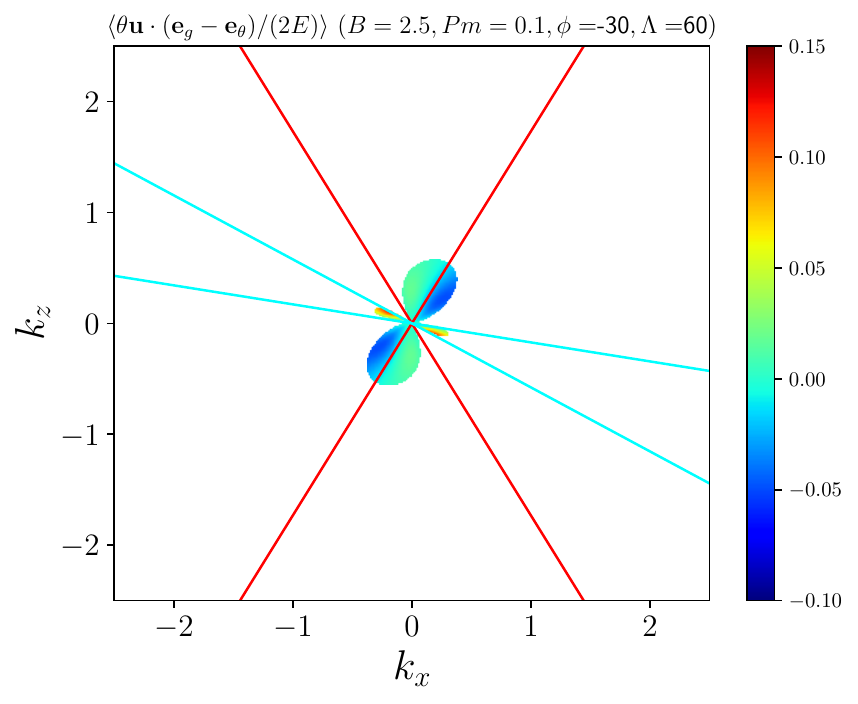}}
    \subfigure[]{\includegraphics[trim=0cm 0cm 0cm 0cm,clip=true,
    width=0.33\textwidth]{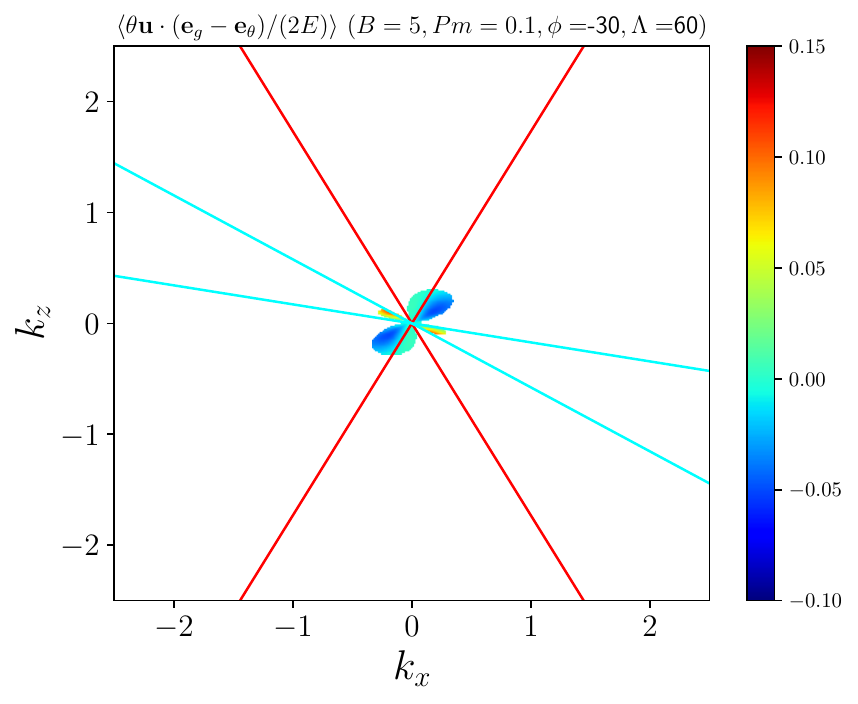}}
    
    \begin{picture}(0,0)
        \put(-140,30){\tiny${\Omega^{\perp}}$}
        \put(30,30){\tiny$\Omega^{\perp}$}
        \put(200,30){\tiny$\Omega^{\perp}$}
        \put(-210,30){\tiny$\nabla{\ell}$}
        \put(-40,30){\tiny${\nabla{\ell}}$}
        \put(130,30){\tiny$\nabla{\ell}$}
        \put(-114.7,56){\tiny$\boldsymbol{e}_g$}
        \put(55.4,56){\tiny$\boldsymbol{e}_g$}
        \put(225.4,56){\tiny$\boldsymbol{e}_g$}
        \put(-114.7,78){\tiny$\boldsymbol{e}_{\theta}$}
        \put(55.4,78){\tiny$\boldsymbol{e}_{\theta}$}
        \put(225.4,78){\tiny$\boldsymbol{e}_{\theta}$}
    \end{picture}
    
\caption{Energetic contributions to instability on the ($k_x,k_z$)-plane for $\Lambda=60^\circ$ and $\phi=-30^\circ$, $B_0=1,2.5$ and $5$ (increasing in columns as we go from left to right) all with $S=2$, $N^2=10$, $\mathrm{Pr}=0.01$ and $\mathrm{Pm}=0.1$. Top row: growth rate. Second row: Reynolds stress contribution. Third row: Maxwell stress contribution. Fourth row: baroclinic contribution. Stable modes with $\Re[s]\leq 0$ are indicated in white for clarity.}
\label{Energy_figs_Lam60phi-30}
\end{figure*}


\subsection{Numerical analysis of linear mode energetics}
\subsubsection{Unstable modes energetics: variation with $B_0$}

We present results from computing the contributions to the growth rate from the three source terms on the right hand side of Eq.~\ref{totalE}. In particular, we determine the contributions to the growth rate from the Reynolds stress, Maxwell stress, and baroclinic driving terms in Eqs.~\ref{RS}, \ref{MS} and \ref{baro_energy} as a visual tool to  understand better the mechanisms driving the various instabilities, as well as the role of the magnetic field. Each of these are divided by $2\mathcal{E}$ in order to compute their contribution to $\sigma$ for the reason explained in Eq.~\ref{Ereason}. All of the figures in this section use our standard choice of parameters, $\mathrm{Pr} = 10^{-2}$, $N^2 = 10$ and $S =2$ unless stated otherwise.

Figs.~\ref{Energy_figs_Lam-30phi30} and  \ref{Energy_figs_Lam60phi-30}
show pseudocolour plots for various $\Lambda$ and $\phi$ of the growth rate (first row) along with the contributions to it from Reynolds stresses (second row), Maxwell stresses (third row) and baroclinic source terms (fourth row) on the $(k_x,k_z)$-plane, for various magnetic field strengths $B_0=1, 2.5$ and $5$. Rows two to four represent the first three terms in Eq.~\ref{totalE}, the sum of these, together with the three diffusive terms (not plotted) in Eq.~\ref{totalE} has been verified to match the growth rate $\sigma$ to machine precision. In contrast to Figs.~\ref{Lobes1}--\ref{Lobes3} they use a linear colour scale since the various contributions plotted can take either sign, as we observe in these figures. Overall, these figures allow us to explore how variations in field strength for $(B_0=1,2.5,5)$ and rotation profile (through $\Lambda$ and $\phi$) alter the instabilities whilst simultaneously probing which energy source terms are responsible. 

In Figs.~\ref{Energy_figs_Lam-30phi30} we first analyse the configuration at the equator with mixed shear ($\Lambda=-30~^\circ$, $\phi=30^\circ$) explored earlier in Fig.~\ref{Lobes1}. This configuration is GSF unstable in the hydrodynamic case and remains unstable for weak fields. Strong fields tend to inhibit instability for $k\sim 1$ and to shrink the unstable lobes, in addition to changing their orientation. For $B_0\leq 2.5$, Reynolds stresses are the primary drivers of instability for most $(k_x,k_z)$, indicating that unstable modes are primarily driven by extracting kinetic energy from the differential rotation. As $B_0$ is increased further, Maxwell stresses play an increasingly important role, until they dominate for $B_0=5$, indicating that shear flow kinetic energy is extracted and input into perturbation magnetic energy. The different locations of the peaks in Reynolds and Maxwell stresses -- and the increasingly stabilising effects (negative values shown) of Maxwell stresses where the Reynolds stresses are maximal -- are consistent with the changes in orientations of the unstable lobes as $B_0$ is increased, from initially being between $\hat{\boldsymbol{\Omega}}^\perp$ and $\nabla \ell$ to become closer to $\nabla \ell$ for the strongest fields. For this latitude and flow baroclinic driving terms are typically subdominant, but they still contribute non-negligibly to driving instabilities for weaker fields. The effect of the field in reducing the maximum growth rate observed in Fig.~\ref{matlablat0} is also confirmed here.

We next look at a case with latitude $30^\circ$ with mixed shear $(\Lambda =60^\circ, \phi =-30^\circ)$ in Fig.~\ref{Energy_figs_Lam60phi-30} as first studied in Fig.~\ref{Lobes2}. This configuration is GSF unstable hydrodynamically and remains unstable for weak fields. We saw from Fig.~\ref{matlablat30} that the field acts to monotonically stabilise the system with increasing $B_0$, which is consistent with Fig.~\ref{Energy_figs_Lam60phi-30}. We again observe that the primary lobes of instability are driven primarily by Reynolds stresses for $B_0\leq 2.5$, but become increasingly driven by Maxwell stresses for stronger fields. We also observe the positive Reynolds stress contributions are mainly confined to within the hydrodynamically unstable wedge delineated by the lines $\hat{\boldsymbol{\Omega}}^{\perp}$ and $\nabla{\ell}$, and are maximal approximately halfway between these. The increasing importance of Maxwell stresses and the shift in orientation of the lobes indicates the transition in the dominant instability from GSF to MRI. Notice that the Maxwell stress generally has a preferred wavevector magnitude, evident by the darkest red (most unstable) modes being located in the centre of the lobes. We also observe the unstable region shrinking as the MRI enables instability for smaller and smaller $k$ for appropriately oriented modes. The baroclinic term is unimportant for the primary lobes, as is indicated by the bottom panels.

The secondary lobes evident in Fig.~\ref{Energy_figs_Lam60phi-30} are hydrodynamically unstable oscillatory modes within the wedge defined by $\boldsymbol{e}_g$ and $\boldsymbol{e}_\theta$. The bottom panels of this figure confirm that these modes are baroclinically driven since $\sigma$ approximately equals its baroclinic contribution, with Reynolds and Maxwell stresses playing negligible roles in driving them. The growth rates and unstable mode wavevectors are mostly unaffected by the magnetic field, except that these become weakly destabilised magneto-inertial-gravity waves rather than inertia-gravity waves when the field is sufficiently strong.

We have found similar trends as $B_0$ is varied are found for $\Lambda$ and $\phi$ that are adiabatically Solberg-H\o iland unstable in the hydrodynamic case, and for cases at the poles that are hydrodynamically adiabatically stable. 

\begin{figure*} 
    \subfigure[]{\includegraphics[trim=0cm 0cm 0cm 0cm,clip=true,
    width=0.49\textwidth]{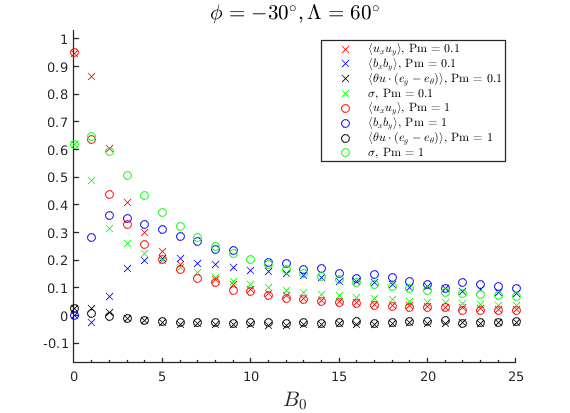}}
    \subfigure[]{\includegraphics[trim=0cm 0cm 0cm 0cm,clip=true,
    width=0.49\textwidth]{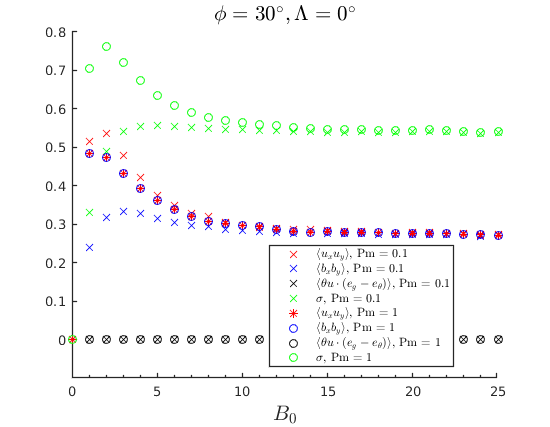}}
    
    \subfigure[]{\includegraphics[trim=0cm 0cm 0cm 0cm,clip=true,
    width=0.49\textwidth]{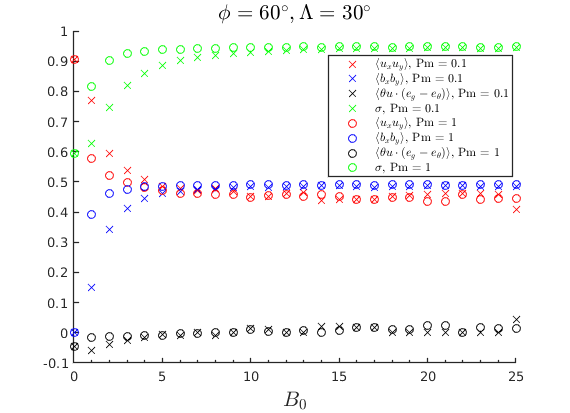}}
    \subfigure[]{\includegraphics[trim=0cm 0cm 0cm 0cm,clip=true,
    width=0.49\textwidth]{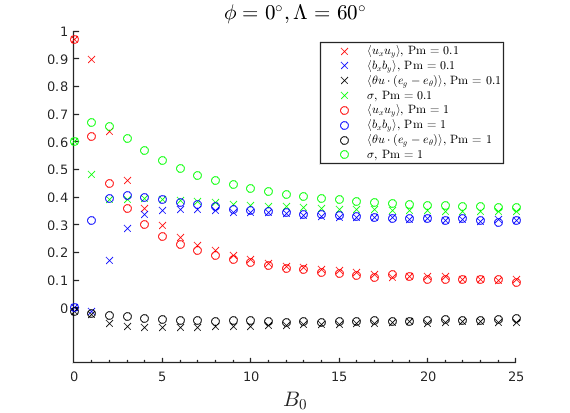}}

    \subfigure[]{\includegraphics[trim=0cm 0cm 0cm 0cm,clip=true,
    width=0.49\textwidth]{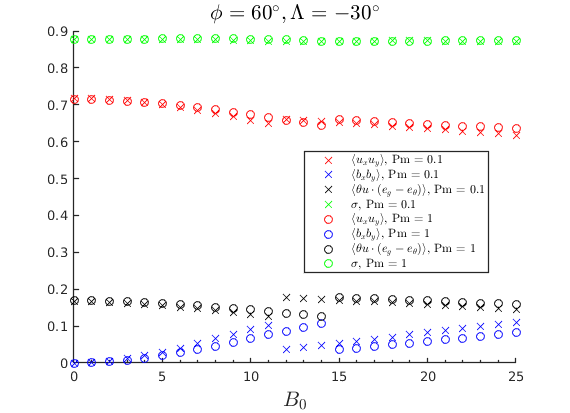}}
    \subfigure[]{\includegraphics[trim=0cm 0cm 0cm 0cm,clip=true,
    width=0.49\textwidth]{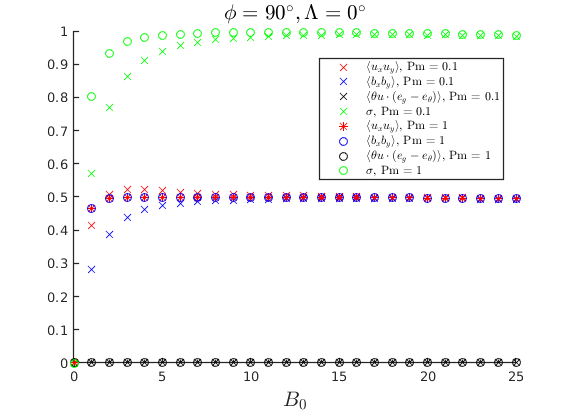}}

  \caption{Energetic contributions to instability for the fastest growing mode (optimised over $k_x$ and $k_z$) as a function of the magnetic field strength $B_0$ for various $\Lambda$ and $\phi$ cases. These show 
  the growth rate, and the contributions to it from Reynolds stresses ($\langle u_{x}u_{y}\rangle$), Maxwell stresses ($\langle b_{x}b_{y}\rangle$) and baroclinic source terms ($\langle \theta \boldsymbol{u} \cdot (\boldsymbol{e}_g - \boldsymbol{e}_{\theta}) \rangle$ against field strength $B_0$. All panels show $\mathrm{Pm}=0.1$ and $\mathrm{Pm}=1$, and the other parameters are  $\mathrm{Pr}=10^{-2}$, $\mathcal{N}^2=10$ and $\mathcal{S}=2$.}
  \label{Energy_figs_vsB0}
\end{figure*}

\subsubsection{Fastest growing mode energetics: variation with $B_0$}

We next turn to analyse how the energetic contributions vary with $B_0$ for the fastest growing modes, obtained by optimising over $k_x$ and $k_z$ for each case. Results are shown in Fig.~\ref{Energy_figs_vsB0} for various latitudes and differential rotations. We study both $\mathrm{Pm}=1$ and  $\mathrm{Pm}=0.1$ in order to investigate the role magnetic diffusivity plays in these results.

Panel (a) of Fig.~\ref{Energy_figs_vsB0} analyses a case with $\phi=60^\circ$, $\Lambda = -30^\circ$ that is adiabatically Solberg-H\o iland unstable in the hydrodynamic case. We find the growth rate in this case is essentially independent of $B_0$, as predicted from Figs.~\ref{adiMATLAB} and ~\ref{matlablat30}. The primary result of changing $B_0$ is to decrease the range of unstable $k_x$ and $k_z$ as we have confirmed in Fig.~\ref{Lobes3}.
This case is driven by the Reynolds stress for all $B_0$ considered, since the red symbols provide a larger contribution to the total growth rate, for both $\mathrm{Pm}$ plotted. Magnetic diffusion does not play an important role here, confirmed by the negligible role of $\mathrm{Pm}$. The baroclinic driving term is the secondary contributor to instability for the smaller $B_0$, but it appears that it may be superseded for $B_0\gtrsim 25$ by Maxwell stresses. There is a jump from one unstable mode to another around $B_0=12$ where baroclinic and Maxwell stress terms approximately balance, where the $k$ and $\theta_k$ values instantly switch.

Panel (b) of Fig.~\ref{Energy_figs_vsB0} analyses a case with cylindrical differential rotation ($\Lambda=0$, which is neutrally hydrodynamically GSF stable for all latitudes with $S=2$. This case is significantly destabilised by even weak magnetic fields as we have seen in Figs.~\ref{matlablat30}, \ref{matlablat60} and \ref{matlablat90}, due to the MRI. This instability is driven by an approximately equal combination of Reynolds and Maxwell stresses, which perfectly coincide for large $B_0$. This can be termed ``Alfv\'{e}nisation'' of the instability for sufficiently strong fields. The complete lack of any baroclinic driving is evident in this figure, and is also observed in panel (f) which also has $\Lambda =0$, as is expected for any cylindrical rotation profile.

Panel (c) probes instabilities at the poles by considering $\phi=60^\circ$ and $\Lambda =30^\circ$. Figs.~\ref{matlabadi} and \ref{matlablat90} indicated that this latitude is widely unstable to adiabatic magnetic instabilities, in stark comparison to the hydrodynamic results \cite{Dymott2023}, which found no adiabatic instability there. This instability is again the MRI, and it is driven by an approximately equal balance of Reynolds and Maxwell stresses indicating ``Alfv\'{e}nisation'' once again. The growth rate increases by around $35\%$ between $B_0=0$ and $B_0=5$, after which increases in $B_0$ lead to only marginal increases in $\sigma$.

Panel (d) explores a shellular rotation profile with $\phi=0$ and $\Lambda=60^\circ$. These cases were explored hydrodynamically in paper 2.
This figure indicates that the instability is initially driven almost entirely by Reynolds stresses when $B_0\sim 0$, but Maxwell stresses dominate for $B_0\gtrsim 5$. The introduction of magnetic fields weakens the instability and reduces $\sigma$ (after a small rise for $B_0\sim 1$) over the $B_0=0$ case. A plateau is reached for $\sigma$ by $B_0\gtrsim 15$, where the instability is primarily driven by Maxwell stresses. Once again, the baroclinic driving term is very weak in this case for any $B_0$.

Panel (e) shows the behaviour of the fastest growing mode from the parameters of Fig.~\ref{Energy_figs_Lam60phi-30} with $\phi=-30^\circ$ and $\Lambda=60^\circ$. As $B_0$ is increased $\sigma$ is drastically reduced. Up to $B_0\approx5$ for $\mathrm{Pm}=0.1$, and $B_0=3$ for $\mathrm{Pm}=1$, Reynolds stresses are the dominant contributor to the growth rate, but as the growth rate decreases with increasing field strength Maxwell stresses become the dominant contributor with these lines converging towards each other. We may achieve ``Alfv\'{e}nisation'' again for sufficiently large $B_0$, but this is not observed by $B_0=25$. As was observed in Fig.~\ref{Lobes2} the $\mathrm{Pm}=1$ case is consistently more unstable than the $\mathrm{Pm}=0.1$ case, however as the growth rate tends to zero this difference becomes marginal.

In this section we have analysed the unstable mode energetics as $B_0$, Pm and the properties of the differential rotation were varied. We have found that the fastest growing modes are always driven predominantly by a combination of Reynolds and Maxwell stresses for non-zero $B_0$ and that baroclinic driving is negligible except for the subdominant secondary lobes. For strong enough magnetic fields, in many cases in which the MRI operates, the contributions of Reynolds and Maxwell stresses equalise. Overall, these results confirm that even a weak magnetic field can drastically alter the stability of differentially rotating flows in stellar radiation zones.

\section{Applications to the Sun and red giant stars}
\label{Applications}

We now turn to estimate parameter values for the solar tachocline as a potential application of this work. Recall that we defined our lengthscale $d$ as
\begin{align}
    d = \left(\frac{\nu \kappa}{\mathcal{N}^2}\right)^\frac{1}{4},
\end{align}
since this describes the scales of the dominant hydrodynamic GSF modes. In the solar tachocline \citep[e.g.][]{Gough2007,Caleo2016}, we find $\nu =2.7 \times 10 ^1 \mathrm{cm}^2 \mathrm{s}^{-1}$, $\kappa=1.4 \times 10 ^7 \mathrm{cm}^2 \mathrm{s}^{-1}$, hence $\mathrm{Pr}=2\times 10^{-6}$ and $\mathcal{N}= 8 \times 10^{-4} \mathrm{s}^{-1}$. This produces a length scale\footnote{Please note the unfortunate typo in paper 1, where this was written as km instead! No other values in paper 1 need modifying and slightly different numbers were used from stellar models for the various parameters there than the ones we quote here.} $d \approx 49.3 \mathrm{m}$. The linear GSF modes thus have very short length-scales approx $10^{-5}$ times smaller than the tachocline thickness. The dimensional wavenumber $k_{dim} = k/d$, using our dimensionless wavenumber $k$. Note that $\eta=4.1\times 10^2\mathrm{cm}^2\mathrm{s}^{-1}$ in the tachocline, so $\mathrm{Pm}=0.065$ and $\mathrm{Pr}/\mathrm{Pm}=3\times 10^{-5}$ there. Hence, we are in the regime of rapid thermal diffusion relative to viscous and ohmic diffusion in the tachocline, as we discussed in \S~\ref{diffusiveanalysis} and \ref{diffusiveanalysis2}.

The magnetic field strength and structure in the tachocline is highly uncertain. Nevertheless, any poloidal magnetic field is probably in the range $0.5 \mathrm{G}$ to $5\mathrm{kG}$ \citep[e.g.][and we are not aware of substantially stronger subsequent constraints]{MW1987}. The field there is likely to be mostly toroidal, but only poloidal fields enter our stability analysis for axisymmetric modes. The arguments of \citet{GM1998} for the maintenance of the tachocline also suggest a minimum poloidal field of $1$ G is required there.

Our dimensionless magnetic field $\boldsymbol{B}$ is written in Alfv\'{e}n speed units; therefore it has units $d\,\Omega$ where $\Omega = 2\pi /P_{\mathrm{rot}}$, and $P_{\mathrm{rot}}=27$ days is the Sun's mean rotation period. The corresponding physical magnetic field magnitude $B_{dim}$ from the dimensional Alfv\'{e}n speed $V_A = B_{dim}/\sqrt{\mu_0 \rho}$
\begin{align}
    B_{dim} = B_0 d \Omega \sqrt{\mu_0 \rho}, \approx 2.1 \times 10^{-6} B_0 \,\mathrm{T} \approx 0.021 B_0 \,\mathrm{G},
\end{align}
using $\Omega=2.7 \times 10^{-6} \mathrm{s}^{-1}$ (implying $\mathcal{N}^2/\Omega^2\approx 8.7\times 10^4$), $\rho = 210 \,\mathrm{kg} \mathrm{m}^{-3}$ and $\mu_0 = 4\pi \times 10^{-7}$ in SI units. This means that a field of 1 G corresponds to a dimensionless $B_0\approx 46$ in our units if $d$ is the relevant length-scale. Note that $d$ was defined based on the diffusive hydrodynamic GSF modes, and we have found the MRI to potentially have much larger wavelengths. However, the small length scales which GSF modes prefer does make them very vulnerable to even a rather weak magnetic field.

On the other hand if we want to consider a field of 1 kG, this requires $B_0=4.6\times 10^4$ in dimensionless units, which is much larger than we have so far considered here. The fields we have primarily explored in this work are at the weaker end, with $B_0\lesssim 25$, corresponding to fields weaker than approximately 0.5 G in the tachocline. This choice was partly made to permit us to explore the modification of hydrodynamic diffusive rotational instabilities by a weak field, and was partly made because we found that for larger $B_0$ the GSF mode is primarily stabilised and the dominant instability by far is the  MRI.

\begin{figure*}
    \subfigure[$\sigma$ at $\Lambda + \phi = 30^\circ$]{\includegraphics[trim=0cm 0cm 1.3cm 0.88cm,clip=true,width=0.32\textwidth]{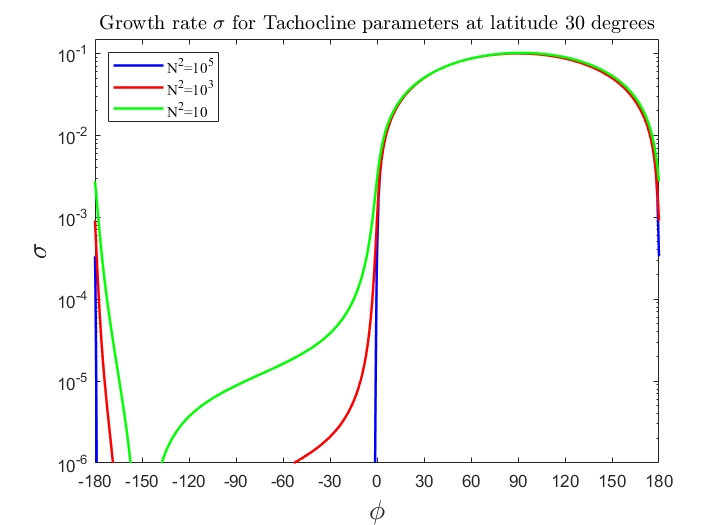}}
    \subfigure[$|k|$ at $\Lambda + \phi = 30^\circ$]{\includegraphics[trim=0cm 0cm 1.3cm 0.895cm,clip=true,width=0.32\textwidth]{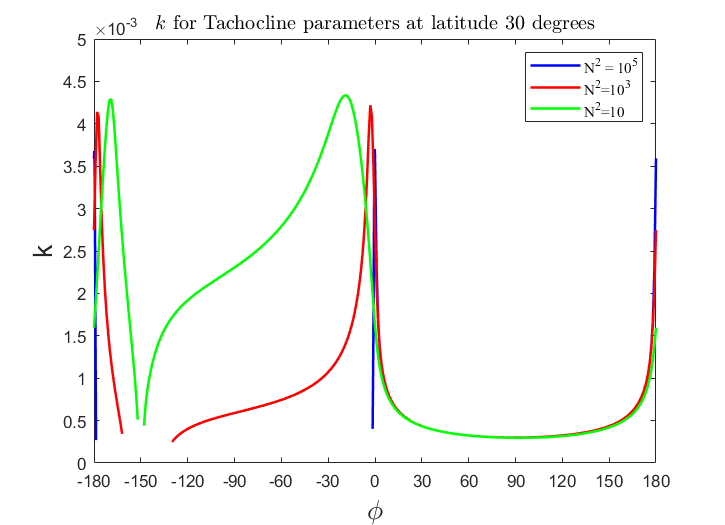}}
    \subfigure[$\theta_k$ at $\Lambda + \phi = 30^\circ$]{\includegraphics[trim=0cm 0cm 1.3cm 0.87cm,clip=true,width=0.32\textwidth]{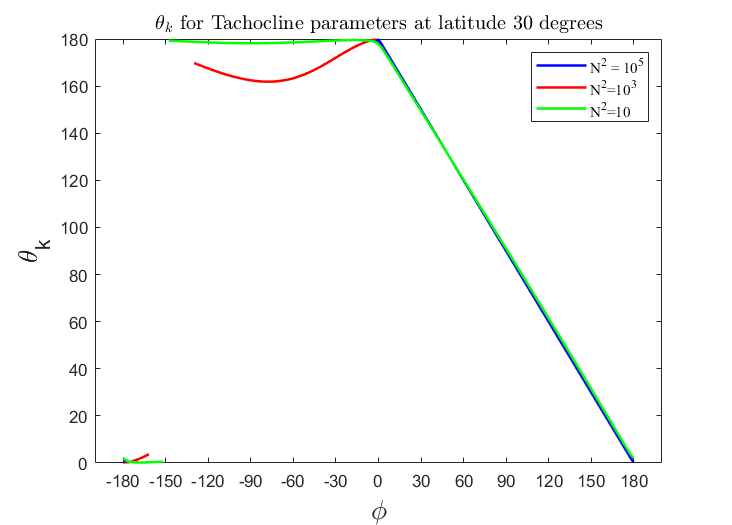}}
  \caption{Properties of the fastest growing modes for tachocline parameter values at a latitude $\beta=\Lambda+\phi=30^\circ$.
We assume $B_0=1000$, $S=0.2$, $\mathrm{Pm}=0.05$, $\mathrm{Pr}=10^{-6}$, and explore $N^2\in[10,10^3,10^5]$.
Note the growth rate is on a log scale, because for $\phi>0$ we get MRI that grows much faster than the GSF modes that may exist for $\phi<0$.}
\label{tacho}
\end{figure*}

Motivated by the values in the solar tachocline, we compute the linear growth rates, wavenumbers and orientations numerically and display them on Fig.~\ref{tacho} at a latitude of $30^\circ$ for a moderately strong field with $B_0=10^3$ for the solar-like values $S=0.2$, $\mathrm{Pm}=0.05$ and $\mathrm{Pr}=10^{-6}$ (so that $\mathrm{Pr}/\mathrm{Pm}=2\times 10^{-5}$). We consider three different values of  $N^2\in[10,10^3,10^5]$ to account for the variation of values in the solar radiation zone as we approach the radiative-convective boundary. Deep down, the higher value is appropriate \citep[and corresponds to the value in][]{Gough2007}, but $N^2$ passes through zero as the convection zone is approached, motivating the smaller values we consider here \citep[see e.g.~Fig.~1 in][]{BO2010}. We observe that MRI operates for $\phi>0$ with modest growth rates $\sigma\sim 0.1$ and $k\sim 0.0005$ (corresponding to wavelengths of order $600-60,000$ km, much smaller than the tachocline thickness), and that it is not affected by variations in $N^2$. This is because the unstable modes orient themselves to avoid doing work against gravity such that $b\sim 0$. On the other hand, the instabilities  for $\phi<0$ are substantially modified by varying $N^2$. These instabilities are GSF modes inhibited by magnetic tension. For $N^2=10^5$, relevant for deeper parts of the solar tachocline and radiative interior, GSF is eliminated by the magnetic field. This agrees with some of the conclusions of \cite{Caleo2016,Caleo2016a}. However, closer to the radiative/convective interface in the tachocline region itself, our smaller values of $N^2$ are appropriate. For $N^2=10$ and $10^3$, the GSF instability operates but with a much weaker growth rate than the MRI modes in operation when $\phi>0$. Fig.~\ref{tacho} suggests that MRI may be more important than GSF for turbulent transport in the solar radiative interior whenever $\phi>0$, but that more weakly growing GSF modes could operate for local rotation profiles with $\phi<0$.

In the core of red giant stars, whose core-envelope differential rotations remain poorly understood, as considered in paper 1 and using the numbers there, $d\sim 100 \mathrm{m}$, $\Omega\sim 10^{-7}\mathrm{s}^{-1}$. This produces $B_{dim}\sim 1.12 B_0 \sqrt{\rho/10^5 \mathrm{gcm}^{-3}}$ G. Hence in that problem $B_{dim}\sim B_0$ G in the cores of red giant stars. Since there have been constraints on fields in these from asteroseismology of order 40 to 610 kG \citep{RGmag2023}, this suggest we should consider $B_0\gtrsim 10^3$ in red giant stars also. Hence, MRI is expected to be more important than GSF, depending on the rotation profile (particularly for $\phi>0$), but perhaps not for $\phi<0$.

\section{Conclusions}
\label{Conclusions}

We have presented a comprehensive theoretical analysis of local diffusive instabilities of differential rotation in magnetised radiation zones of stars and planets, building upon the hydrodynamical studies of \citet{barker2019,barker2020,Dymott2023}. Understanding the properties of these instabilities, and ultimately their nonlinear behaviour, is essential because they have been proposed to play important roles in angular momentum transport and chemical mixing in stars \citep[e.g.][]{Caleo2016,aerts2019}, and they may even play a role in the solar dynamo \citep{ParfreyMenou2007,V2024}, but many aspects of them are currently very poorly understood. Our focus has been on the effects of a poloidal magnetic field on the properties of linear axisymmetric instabilities of differential rotation, which are governed by a quintic dispersion relation first derived by \cite{Menou2004}. We have performed a detailed analysis of the dispersion relation, firstly for non-diffusive instabilities, reproducing prior work on the stratified MRI \citep[e.g.][]{Balbus1995}, before comprehensively analysing diffusive instabilities in various limits analytically and numerically \citep[see also][]{Caleo2016,Caleo2016a}. 

In strongly stably stratified regions of stars, the fastest growing mode displacements are along stratification (i.e.~approximately spherical) surfaces and correspond with operation of the MRI. However, rapid thermal diffusion can eliminate the stabilising effects of buoyancy if $\mathrm{Pr}/\mathrm{Pm}$ and $\mathrm{Pr}$ are sufficiently small. In this limit
MRI operates and can change the properties of the unstable modes depending on the differential rotation. We have obtained new analytical and numerical results on the various instabilities in this triply-diffusive system as a function of the differential rotation profile and magnetic field strength.

Our analytical and numerical results have highlighted that even a weak magnetic field can considerably modify the local instabilities of differentially rotating flows \citep[e.g.][and many prior works]{BH1998}. We have found that for differential rotations with (angle from the local angular velocity gradient to the effective gravity direction) $\phi>0$ in the northern hemisphere (and vice versa in the southern hemisphere because the relevant parameter is the sign of $\beta \phi$), MRI may dominate over the magnetic modification of hydrodynamic GSF instabilities. However, for $\phi<0$ there, hydrodynamic GSF modes could still be important though they are weakened by magnetic tension for moderately strong fields. We found that even weak fields destabilise hydrodynamically stable regions in parameter space, particularly for nearly cylindrical differential rotation profiles.

We have analysed in detail the properties of axisymmetric modes, including how the growth rates and wavevectors depend on the strength of the magnetic field, magnetic Prandtl number Pm, and local differential rotation profile. We have analysed in detail the energetics of the various instabilities in our system, first by deriving the energy equation and then by evaluating the various source terms for linear axisymmetric modes. These consist of Reynolds stresses, Maxwell stresses and baroclinic driving terms. We find that the MRI is typically driven by Reynolds and Maxwell stresses in approximately equal proportions (so-called ``Alfv\'{e}nisation'') in a wide range of cases.

We believe that it is important to set up a meaningful time-independent magnetic equilibrium to properly analyse MHD instabilities. We take a different viewpoint to many prior works that attempted to model arbitrary field configurations without ensuring Ferraro's law of isorotation was satisfied \citep[e.g.][]{Balbus1994, Menou2004, Menou2006, ParfreyMenou2007, Caleo2016}. In our model we ensured our basic state was an equilibrium state and verified the local analogue of Ferraro's law of isorotation. This is analogous to the original works of \citet{GSF,Fricke1968} having an additional degree of freedom because they ignored the constraint of thermal wind balance \citep[e.g.][]{Acheson1978,Busse1981}. Similar issues have also plagued studies of the effects of magnetic fields on the vertical shear instability in astrophysical discs \citep[e.g.][in which the latter authors take the same viewpoint as us]{UB1998,LatterPap2018}.

Future work should explore the nonlinear evolution of the instabilities we have analysed here \citep[building upon][]{barker2019,barker2020,Dymott2023,Bindesh2024}, as well as the role of compositional gradients on both the linear \citep{KS1983} and nonlinear properties of this problem. Further global simulations tailored to study these instabilities would also be worthwhile.

\section*{Acknowledgements}
RWD was supported by an STFC studentship (2443617). AJB was supported by STFC grants ST/S000275/1 and ST/W000873/1. SMT would like to acknowledge support of funding from the European Union Horizon 2020 research and innovation programme (grant agreement no. D5SDLV-786780). Additionally, we would like to thank the Isaac Newton Institute for Mathematical Sciences, Cambridge, for support and hospitality during the programmes DYT2 and “Anti-diffusive dynamics: from sub-cellular to astrophysical scales”, supported by EPSRC grant no EP/R014604/1. We would like to thank the referee for a constructive and timely report.

\section*{Data Availability}
The data underlying this article will be shared on reasonable request to the corresponding author.


\bibliographystyle{mnras}
\bibliography{mybib}


\bsp
\label{lastpage}
\end{document}